\documentclass[12pt, a4paper]{article}
\pdfoutput=1
\usepackage{amsmath}
\usepackage{amsfonts}
\usepackage{amssymb}
\usepackage{latexsym}
\usepackage{mathrsfs,cite}
\usepackage{twistor}
\usepackage{graphicx}
\usepackage[colorlinks=true, linkcolor=black, citecolor=black, urlcolor=blue, linktocpage=true]{hyperref}

\newcommand{\V}{\, V}
\newcommand{\cR}{\mathcal{R}}
\renewcommand{\d}{\mathrm{d}}
\newcommand{\sA}{\mathscr{A}}
\newcommand{\odel}{\bar{\delta}}
\newcommand{\ie}{i\!-\!1}
\newcommand{\je}{j\!-\!1}

\makeatletter
\let\@keywords\@empty
\let\@subject\@empty
\providecommand{\keywords}[1]{\gdef\@keywords{#1}}
\providecommand{\subject}[1]{\gdef\@subject{#1}}
\def\thetitle{\@title}
\def\theauthor{\@author}
\def\thesubject{\@subject}
\def\thedate{\@date}
\def\thekeywords{\@keywords}
\usepackage[font=small,labelfont=bf,width=\textwidth]{caption}

\numberwithin{equation}{section}
\let\oldbfseries=\bfseries
\let\oldmdseries=\mdseries
\let\oldnormalfont=\normalfont
\renewcommand{\bfseries}{\oldbfseries\boldmath}
\renewcommand{\mdseries}{\oldmdseries\unboldmath}
\renewcommand{\normalfont}{\oldnormalfont\unboldmath}

\title{Scattering Amplitudes and Wilson Loops\\ in Twistor Space}   
\author{Tim Adamo$^1$, Mathew Bullimore$^2$, Lionel Mason$^1$ and David Skinner$^3$}

\begin{document}

\thispagestyle{empty}

\begin{center}%
\begingroup\Large\bfseries\thetitle\par\endgroup
\vspace{1cm}%

\begingroup\scshape\theauthor\par\endgroup
\vspace{5mm}%

\begingroup\itshape
$^1$The Mathematical Institute,\\
24-29 St. Giles',\\
Oxford OX1 3LB, United Kingdom
\vspace{3mm}

$^2$Rudolf Peierls Centre for Theoretical Physics,\\
1 Keble Road,\\
Oxford OX1 3NP, United Kingdom
\vspace{3mm}

$^3$Perimeter Institute for Theoretical Physics,\\
31 Caroline Street North,\\
Waterloo, Ontario N2L 2Y5, Canada
\par\endgroup
\vspace{5mm}

\abstract{This article reviews the recent progress in twistor approaches to  Wilson loops, amplitudes and their duality for $\cN=4$ super Yang-Mills.  Wilson loops and amplitudes are derived from first principles using the twistor action for maximally supersymmetric Yang-Mills theory.  We start by deriving the MHV rules for gauge theory amplitudes from the twistor action in an axial gauge in twistor space, and show that this gives rise to the original momentum space version given by Cachazo, Svr{\v c}ek and Witten.  We then go on to obtain from these the construction of the momentum twistor space loop integrand using (planar)  MHV rules and show how it arises as the expectation value of a holomorphic Wilson loop in twistor space.  We explain the connection between the holomorphic Wilson loop and certain light-cone limits of correlation functions.  We give a brief review of other ideas in connection with amplitudes in twistor space: twistor-strings, recursion in twistor space, the Grassmannian residue formula for leading singularities and amplitudes as polytopes. This article is an invited review for a special issue of \textit{Journal of Physics A} devoted to `Scattering  Amplitudes in Gauge Theories'.}    

\end{center}

\newpage

\tableofcontents
\newpage

\section{Introduction}
\label{intro}

In recent studies of scattering amplitudes and correlation functions
in $\cN=4$ super Yang-Mills (SYM), twistor variables have become a
powerful tool. There are technical reasons for this: the twistor data
for both scattering amplitudes and null polygonal Wilson loops are
unconstrained, twistor space makes manifest symmetry under the
superconformal group and, in particular, maximal supersymmetry is most
naturally and straightforwardly expressed off-shell in twistor space.
As a consequence, twistors have emerged in many approaches to
scattering amplitudes and Wilson loops. Twistors
play an important role in the Grassmannian integral formulae
of~\cite{ArkaniHamed:2009dn,Mason:2009qx}, various formulations of the
correspondence between Wilson loops and scattering
amplitudes~\cite{Alday:2010ku,Mason:2010yk,Caron-Huot:2010ek,Bullimore:2011ni,Drummond:2010zv},
the computations of~\cite{Goncharov:2010jf} using symbols to simplify
the 2-loop 6-particle MHV remainder function, the Y-system for the
amplitude at strong coupling~\cite{Alday:2010vh}, the all-loop
extension of the BCFW recursion relation~\cite{ArkaniHamed:2010kv},
and even in studying certain correlation
functions~\cite{Adamo:2011dq,Eden:2011}.

Although twistors are clearly a useful set of variables for these
problems, in this review we will take the position that the reason
scattering amplitudes in $\cN=4$ SYM look simplest when expressed in
terms of twistors is because \emph{the theory itself} is simple
there. That is, rather than computing amplitudes in momentum space or
null Wilson loops in space-time and then trying to discover hidden
structures and simpler expressions by merely \emph{translating} the
results into twistor space, we instead seek to understand how to
describe quantum field theory itself in twistor space. We shall see
that this can indeed be done, and that twistor space scattering
amplitudes and Wilson loops are beautiful objects \emph{in their own
  right}, and are much more readily computable in twistor space than
on space-time.  The calculation from first principles of elementary
multiparticle tree amplitudes in twistor space either directly or via
the holomorphic Wilson loop formulation is essentially a straightforward combinatoric one or two line computation. The corresponding calculations on space-time require a tour de force of computation or computer algebra, and indeed their quantum consistency remains controversial \cite{Belitsky:2011zm}.  This can therefore be taken as a nontrivial model for Penrose's original proposal that twistor space should provide a more natural arena for physics than space-time \cite{Penrose:1989pb}.

Maximally supersymmetric Yang-Mills (SYM) theory is expressed on twistor space via an action functional on twistor space that, for $\cN=4$ SYM, was first introduced and studied in~\cite{Mason:2005zm,Boels:2006ir,Boels:2007qn}\footnote{See also~\cite{Mason:2007sd,Mason:2008jy} for a tentative twistor 
action for supergravity.}. This action was first introduced as a field theory explanation for the MHV  formalism, providing a bridge between the standard, space-time action when expressed in one  gauge and the MHV formalism when expressed in an axial gauge that is inaccessible from space-time \cite{Boels:2007qn,Jiang:2008xw}.  However, this early work was only able to arrive at  the MHV formalism by using twistor wave-functions corresponding to momentum eigenstates.   These obscure the superconformal symmetry and so much of the power of the twistor  representation is lost. Recent work~\cite{Bullimore:2010pj,Mason:2010yk,Adamo:2011cb} has 
now led to significant advances that make it possible to develop the resulting quantum field theory directly in twistor space. The twistor space MHV formalism is much more efficient than that in momentum space, as it fully exploits the superconformal invariance of $\cN=4$ SYM.\footnote{As we will see, the twistor space MHV formalism gives very efficient computations for tree-level and IR finite loop amplitudes, but general (divergent) loop amplitudes still require a regularization mechanism; see \S \ref{twistorspace}.} 

The main focus of the review will be to illustrate the use of the twistor action by computing scattering amplitudes and correlation functions of certain Wilson loops. The interest in  these particular objects comes from a conjecture of Alday \& Maldacena~\cite{Alday:2007hr} at  strong coupling, and of Drummond, Korchemsky \& Sokatchev~\cite{Drummond:2007aua} at weak 
coupling, claiming that, suitably interpreted, the correlation function of a piecewise null Wilson loop is equal to the ratio of the planar all-loop MHV scattering amplitude, divided by the MHV tree  amplitude\footnote{The relation between null Wilson lines and scattering amplitudes of course has  a long history, and is important for example in understanding the exponentiation properties of 
infra-red divergences of scattering amplitudes~\cite{Korchemsky:1985xj,Collins:1989bt,Magnea:1990zb}.}. On space-time this conjecture has been checked for a number of low-lying examples, both at weak and strong coupling~\cite{Brandhuber:2007yx,Drummond:2007aua,Drummond:2007cf,Drummond:2007bm,Drummond:2008aq}. 

Reformulating the Wilson loop in twistor space~\cite{Mason:2010yk} led to a conjecture relating a supersymmetric and holomorphic version of the Wilson loop in twistor space to the full super-amplitude ({\it i.e.}, with arbitrary external helicities). This conjecture has now been proved at the 
level of the loop integrand in~\cite{Bullimore:2011ni} using a holomorphic version of the Migdal-Makeenko loop equations to recover the all-loop BCFW recursion.  The extension of these ideas  to the correspondence between Wilson loops and other correlators~\cite{Alday:2010zy} can also  be realized and extended supersymmetically in twistor space~\cite{Adamo:2011dq}. 
There were almost simultaneous proposals for space-time versions of these ideas~\cite{Caron-Huot:2010ek,Eden:2011}, but it is considerably more difficult to calculate examples of amplitudes with this approach, nor has it been possible to develop clear proofs.  On twistor space, the calculations of examples from first principles is as straightforward as any available technique and the proofs are now clear.

Although we still do not have a completely systematic regularization procedure for divergent integrals on twistor space, clear approaches certainly exist.  In particular the Coulomb branch mass regularization of~\cite{Alday:2009zm,Henn:2010bk,Hodges:2010kq,Mason:2010pg} is compatible with the twistor framework. For this reason, we confine ourselves in this review to studying the \emph{integrand}~\cite{ArkaniHamed:2010kv} of  the scattering amplitude or Wilson loop. This object is well-defined in a planar theory, and allows us to `freeze' the loop momenta at some generic points. For generic external momenta, the loop integrand is finite and superconformally invariant data associated with the scattering amplitude.  It is the sum of all Feynman diagrams before the loop integrations are actually carried out.   The familiar infrared divergences of $\cN=4$ super Yang-Mills only arise when these integrations are performed, and it is at that point that one must move out along the Coulomb branch in order to regulate the amplitudes.
We remark that it has been argued that there are quantum inconsistencies in the {\em space-time} definition even of the loop integrand (before integration) for the supersymmetric Wilson loop\cite{Belitsky:2011zm}.   These arise from ambiguities in the definition of the space-time Wilson loop as it cannot be defined off-shell.   On the other hand, we will see that the twistor space holomorphic Wilson-loop is defined completely off-shell so that issue does not arise in twistor space.   Indeed, as we shall see, direct concrete calculation gives the correct answer for the loop integrand as proved by other means~\cite{Bullimore:2010dz,He:2010ju}.

\smallskip

The review is structured as follows. After providing a brief introduction to twistor geometry and the 
Penrose transform, in section~\ref{NC} we describe the distributional twistor wave-functions which 
provide a useful calculus that has lead to many of the recent advances.   In section~\ref{TA} we 
then introduce the twistor action and develop the Feynamn rules that arise when an axial gauge 
condition is imposed.  In section~\ref{twistorspace} these are used to construct amplitudes in 
twistor space. Upon transforming to momentum space, the twistor action thus provides a 
\emph{derivation} of the original momentum space MHV rules.  (See also~\cite{Brandhuber:2011ke} for a review of the MHV formalism in momentum space.) The relationship between 
Wilson loops and amplitudes follows by encoding the momenta of particles taking part in a 
scattering process into edges of a  polygon; the fact that the polygon is closed reflects 
momentum conservation, while the fact that the particles are massless means the polygon's edges 
are null.  This momentum space data can then be translated into integrands for the scattering 
amplitudes on momentum twistor space, as covered in section \ref{momtwist}.  In 
section~\ref{HWL} we then review how to construct a supersymmetric and holomorphic Wilson 
loop in twistor space, and explain why its expectation value (evaluated with respect to the twistor 
action) is equivalent to the amplitude integrands in section \ref{HWL}.   We also explain the extension of these ideas to a correspondence between correlation 
functions of local, gauge invariant operators and Wilson loops~\cite{Alday:2010zy,Eden:2010zz,Eden:2010ce,Adamo:2011dq,Eden:2011}.

Recent developments in the study of amplitudes using twistor methods have moved quite rapidly, 
and it is beyond the scope of this review to give each aspect of these advances a fair treatment.  
Nevertheless, we also provide short (and very incomplete) summaries of  a number of key topics:  the current status of 
twistor-string theory~\cite{Witten:2003nn,Berkovits:2004hg,Mason:2007zv,Dolan:2007vv,Dolan:2009wf,Spradlin:2009qr,Bullimore:2009cb,ArkaniHamed:2009dg,Dolan:2010xv,Bourjaily:2010kw,Corn:2010uj,Skinner:2010cz}, BCFW recursion in twistor space~\cite{Mason:2009sa,ArkaniHamed:2009si,Skinner:2010cz}, the Grassmannian integral 
representation~\cite{ArkaniHamed:2009dn,Mason:2009qx,Bullimore:2009cb,Kaplan:2009mh,ArkaniHamed:2009vw,ArkaniHamed:2009sx,Nandan:2009cc,Drummond:2010uq,Ashok:2010ie}, the relationship between amplitudes, polytopes and local forms. Some of these ideas are covered from a different perspective in other chapters of this review~\cite{Bargheer:2011mm}.

\smallskip

We hope that this review will be sufficient, not only to impress the reader with the power of recent twistor methods, but also with their elegance and simplicity.   There are some sacrifices made for this simplicity.  Twistor theory is chiral, non-local and unitarity is not manifest.   Nevertheless, the simplicity is suggestive that we are making progress towards Penrose's original goal of reformulating physics on twistor space to the point where the twistor formulation supercedes that on space-time.  To fulfill this goal, many more insights will be required, but the rapid progress of the last few years suggests that this may yet appear in the forseeable future.

\subsubsection*{Brief guide to the background literature}
There are now a number of textbooks now on twistor theory \cite{Penrose:1986ca,WardWells, HuggettTod,MWbook,Dunajski:2009} of which \cite{HuggettTod} represents a first primer.  Supertwistors are covered in the textbook \cite{Manin:1988ds}.  An early review of twistor-strings and amplitudes is \cite{Cachazo:2005ga} whereas the more recent lecture notes \cite{Wolf:2010av} give  a detailed introduction to some of the material in this review.  Witten's original twistor-string paper \cite{Witten:2003nn} also provides much excellent background and more.

\section{Twistor Theory Basics}
\label{NC}


In this review, twistor space will mean the Calabi-Yau supermanifold $\PT\cong\CP^{3|4}$. When we wish to refer to the bosonic twistor space only, we will use the name $\PT_{b}\cong\CP^{3}$.
$\PT$ may be described by homogeneous coordinates
\be{eqn: tsc} 
Z^{I}=(Z^{\alpha},\chi^{a})=(\lambda_{A},\mu^{A'},\chi^{a}), 
\ee 
where $\lambda_{A}$ and $\mu^{A'}$ are 2-component complex Weyl spinors, and $\chi^a$  is an 
anti-commuting Grassmann coordinate\footnote{These conventions, first adopted in \cite{Witten:2003nn}, are not far off  the \emph{dual} twistor space conventions of standard twistor 
textbooks such as\cite{HuggettTod,WardWells}.}, with $a=1,\ldots,4$ indexing the $\cN=4$ 
R-symmetry. Being homogeneous coordinates, the $Z^I$s are defined only up to the equivalence 
relation $Z^I\sim r Z^I$ for any non-zero complex number $r$ that rescales all components 
equally.

\begin{figure}[t]
\centering
\includegraphics[width=100mm]{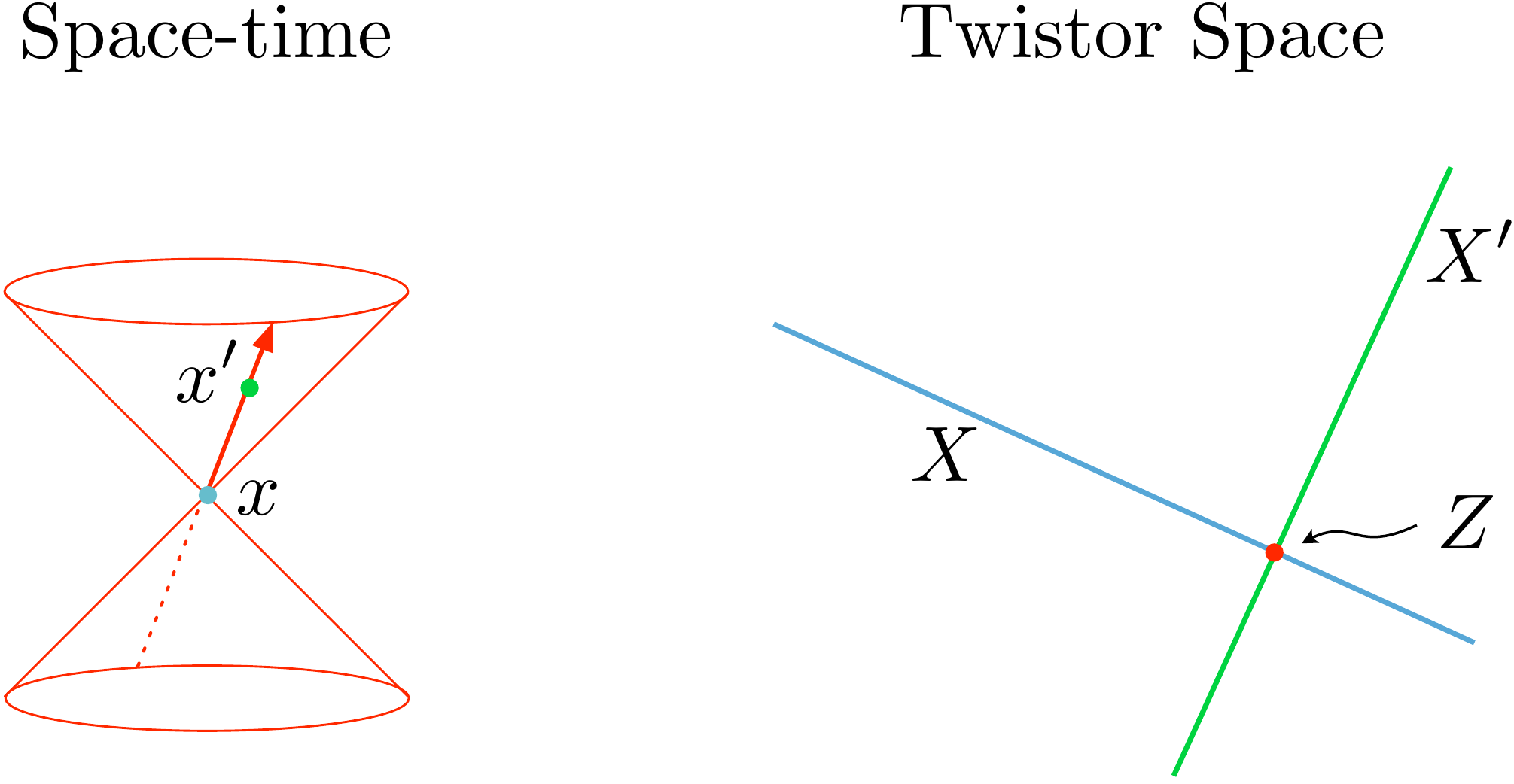}
\caption{\emph{Points in space-time correspond to complex lines in twistor space. Two space-time points are null separated if and only if their corresponding twistor lines intersect.}}
\label{tcorr}
\end{figure}

The basic geometric correspondence with (complexified) chiral super Minkowski space $\M^{4|8}$ 
is that a point $(x,\theta)\in\M^{4|8}$ corresponds to a complex line\footnote{That is, $X$ is a 
linearly embedded Riemann sphere.} $X$ in twistor space. Furthermore, two space-time points 
$(x,\theta)$ and $(x',\theta')$ are null separated if and only if the corresponding twistor lines $X$ 
and $X'$ intersect, as illustrated in figure~\ref{tcorr}. In this way, the complex structure of $\PT$ 
(i.e., knowledge of where the complex lines are) determines and is determined by the conformal 
structure (i.e., knowledge of the null cones) in space-time.

This geometric correspondence is encapsulated in the \emph{incidence relations}
\be{eqn: inc}
	\mu^{A'} =ix^{AA'}\lambda_{A}\, , \quad \chi^a =\theta^{Aa}\lambda_{A},
\ee 
where we can interpret $\lambda_A$ as homogeneous coordinates on the Riemann sphere $X$, and $(x,\theta)$ then tell us how this Riemann sphere is embedded in $\PT$. If two lines $X$ and $X'$ intersect at the point $Z=(\lambda,\mu,\chi)$, then as well as~\eqref{eqn: inc} we have
\be{eqn: inc2}
	\mu^{A'}=i{x'}^{AA'}\lambda_A\, , \quad \chi^a ={\theta'}^{Aa}\lambda_{A},
\ee
and subtracting gives $(x-x')^{AA'}\lambda_A=0$ and $(\theta-\theta')^{Aa}\lambda_A=0$ so that
$(x-x')^{AA'}=\tilde\lambda^{A'}\lambda^A$ and $(\theta-\theta')^{Aa} = \eta^a\lambda^A$ for some $\tilde\lambda$ and some $\eta$. In complex chiral superspace, as we vary the possible choices of $(\tilde\lambda,\eta)$, the possible vectors $(\tilde\lambda\lambda,\eta\lambda)$ span a totally null complex $2|4$-dimensional plane known as a (super) $\alpha$-plane. Thus, for every point $Z\in\PT$, the incidence relation assigns an $\alpha$-plane in $\M^{4|8}$.

\smallskip

One of the reasons twistor space is useful when describing $\cN=4$ SYM is that it carries a 
particularly natural action of the (complexified) superconformal group PSL$(4|4,\C)$. Acting on the 
homogeneous coordinates $Z^I$, this group is generated by
\be{sl4liealg}
	J^I_{\ J} = Z^I\frac{\del}{\del Z^J}\ ,
\ee
except that the overall homogeneity operator $\sum_I Z^I\del/\del Z^I$ and the fermionic 
homogeneity operator $\sum_a \chi^a\del/\del\chi^a$ should each be 
removed\footnote{Any object in $\cN=4$ SYM should have overall homogeneity zero in each 
twistor, while the amplitudes are graded by their fermionic homogeneity, known as MHV degree.}. 
In particular, the super Poincar\'e group is generated by
\begin{equation}
\begin{aligned}
	&P_{AA'} = \lambda_A\frac{\del}{\del\mu^{A'}}
	\qquad 
	&&J_{AB} = \frac{1}{2}\left(
	\lambda_A\frac{\del}{\del\lambda^B} + \lambda_B\frac{\del}{\del\lambda^A}\right)
	\qquad
	&&J_{A'B'} = \frac{1}{2}\left(
	\mu_{A'}\frac{\del}{\del\mu^{B'}}+\mu_{B'}\frac{\del}{\del\mu^{A'}}\right)\\
	&Q_{Aa} = \lambda_A\frac{\del}{\del\chi^a}
	&&\widetilde Q_{A'}^{\ a} = \chi^a\frac{\del}{\del\mu^{A'}}
	&&R^a_{\ b} = \chi^a\frac{\del}{\del\chi^b}\ ,
\end{aligned}
\end{equation}
while the superconformal generators also include
\begin{equation}
\begin{aligned}
	&K^{AA'} = \mu^{A'}\frac{\del}{\del\lambda_A}\qquad
	&&D =\frac{1}{2}\left(
	\lambda_A\frac{\del}{\del\lambda_A}-\mu^{A'}\frac{\del}{\del\mu^{A'}}\right)\\
	&S^{Aa} = \chi^a\frac{\del}{\del\lambda_A}
	\qquad 
	&&\widetilde S^{A'}_{\ a} = \mu^{A'}\frac{\del}{\del\chi^a}\ .
\end{aligned}
\ee
These show that $\lambda$ is inert under a translation in chiral superspace, while $(\mu,\chi)$  transform as
\be{translation}
	\mu^{A'}\to \mu^{A'} + i y^{AA'}\lambda_A\qquad
	\chi^a\to\chi^a + \theta^{aA}\lambda_A\ .
\ee
Thus $(\mu,\chi)$ have a rather different status to $\lambda$ under the super Poincar\'e group, 
and for this reason, $(\mu,\chi)$ are sometimes known as the `primary' part of the supertwistor, 
while $\lambda$ is the `secondary' part. Their roles are interchanged under a special conformal transformation.

\smallskip

So far, we have explained the correspondence for complexified space-time. We now wish to 
impose various reality conditions on our space-time and twistor space. In Minkowski signature, the 
conformal group is SU(2,2) as a real form of SL$(4;\C)$. This real subgroup 
preserves a pseudo-Hermitian metric $g_{\alpha\bar\beta}$ of signature (2,2) on non-projective 
twistor space, that we may write as
\be{Minktw}
	g_{\alpha\bar\beta}Z^\alpha\bar Z^{\bar\beta} 
	= \lambda_A\bar\mu^A+\mu^{A'}\bar\lambda_{A'}
\ee
where $\bar\lambda_{A'}$ is the Lorentzian complex conjugate of the spinor $\lambda_A$ 
(similarly for $\bar\mu^A$ and $\mu^{A'}$). If we define 
\be{Minkdual}
	\bar Z_{\alpha}\equiv g_{\alpha\bar\beta}\bar Z^{\bar\beta} = (\bar\mu^A,\bar\lambda_{A'})
\ee
then we can equivalently view Lorenztian complex conjugation as an anti-holomorphic map from 
twistor space to dual twistor space. On the projective space, the value of $Z\cdot\bar Z$ is 
meaningless, but the sets
\be{PT+-}
	\PT^+:=\left\{ Z\,|\,Z\cdot \bar Z>0\right\}\qquad
	\PN:=\left\{ Z\,|\,Z\cdot\bar Z=0\right\}\qquad
	\PT^-:=\left\{ Z\,|\,Z\cdot\bar Z<0\right\}
\ee
are preserved under the scaling $Z\sim rZ$. In particular, if a twistor line $X$ lies entirely in $\PN$, 
then from~\eqref{eqn: inc} we have
\be{Minkincidence}
	0= i(x-x^\dagger)^{AA'}\lambda_A\bar\lambda_{A'}\qquad\hbox{for all }\lambda\ ,
\ee
which is possible if and only if the matrix
\be{}
	x^{AA'} = \sigma^{AA'}_\mu x^\mu = \frac{1}{\sqrt 2}
	\begin{pmatrix}
		t+z & x-iy\\
		x+iy & t-z
	\end{pmatrix}
\ee
is Hermitian. Thus the corresponding point $x$ lies in real Minkowski space. Conversely, a point $Z\in\PN$ corresponds to a unique real null ray (the intersection of the complex $\alpha$-plane with the Minkowski real slice).

In Euclidean signature, we instead equip twistor space with an anti-holomorphic map
$Z^\alpha\rightarrow \hat Z^\alpha$ that satisfies $\hat{\hat Z}=-Z$.  This conjugation may be given 
explicitly by $Z^\alpha=(\hat\lambda_A,\hat \mu^{A'})=
(-\bar\lambda_1,\bar\lambda_0,-\bar\mu^1,\bar\mu^0)$.  The fact that the conjugation squares to 
$-1$ shows that there are no (non-zero) real twistors in Euclidean signature. A point in real 
Euclidean space corresponds to a twistor line that is mapped to itself by this conjugation (with the 
conjugation acting as the antipodal map on $X\cong\CP^1$). Thus, given the Euclidean 
conjugation, there is a `preferred' twistor line through any point $Z$ -- namely the line joining $Z$ 
to  $\hat Z$. This line is clearly preserved under the conjugation,  so to any twistor $Z$ we can 
always associate a unique point in Euclidean space. Explicitly, this point is
\be{}
	x^{AA'} = \frac{\mu^{A'}\hat\lambda^A-\hat\mu^{A'}\lambda^A}{\la\lambda\hat\lambda\ra}
	=\hat x^{AA'} \ .
\ee
Said differently, in Euclidean signature, there 
is a non-holomorphic fibration $\CP^3\to S^4$ whose fibres are the twistor lines $(Z\hat Z)$. The 
Euclidean structure was introduced by Atiyah, Hitchin \& Singer\cite{Atiyah:1978wi} and was used in the ADHM 
approach to the construction of instantons~\cite{Atiyah:1978ri}.

Finally, in $(++--)$ space-time signature, the superconformal group is PSL$(4|4;\R)$, so we simply 
take all the twistors to be real (and drop the factors of $i$ from the incidence relations). This 
signature was exploited by Witten~\cite{Witten:2003nn} in his `half-Fourier transform' to readily 
transform scattering amplitudes on (2,2)-signature momentum space on to twistor space.

\subsection{The Penrose transform and cohomology}

The Penrose transform relates helicity $h$ solutions of the zero-rest-mass (z.r.m.) free field
equations on a region $U'\subset\M$  to {\em cohomology classes} of functions of homogeneity 
degree $2h-2$ over a corresponding region $U\subset\PT_b$, where $U$ is the region swept out 
by the twistor lines corresponding to the points of $U'$. 

Cohomology classes can be represented in a variety of ways. In this review we will use the 
Dolbeault representation in which the cohomology classes are described by $\dbar$-closed 
$(0,1)$-forms modulo $\dbar$-exact ones. The Penrose transform is then expressed as the 
isomorphism
\be{eqn: ptp}
	H^{1}(U,\cO(2h-2))\cong 
	\left\{\mbox{On-shell z.r.m. fields on $U'$ of helicity}\; h\right\},
\ee 
where
\be{Dol-def} 
	H^{1}(U,\cO(n)):= \frac{\{ f\in\Omega^{0,1}(n)\,|\,\dbar f=0\}}{\{f\,|\,f=\dbar g\}} \, .
\ee 
and $\Omega^{0,1}(n)$ is the space of smooth (0,1)-forms\footnote{A $(p,q)$-form has degree $p$ 
in the differentials of the holomorphic coordinates and degree $q$ in the differentials of 
anti-holomorphic coordinates.} on $U$ that are homogeneous of degree $n$ (i.e., $f(rZ)=r^nf(Z)$).

This transform is most easily realized by the integral formula
\be{Pintdirect}
	\phi_{A_1 A_2 \ldots A_{|2h|}} 
	= \frac{1}{2 \pi i} \int_X\lambda_{A_1} \lambda_{A_2} \ldots \lambda_{A_{|2h|}}\,
	f (ix^{AA'}\lambda_{A},\lambda_{A}) \wedge\D\lambda \, , 
\ee
when $h\leq0$, and
\be{Pintpot}
	\phi_{A'_1 A'_2 \ldots A'_{2h}} 
	= \frac{1}{2 \pi i} \int_X\frac\p{\p\mu ^{A'_1}} \frac \p{\p\mu ^{A'_2}} \cdots 
	\frac\p{\p\mu^{A'_{2h}}}\, g (ix^{AA'}\lambda_{A}, \lambda_{A} )\wedge \D\lambda\, , 
\ee
when $h>0$. In these formulae, $\D\lambda=\lambda_{C} \rd \lambda^{C}$ and in~\eqref{Pintpot} 
the derivatives are really Lie derivatives acting on the forms. The fact that these integral
formulae give rise to solutions of the massless field equations can be
seen by directly differentiating under the integral sign and using the
fact that 
\be{}
\frac{\p f}{\p x^{AA'}}=i\lambda_{A}\frac{\p f}{\p\mu^{A'}}\, .
\ee
The Penrose isomorphism~\eqref{eqn: ptp} states that all solutions of the z.r.m. field equations arise this way. See for example \cite{WardWells} for a proof\footnote{Traditionally,
  twistor theorists have used the \v{C}ech 
  cohomology, which involves choosing an open cover of $U$ and describing cohomology
  representatives in terms of holomorphic functions defined on
  overlaps. The integral formulae~\eqref{Pintdirect}-\eqref{Pintpot} are then taken to be contour
  integrals.  We have not used that formulation because it is
  difficult to express the full invariances of the theory, and the
  combinatorics of the open covers becomes complicated.}.

The beauty of the construction is that it transforms the differential equations that on-shell fields 
satisfy in space-time into pure holomorphy in twistor space. As well as the clean representation of 
the superconformal group, this provides a second reason that twistors are useful in describing 
scattering amplitudes: the constraint that the external states be on-shell is automatically satisfied 
by using \emph{arbitrary} holomorphic wave-functions in twistor space.

\medskip

We can easily construct an action on twistor space whose field equations yield the above 
cohomology classes. Let $(f, \tilde f)$ be a pair of smooth $(0,1)$-forms on $U$ of respective 
homogeneities $2h-2$ and $-2h-2$, and consider the action 
\be{freeact} 
	S[f,\tilde f]=\int_U \tilde f\wedge \dbar  f \wedge \D^{3}Z
\ee 
where  $\D^{3}Z=\varepsilon_{\alpha\beta\gamma\delta}Z^\alpha\rd Z^\beta\rd Z^\gamma \rd Z^\delta/4!$.  The field equations are $\dbar f=0=\dbar \tilde f$, while the fields are defined only up to 
the gauge freedom  $f\rightarrow f + \dbar g$, etc.  Thus the on-shell fields of~\eqref{freeact} 
correspond to elements of the cohomology group $H^1(U,\cO(\pm2h-2))$ and therefore to z.r.m. 
fields of helicity $\pm h$ on space-time. Notice that even when $h=0$ (space-time scalars), the 
twistor action is still subject to gauge redundancy. This is because we are describing a four-
dimensional theory in terms of a six (real) dimensional space.

\smallskip

Standard choices for $U$ are the sets $\PT^\pm$ introduced in~\eqref{PT+-}. In complex 
space-time, these correspond to the future/past tubes $\M^{\pm}$: the sets on which the imaginary 
part of  $x^{AA'}$ is past or future pointing time-like, respectively.  This follows from the fact that if 
we take $x=u+iv$ and substitute into the incidence relation, then $Z\cdot \bar
Z= -v^{AA'}\bar\lambda_{A'}\lambda_{A}$, which has a definite sign when $v$ is time-like. The 
sign itself depends on whether $v$ is future or past pointing. The significance of this is that a field 
of positive frequency, whose Fourier transform is supported on the future light-cone in momentum space, automatically extends over the future tube because $\e^{ip\cdot x}$ is rapidly 
decreasing there, bounded by its values on the real slice. Another frequently used set is $U=\PT'$ 
on which $\lambda_{A}\neq 0$. This corresponds to excluding the light-cone of the `point at 
infinity' in space-time and includes all momentum eigenstates.

\subsubsection*{\textit{The supersymmetric extension}}

On $\cN=4$ supertwistor space, the transform has a straightforward supersymmetrization to the superfield 
\be{superfield}
	\cA(Z,\bar Z,\chi)=
	a(Z,\bar Z)+\chi^{a}\tilde{\psi}_{a}(Z,\bar Z)+\frac{\chi^{a}\chi^{b}}{2}\phi_{ab}(Z,\bar Z)+
	\frac{\epsilon_{abcd}}{3!}\chi^{a}\chi^{b}\chi^{c}\,\psi^{d}(Z,\bar Z)+
	\frac{\epsilon_{abcd}}{4!}\chi^a\chi^b\chi^c\chi^{d}\, g(Z,\bar Z)   
\ee 
where $a$, $\tilde{\psi}$, $\phi$, $\psi$, and $g$ have homogeneity $0$ $-1$, $-2$, $-3$ and $-4$
respectively, corresponding on-shell to zero-rest mass fields 
$(F_{A'B'},\widetilde\Psi_{aA'}, \Phi_{ab}, \Psi_{A}^a, G_{AB})$ on space-time. This is the 
supermultiplet appropriate to $\cN=4$ SYM.  The supersymmetric action is an Abelian Chern-
Simons action
\be{CS1}
	S[\cA]=\int \cA\wedge \dbar \cA \wedge \D^{3|4}Z
\ee
where $\D^{3|4}Z=\D^{3}Z\, \rd^4\chi$. It is easily seen that
integrating out the fermionic coordinates gives the appropriate actions
for each component field. Thus $\cA$ is an off-shell representation of the complete $\cN=4$ 
supermultiplet on twistor space.

On-shell, the field equations modulo gauge redundancy of~\eqref{CS1} show that $\cA$ 
represents an element of the cohomology group $H^1(U,\cO)$ on a region of supertwistor space.
The integral formulae \eqref{Pintpot}-\eqref{Pintdirect} extend directly to this supersymmetric
context to give on-shell superfields on space-time that incorporate derivatives of the component 
fields. Specifically, one has
\be{Susy-Pint1}
\begin{aligned}
	\cF_{A'B'}:&=
	\int _{X } \frac{\del^2}{\del\mu ^{A'}\del\mu^{B'}} \,
	\cA(ix^{AA'}\lambda_{A}, \lambda_{A},\theta^{Aa}\lambda_{A} )\wedge \D\lambda\\
 	&=F_{A'B'}+\theta^{Aa}\p_{AA'}\left[\widetilde\Psi_{aB'}+ \theta^{Bb}\p_{BB'}\left(\frac{\Phi_{ab}}{2}+
    \theta^{Cc}\varepsilon_{abcd}\left(\frac{\Psi_{C}^d}{3!}+
    \theta^{Dd}\frac{G_{CD}}{4!}  \right) \right)\right] \nonumber\\
\end{aligned}
\ee
and
\be{Susy-Pint2}
\begin{aligned}
	\cF_{ab}:&=
	\int _{X }  \frac{\del^2}{\del\chi^a\del\chi^b}\,
	\cA (ix^{AA'}\lambda_{A}, \lambda_{A},\theta^{Aa}\lambda_{A} )\wedge \D\lambda\\
	&=\Phi_{ab} +\theta^{Cc}\varepsilon_{abcd}(\Psi_{C}^d+\theta^{Dd}\frac{G_{CD}}2)\ .
\end{aligned}
\ee
These fields together $\cF_{aA'}$ (which has a formula as above with a mixed $\mu$ and $\chi$ derivative) have the interpretation as being the non-zero parts of
the curvature
\be{susy-curv}
\cF=\cF_{A'B'}\varepsilon_{AB}\rd x^{AA'}\wedge \rd
x^{BB'}+\cF_{aA'} \varepsilon_{AB}\rd x^{AA'}\wedge\rd\theta ^{Ba}+ \cF_{ab}\varepsilon_{AB}\rd \theta^{Aa}\wedge \rd 
\theta^{Bb}
\ee
of the on-shell space-time superconnection
\begin{multline}
\sA=\left[A_{AA'} +\theta^{a}_{A}\left(\widetilde\Psi_{aA'}+
    \theta^{Bb}\p_{BA'}\left(\frac{\Phi_{ab}}{2} +\varepsilon_{abcd}
\theta^{Cc}\left(\frac{\Psi_{C}^{d}}{3!}+\theta^{Dd}\frac{G_{CD}}{4!}
\right)\right)\right) \right] \rd x^{AA'}
\\ +
\left[\Phi_{ab}+\varepsilon_{abcd}\theta^{Bc}\left(\frac{\Psi^{d}_{B}}{2} +
  \theta^{Cd}\frac{G_{BC}}{3!}\right)\right)  \theta^{a}_{A}\rd\theta^{Ab}\, .
\label{superconn}
\end{multline}
This superconnection can be obtained directly and geometrically from $\cA$ via
the Penrose-Ward transform, which treats $\cA$ geometrically as a deformation
of the $\dbar$-operator on a line bundle and obtains $\sA$ as a (super)-connection on a 
corresponding line bundle on space-time.  This interpretation will follow from our treatment of  
the Yang Mills equations in subsequent sections.

\subsection{Distributional forms for twistor wave functions}
\label{Dist}
In this section, we describe twistor representatives for various commonly used wave-functions. In 
our Dolbeault framework, it will be convenient to work with distribution-valued forms. We 
first describe twistor wave-functions for momentum eigenstates.  We then 
describe the {\em elementary state} on twistor space that corresponds to the fundamental solution 
of the wave equation.  Finally, we describe {\em elemental states} that are supported at a point in 
twistor space (i.e., they are twistor eigenstates). These elemental states form the main basis of the calculus that we actually use in the rest of 
the review.

\smallskip

On the complex plane with coordinate $z=x+iy$, the delta function supported at the origin 
is naturally a $(0,1)$-form which we denote by
\be{delta-bar}
	\bar\delta^1(z):=
	\delta(x)\delta(y)\, \d \bar z =\frac1{2\pi i}\,\d \bar z\frac{\del}{\del\bar z}\, \frac1z\ ,
\ee
the second equality being a consequence of the standard Cauchy kernel for the $\dbar$-operator.  
This second representation makes clear the homogeneity property 
$\bar\delta(r z)=r^{-1}\bar\delta(z)$. (Note that there is no absolute-value sign here, in contrast to 
the case of real $\delta$-functions.)

We can extend this to the Riemann sphere by defining
\be{spin-delta}
	\bar\delta^1_m(\lambda, p) :=
	\int \frac{\rd s}{ s^{m+1}}\,\bar\delta^2(s\lambda_{A}+p_{A})\, .
\ee
This has support only when $p_A\propto\lambda_A$ with some constant of proportionality that we 
integrate over. Thus it has support only when $p$ and $\lambda$ coincide projectively. One can 
check that $\bar\delta^1_m(\lambda,p)$ has homogeneity $m$ in $\lambda$ and $-m-2$ in $p$, so that
\be{}
	g(p) = 
	\int_{\CP^1} \bar\delta^1_m(\lambda,p) g(\lambda)\wedge\D\lambda
\ee
for any function $g$ of homogeneity $-m-2$.  

This idea can be used to give the twistor cohomology class for an on shell momentum
eigenstate with momentum $p_{AA'}=p_A\tilde{p}_{A'}$ 
\be{}
	f_{p,-m-2}(\mu,\lambda)=\int\frac{\rd s}{ s^{m+1}}\,
	\e^{s\mu^{A'}\tilde{p}_{A'}} \bar\delta^2(s\lambda_{A} +p_{A})\, . 
\ee
It is easily seen that this evaluates via \eqref{Pintdirect} and \eqref{Pintpot} to
give the appropriate momentum eigenstates 
\be{}
p_{A_1}\ldots p_{A_m}\e^{ip\cdot x}\, ,
\qquad \tilde{p}_{A'_1}\ldots \tilde{p}_{A'_{-m}}\e^{ip\cdot x}\, 
\ee
for the zero rest mass fields of momentum $p$ and helicity $m/2$, depending on the sign of $m$.

\smallskip

In $\cN=4$ SYM, an on-shell supermultiplet with definite momentum $p_{AA'} = p_A\tilde p_{A'}$ 
and definite supermomentum $p_A\eta_a$ may be represented by the supertwistor space cohomology class
\be{super-mom-rep}
	f_{p;\eta}(Z;\chi):=
	\int \frac{\rd s}{ s}\,\e^{s(\mu^{A'}\tilde{p}_{A'}+\eta_{a}\chi^{a})}
	\bar\delta^2(s\lambda_{A} + p_{A})\ .  
\ee
Inserting this into~\eqref{Susy-Pint1} gives the chiral Minkowski space wave-function
\be{superfield}
	F_{A'B'}(x,\theta)
	= \tilde{p}_{A'}\tilde{p}_{B'} \e^{ip\cdot x+ i\eta_{a}\theta^{Aa}p_{A}}
\ee
for the on-shell supermultiplet. As in~\eqref{Susy-Pint1}, in expanding out the $\theta$s, all except 
the  helicity $-2$ parts of the multiplet appear in a differentiated form; the helicity $-1$ part is  
differentiated once, and the higher ones twice. 

\smallskip

Our second class of examples are `elementary states'.  These are states
that are singular on a line in twistor space and correspond to fields that are singular on the 
lightcone of the corresponding point in space-time.  The simplest of these is for homogeneity $-2$ 
(space-time scalar), where we can set
\be{} 
	\phi=\frac1{\mu^{0'}}\,\dbar\left( \frac1{\mu^{1'} }\right)\ .
\ee 
Evaluation on space-time via \eqref{Pintdirect} can be seen to give the
the fundamental solution $\Phi=1/x^2$ to the wave equation.  The
general elementary state is obtained from this example by
differentiating with respect to $Z$ or the basis parameters and multiplying
by monomials in the twistor coordinates.

\smallskip

While the above representatives are useful for converting twistor expressions directly to on-shell 
momentum space, part of the power of the twistor representation is its 
underlying conformal invariance and this is not manifest if we use momentum eigenstates as 
external wavefunctions.  We therefore introduce a version of an `elemental
state' which may be viewed as `eigenstates of definite twistors'.

Following the strategy of~\eqref{spin-delta}, to obtain $\delta$-functions on projective
twistor space, we first introduce the Dolbeault $\delta$-functions on\footnote{Recall that a 
$\delta$-function for a Grassmann variable $\chi$ is simply $\delta(\chi) = \chi$, as follows from the Berezinian integration rules $\int \chi\rd\chi=1$ and $\int 1\,\rd\chi=0$, so that $\int f(\chi)\,\chi\rd\chi=f(0)$.}
 $\C^{4|4}$ 
\be{delta44}
	\bar\delta^{4|4}(Z)=\prod_{\alpha=0}^3 \, \bar \delta(Z^\alpha)
	\prod_{a=1}^4 \chi^a\ . 
\ee
This is a $(0,4)$-form on $\C^{4|4}$ of homogeneity zero, having support only where $Z^I=0$. We 
now define projective $\delta$-functions by 
\be{projdelta3 4}
	\bar{\delta}^{3|4}(Z_{1},Z_{2})
	:=\int_{\C}\frac{\d s}{s}\,\bar{\delta}^{4|4}(Z_{1}+sZ_{2})\ .
\ee  
This is a (0,3)-form on $\PT_1\times\PT_2$, homogeneous of degree zero in each entry and
 antisymmetric under $Z_1\leftrightarrow Z_2$. It satisfies
\begin{equation*}
	f(Z_1)=\int_{\PT} f(Z_2)\, \bar{\delta}^{3|4}(Z_1,Z_2)\wedge \D^{3|4}Z_2
\end{equation*}  
as befits a delta function. 

The elemental state will be taken to be the part of $\bar{\delta}^{3|4}(Z,Z_1)$ that is a 
$(0,1)$-form in $Z$ and a $(0,2)$-form in $Z_{1}$.  Thus, an elemental state does not correspond 
to an ordinary real or complex valued z.r.m. field, but rather one that takes values in $(0,2)$-forms 
on the twistor space of the auxilliary $Z_1$ variable.  It is peculiar in a number of ways that are
perhaps best illustrated by evaluating it on space-time via~\eqref{Pintdirect}. Applying this to the 
coefficient of $(\chi)^2(\chi_1)^2$ in~\eqref{projdelta3 4} we obtain the space-time field 
\be{elemental-spt} 
	\phi_{Z_1}\!(x)=\bar\delta^2(\mu_1^{A'}-ix^{AA'}\lambda_{1\, A}) 
\ee
which we understand as a distribution on complexified Minkowski space $\M$ with values in
the $(0,2)$-forms in the $Z_1=(\lambda_1,\mu_1)$ variables.  It should be emphasized
that because it is a $\delta$-function, $\phi_{Z_{1}}\!(x)$ is clearly not holomorphic on
$\M$.  It nevertheless satisfies the holomorphic wave equation made
from just the holomorphic derivatives, and similarly the
anti-holomorphic wave equation made from anti-holomorphic derivatives.
However, when restricted to a real space-time slice, $\phi_{Z_1}\!(x)$ does not satisfy any the
wave equation as the cross terms between holomorphic and
anti-holomorphic derivatives in the real wave equation do not
vanish.

\subsubsection{Elemental states and amplitudes}\label{elemental-amp}

By the Penrose transform, finite norm on-shell external wave-functions are represented on twistor
space by cohomology classes $H^{1}(\PT',\cO(2h-2))$.  Since amplitudes are multilinear 
functionals of these external wave functions, on twistor space the kernel for an $n$-particle 
amplitude should be dual to the $n$-fold product of such $H^1$s.   Although we could use the 
Hilbert space structure on $H^{1}(\PT',\cO(n))$, this is actually a rather non-trivial duality that is 
non-local on twistor space and depends crucially on the choice of space-time signature.   Instead,  we will use the natural duality between $(1,0)$-forms and $(2,0)$-forms with compact support  that is manifested by
\begin{equation*}
	(\phi,\alpha)\mapsto \int_{\PT'}\D^{3|4}Z\wedge \phi\wedge\alpha \,, \qquad (\phi,\alpha)\in (\Omega^{(0,1)},\Omega^{(0,2)}_c) \ .
\end{equation*}

Thus, we would like to say that an $n$-particle amplitude is represented by the twistor space 
kernel
\be{eqn: kernel}
	A(Z_{1},\ldots, Z_{n})\in \Omega^{(0,2n)}_{c}(\times_{i=1}^n \PT'_{i},\cO) \ ,
\ee
a $(0,2n)$-form with compact support on $n$ copies of $\PT'$. The fact that this is to be paired with cohomology classes means that it should make no difference if we add an exact form to $A(1,\ldots,n)$, and for the amplitude to not depend on the choice of representative for the wave functions we should have that the amplitude is $\dbar$-closed.  Thus in an ideal world, the amplitude would be an element of  $H^{2n}(\times_{i=1}^n\PT)$.  However this picture is not attainable for a number of reasons. Firstly the relevant cohomology groups vanish.  Secondly, collinear 
divergences of massless amplitudes mean (even at tree-level) that if we wedge such an $A(Z_1,\ldots,Z_n)$ with finite-norm external wave-functions and 
integrate over $n$ copies of twistor space, the resulting object will diverge. In the explicit formulae that we have for $A(Z_1,\ldots,Z_n)$, we can compute its exterior derivative, and we find that it fails to vanish where at the diagonals where pairs of adjacent twistors come together.   In effect, it is an element of $H^{2n}(\times_{i=1}^n\PT_i)$ that has simple poles at these adjacent diagonals (see \cite{Atiyah:1981ey} for a discussion of such objects).   In 
practice, at tree-level we simply treat $A(Z_1,\ldots,Z_n)$ as a $(0,2n)$-form and choose `generic' 
external twistors to avoid these singularities in the same manner as one usually chooses 
generic external momenta.

The effect of integrating a momentum eigenstate form as in
\eqref{super-mom-rep} against such an amplitude kernel will be the same as
inserting that momentum eigestate into the formula for an amplitude in
the first place.  As explained above, finite-norm twistor wave-functions do not have
compact support in twistor space, being defined for example either on
$\PT^+$ or $\PT^-$.  Using the above elemental states, we will obtain
amplitudes that are supported on $\PN$ and hence valid when
integrated against a twistor wave function defined on $\PT^+$ or
$\PT^-$.  This is how crossing symmetry becomes manifest on twistor
space.

\subsubsection{A calculus of  distributional forms and the
  propagator} 
\label{prop-delta}

The  $\delta$-functions $\bar\delta^{3|4}(Z_1,Z_2)$ defined above are just the first example of a 
family of projective $\delta$-functions. By integrating against a further parameter, we can obtain 
the $\delta$-function
\be{delta-line}
\begin{aligned}
	\bar{\delta}^{2|4}(Z_{1},Z_{2},Z_3):&=
	\int_{\CP^{2}}\frac{\D^{2} c}{c_{1}c_{2}c_3}\,
	\bar{\delta}^{4|4}(c_1Z_{1}+c_2Z_{2}+c_3Z_3)\\
	&=\int_{\C^2}\frac{\d s}{s}\frac{\d t}{t}\,\bar{\delta}^{4|4}(Z_3+sZ_{1}+tZ_{2}) \\
	&= \int_\C \frac{\d s}{s}\, \bar{\delta}^{3|4}(Z_{1},Z_{2}+s Z_3)\ ,
\end{aligned}
\ee
that has support when $Z_1$, $Z_2$ and $Z_3$ are collinear in projective space. This object is 
manifestly superconformally invariant, weightless in each twistor and antisymmetric 
under exchange of any two.  It has simple poles where any pair of twistors  coincide.  We will see 
below that it can be used to define the propagator for the action \eqref{CS1} in an axial gauge.

We can similarly define a coplanarity delta function
\be{delta-plane}
\begin{aligned}
	\bar{\delta}^{1|4}(Z_{1},Z_{2},Z_3, Z_4)
	:&=\int_{\CP^{3}}\frac{\D^{3} c}{c_{1}c_{2}c_3c_4}\,
	\bar{\delta}^{4|4}(c_1Z_{1}+c_2Z_{2}+c_3Z_3+c_4Z_4) \\
	&=\int_{\C^3}\frac{\d s}{s}\frac{\d t}{t}\frac{\d u}{u}\,
	\bar{\delta}^{4|4}(Z_4+sZ_3+tZ_{2}+uZ_{1})\\
	&= \int_\C \frac{\d s}{s} \bar{\delta}^{2|4}(Z_{1},Z_{2},Z_3+s Z_4)
\end{aligned}
\ee
and the rational object 
\be{superconf}
\begin{aligned}
	\bar{\delta}^{0|4}(Z_{1},Z_{2},Z_3,Z_4,Z_5) 
	:&=\int_{\CP^4}\frac{\D^4 c}{c_1c_2c_3c_4c_5}\,
	\bar{\delta}^{4|4}\!\left(\sum_{i=1}^5c_iZ_i\right)\\
	&=\frac{\left( (1234)\chi_5 + \mbox{cyclic}\right)^4}
	{(1234)(2345)(3451)(4512)(5123) }\ ,
\end{aligned}
\ee
where the second formula is obtained by integration against the delta functions and $(1234)\equiv \epsilon_{\alpha\beta\gamma\delta}Z_1^\alpha Z_2^\beta Z_3^\gamma Z_4^\delta$ (see \cite{Mason:2009qx} for full details).  The $\delta$-functions in~\eqref{delta-plane}-\eqref{superconf} enforce that their arguments lie on 
a common $\CP^2\subset\CP^{3|4}$ or a common $\CP^3\subset\CP^{3|4}$, respectively. They 
are each totally antisymmetric and of homogeneity zero in their arguments.

We will frequently abbreviate $\bar\delta^{0|4}(Z_1,Z_2,Z_3,Z_4,Z_5)$ by $[1,2,3,4,5]$.  It is the
most elementary superconformal invariant that one can form.  In the context of momentum twistors 
it is the standard dual superconformal invariant of \cite{Drummond:2008vq}, sometimes denoted 
by $R_{5;13}$. Our definition makes it clear that these `R-invariants' depend (antisymmetrically) 
on five arbitrary supertwistors (or momentum supertwistors).

\smallskip

These projective $\delta$-functions are not generally $\dbar$-closed, but rather
\be{}
	\dbar\,\bar\delta^{r|4}(Z_1,\cdots,Z_{5-r})
	= (2\pi i) \sum_{i=1}^{5-r}(-1)^{i+1}\,\bar\delta^{r+1|4}(Z_1,\cdots,\widehat{Z_i}
, \cdots, Z_{5-r})\ ,
\ee
where $\widehat{Z_i}$ is ommitted (see \cite{Adamo:2011cb} for a
proof).  The right  hand side necessarily vanishes for $r=3$.  We will use these relations for $r=2$, 
with $Z_3=Z_*$ a fixed reference twistor, to obtain the propagator associated to the
free field action \eqref{CS1}:
\be{propagator}
\Delta(Z_{1},Z_{2})=\bar{\delta}^{2|4}(Z_{1},\rf , Z_{2}).
\ee
This is easily seen to have the correct properties via the relation 
\begin{equation*}
\dbar\Delta(Z_{1}, Z_{2})=2\pi i\left(\bar{\delta}^{3|4}(Z_{1},Z_{2})
  +\bar{\delta}^{3|4}(Z_{1},Z_\rf)+ \bar{\delta}^{3|4}(Z_{2},Z_\rf)\right)\ .
\end{equation*}  
For $Z_{1}, Z_{2}\in\PT'$ and $Z_*\notin\PT'$, the last two terms on the
right hand side of this expression vanish. We are then left with 
\be{}
	\dbar\Delta(Z_{1},Z_{2})=\bar{\delta}^{3|4}(Z_{1},Z_{2})\ ,
\ee 
which is the equation we require of the propagator for the action \eqref{CS1}.  This form of the
propagator is in an axial gauge associated to the reference twistor;
it satisfies the condition 
\be{} 
	\bar Z_*^{\bar\alpha}\frac{\del}{\del\bar Z^{\bar\alpha}}\,\lrcorner\, \Delta=0 
\ee 
that says the (0,2)-form $\Delta(Z_1,Z_2)$ vanishes when contracted into
directions tangent to the lines through $Z_*$.  It follows from the
various definitions that when the integration in \eqref{delta-line}
with respect to $s$ is performed, a $\bar Z_*$ is skewed into the
anti-holomorphic form indices.

As a final remark, note that many integrals can be performed algebraically when these 
distributional delta functions are involved. Useful examples that we will frequently use are
\be{composition}
\begin{aligned}
	\bar\delta^{1|4}(Z_1,Z_2, Z_3,Z_4) 
	&=\int \bar\delta^{2|4}(Z_1,Z_2,Z)\wedge\bar\delta^{2|4}(Z,Z_3,Z_4)\wedge\D^{3|4}Z\\
	[1,2,3,4,5]&=
	\int \bar\delta^{2|4}(Z_1,Z_2,Z)\wedge\bar\delta^{1|4}(Z,Z_3,Z_4,Z_5)\wedge \D^{3|4}Z \ .
\end{aligned}
\ee
Such relations form a key part of the calculus that enables us to readily evaluate Feynman 
diagrams or recursion relations on twistor space.

\section{The Twistor Yang-Mills Action}
\label{TA}

The twistor action is our starting point for studying amplitudes and
correlation functions so we provide a thorough introduction to its
construction and properties here.  Beginning from the Chalmers-Siegel
\cite{Chalmers:1996rq} action for $\cN=4$ SYM on space-time, we proceed to show
how the twistor action is constructed and discuss some of its
features which are apparent in different choices of gauge.

\subsubsection*{\textit{The space-time action}}

We begin by recalling the Chalmer-Siegel action  for ordinary
Yang-Mills theory with no supersymmetry; this allows one to work with
an action that manifestly allows for expansion around the self-dual
sector of Yang-Mills theory.  Consider a bundle
${E'}\rightarrow\M$ with a connection 1-form $A(x)$ taking values
in $\End({E'})$ (the Lie algebra of the complexified gauge group)
and curvature $F=\d A+A\wedge A$.  Rather than consider the ordinary
Yang-Mills action for such a connection, introduce an auxilliary ASD
2-form $G\in\Omega^{2-}(\M,\End({E'}))$ coupled to the connection
via the action \cite{Chalmers:1996rq}: 
\be{eqn: CSA} 
S[A,G]=\int_{\M}\tr\left(G\cdot F-\frac{\varepsilon}2 \ G\cdot G\right)\d^4x , 
\ee 
for a parameter
$\varepsilon$ (here $\cdot$ denotes the natural inner product on
2-forms).\footnote{The parameter $\varepsilon$ is naturally proportional to the 't Hooft coupling, $\lambda$, of the theory.  For a $\SU(N)$ theory, $\lambda=\frac{g^{2}N}{8\pi^{2}}$, and $\varepsilon=-\lambda$.}   Splitting the curvature into its SD and ASD parts,
$F=F^{+}+F^{-}$, this action has field equations \be{eqn: CSFE}
F^{-}=\varepsilon G, \qquad \nabla\wedge G=0, \ee where $\nabla=\d +A$
is the connection corresponding to $A$.  We can easily see that when
$\varepsilon=0$, we have the SD Yang-Mills equations, and generally
\begin{equation*}
\nabla \wedge F^{*}=\nabla\wedge (F^{+}-F^{-})=\nabla\wedge (F-2F^{-})=0
\end{equation*} 
by the Bianchi identity and field equations.  Hence, the Chalmers-Siegel action of \eqref{eqn: CSA} has field equations which are equivalent to the full Yang-Mills equations, and $\varepsilon$ has the natural interpretation of an expansion parameter away from the SD sector of the theory.

This action can be extended to $\cN=4$ SYM by including the additional
fields of the multiplet, which (besides $A(x)$ and $G$) are: the ASD
and SD gluino fermions $\Psi^{a}_{A}$ and $\widetilde{\Psi}_{aA'}$
respectively, and the scalars
$\Phi_{ab}=\frac{1}{2}\epsilon_{abcd}\Phi^{cd}$.  Once again, the
action can be split into its two parts:
\be{eqn: SCS}
S[A,\Psi,\Phi,\tilde{\Psi},G]=S_{\mathrm{SD}}[A,\tilde{\Psi},\Phi,\Psi, G]-\frac{\varepsilon}{2}I[A,\tilde{\Psi},\Phi,\Psi, G],
\ee
where
\begin{eqnarray}
S_{\mathrm{SD}}[A,\tilde{\Psi},\Phi,\Psi, G] & = &
\int_{\M}\tr\left(\frac{1}{2}G\cdot
  F+\Psi^{a}_{A}\nabla^{AA'}\tilde{\Psi}_{aA'}-\frac{1}{8}(\nabla\Phi)^2 
+\tilde{\Psi}_{aA'}\tilde{\Psi}^{A'}_{b}\Phi^{ab}\right)\d^{4}x \nonumber \\
I[A,\tilde{\Psi},\Phi, \tilde{\Psi}, G] & = &
\frac{1}{2}\int_{\M}\tr\left(G\cdot  G +\Psi^{a}_{A}\Psi^{bA}\Phi^{ab}+\frac{1}{4}\Phi^{ac}\Phi_{ab}\Phi^{bd}\Phi_{cd}\right) \d^{4}x.  \label{eqn: symasd}
\end{eqnarray}
The second term contains all the interactions of the full theory that
are absent in the self-dual theory.
As in the bosonic case, \eqref{eqn: SCS} manifestly allows us to
expand around the SD sector via the expansion parameter $\varepsilon$.
In the next subesection, we express this action on twistor space.

\subsection{The SYM Twistor Action}
The Ward construction establishes a correspondence between self-dual
Yang-Mills fields on a 
region $U'$ in
space-time and certain holomorphic vector bundles $E\rightarrow
U\subset \CP^3$
where $U$ is the corresponding
region in bosonic twistor space (c.f., \cite{HuggettTod,WardWells}).  Such
bundles can be expressed in terms of a smooth 
 topologically trivial bundle $E\rightarrow U$ with complex structure
 given by a deformed $\dbar$-operator:  
\be{eqn: dbar}
\dbar_{a}=\dbar +a, \qquad a\in\Omega^{0,1}(\PT_{b}',\End(E)).
\ee
In the self-dual case, the space-time bundle is flat on
$\beta$-planes, and so $E_Z$ can be defined to be the parallel
sections of $E'$ on the $\beta$-plane coresponding to $Z$.  This
varies holomorphically with $Z$ (assuming the space-time field to vary
holomorphically on space-time).  The more remarkable fact of the Ward
transform is that $a$ determines the bundle with connection $(E',A)$
up to gauge on space-time.

The supersymmetric Ward correspondence \cite{WardWells,Manin:1988ds}
similarly gives a one-to-one correspondence between integrable complex
structures on $E\rightarrow U\subset \CP^{3|4}$, now given in terms of a homogeneous $(0,1)$-form $\cA$ on $\PT'$ (i.e.,
$\dbar_{\cA}^{2}=0$), and SD solutions to the field equations of
$\cN=4$ SYM on space-time.  As $\dbar_{\cA}^{2}=0$ are precisely the
field equations for a holomorphic Chern-Simons action, we can account
for the SD portion of the theory on twistor space with the functional
\be{eqn: HCS}
S_{\mathrm{SD}}[\cA]=\frac{i}{2\pi}\int_{\PT'}\D^{3|4}Z \wedge
\tr\left(\cA\wedge\dbar\cA  + \frac{2}{3}\cA\wedge\cA\wedge\cA\right).  
\ee 
Here we require that
$\cA$ depends holomorphically on the fermionic coordinates
$\chi^{a}$ and has no components in the
$\d\bar{\chi}$-directions (c.f., \cite{Witten:2003nn, Boels:2006ir}).  
As in the abelian case,
we can expand $\cA$ in powers of $\chi^{a}$: \be{eqn: Aexp} \cA=
a+\chi^{a}\tilde{\psi}_{a}+\frac12{\chi^{a}\chi^{b}}\phi_{ab}+
\epsilon_{abcd}\chi^{a}\chi^{b}\chi^{c}\left(\frac1{3!}{\psi^{d}} +
  \frac1{4!}\chi^{d} g \right), \ee with
 $a$, $\tilde{\psi}$, $\phi$, $\psi$, and $g$  of
weights $0$ $-1$, $-2$, $-3$, and $-4$ respectively.  
We have already seen that in the $\U(1)$ theory these give the standard
$\cN=4$ super-Yang-Mills multiplet on space-time.

Just as with the Chalmers-Siegel action, we must add an extra
term to the self-dual action to obtain the remaining
interactions of the full theory.  This can be expressed as the functional
\be{eqn: TASD}
I[\cA]=\int_{\M_{\R}}\d^{4|8}x
\log\det\left(\dbar_{\cA}|_{X}\right), 
\ee
where $X$ is the $\CP^{1}\subset \PT$ corresponding to
the point $ (x,\theta)\in\M_{\R}$; $\M_{\R}$ is a real slice of the
complexified space-time; $\dbar_{\cA}|_{X}$ is the
restriction of the deformed complex structure \eqref{eqn: dbar} to the
line $X$; and $\d^{4|8}x$ is the natural supersymmetric
volume form on space-time:
\begin{equation*}
\d^{4|8}x=\frac1{4!}\epsilon_{\mu\nu\rho\sigma}
\d x^{\mu}\wedge\d x^{\nu}\wedge\d x^{\rho}\wedge\d x^{\sigma}\wedge\d^{8}\theta.
\end{equation*}
A motivation for this specific form of this interaction comes from
twistor-string theory, see \cite{Mason:2005zm,Boels:2006ir} for
discussion.  The integrand of $I[\cA]$ is not \emph{a priori}
well-defined  because the determinant $\det(\dbar_{\cA}|_{X})$ is
not an honest function, but in fact a section of a determinant line
bundle over the space of connections on $E\rightarrow\PT'$;
this determinant will pick up anomalous terms under gauge
transformations of the SD portion of the action.  However, such
anomalous transformations can be seen to produce  terms that
necessarily vanish upon performing the  fermionic
integrals.\footnote{This follows essentially because the variations in $\log \det$ are additive and depend on the $\theta$s only through the $\chi$s
which only have four components \cite{Boels:2006ir}.}

Hence, \eqref{eqn: TASD} is a well-defined functional, and we combine
it with \eqref{eqn: HCS} to yield the full twistor action: 
\be{eqn: TA}
S[\cA]=S_{\mathrm{SD}}[\cA]-\frac{\varepsilon}{2}I[\cA]\, .
\ee
The first and second terms have precursors respectively in the work of \cite{Sokatchev:1995nj,Witten:2003nn} and \cite{Abe:2004ep}. 

The gauge freedom $\dbar_{\cA}\rightarrow h\dbar_{\cA}h^{-1}$ of this
action depends essentially arbitrarily on
the full six real bosonic variables of $\PT$ and so is much greater than
that on space-time.  It can be reduced in various ways.  As shown
in \cite{Boels:2006ir,Jiang:2008xw}, it can be reduced to the four
variables of the space-time action \eqref{eqn: symasd} in Euclidean
signature by requiring that the restriction of $\cA$ to any Euclidean real
$\CP^{1}$ be `harmonic': 
\begin{equation*}
\dbar^{*}|_{X}\cA_{0}=0,
\end{equation*}
where $\cA_{0}$ is the component of $\cA$ in the direction tangent to
the Euclidean real $\CP^{1}$s (which fibre $\CP^3$ over Euclidean
space-time).  It restricts the remaining gauge freedom to that of the
standard space-time gauge transformations, as first shown by Woodhouse \cite{Woodhouse:1985id}.  In this gauge the twistor
action $S[\cA]$ reduces to precisely the space-time action \eqref{eqn:
  symasd}, \cite{Boels:2006ir}.  Thus, the twistor action is
non-perturbatively and classically equivalent to the space-time Chalmers-Siegel action in this gauge.

We will see, however, that there are other gauge choices available on
twistor space which are inaccessible from space-time.  The main
example that we will use is an axial gauge.  In general, for any
holomorphic 1-dimensional distribution $D\subset T^{1,0}\PT$, this is
the condition that $\cA|_{\bar{D}}=0$.  As already discussed, we will
implement this by choice of a reference twistor $Z_\rf$ and let $D$ be
the span of $Z_\rf\cdot\p$. We will see that in this gauge the twistor
action yields Feynman rules that have some surprising properties on
twistor space, and, when transformed to momentum space,
precisely reproduce the momentum MHV formalism
of \cite{Cachazo:2004kj}.

\subsection{CSW Gauge and Twistor Space Feynman Rules} 
\label{FR}

To obtain twistor space Feynamn rules, we impose a simple axial gauge
by choosing a twistor at infinity $Z_\rf=(0,\hat{\iota}^{A'},0)$ and
imposing the condition   
\be{eqn: CSWg}
\overline{Z_\rf \cdot \frac{\partial}{\partial Z}}\lrcorner\cA =0,
\ee
where $\cA$ is the fundamental field in our twistor action, given by \eqref{eqn: Aexp}.  This requires $\cA$ to vanish upon restriction to the leaves of the foliation of $\PT\setminus\{Z_\rf\}$ by the lines which pass through $Z_\rf$, and is referred to as the CSW gauge, since it was first introduced in \cite{Cachazo:2004kj}.  

Clearly, \eqref{eqn: CSWg} reduces the number of components of $\cA$ from three down to two, so the cubic Chern-Simons vertex in \eqref{eqn: HCS} vanishes.  This leaves the twistor action in the CSW gauge as:
\begin{equation*}
\frac{i}{2\pi}\int_{\PT'}\D^{3|4}Z\wedge\tr\left( \cA\wedge\dbar\cA\right)-\frac{\varepsilon}{2}\int_{\M_{\R}}\d^{4|8}x \log\det\left(\dbar_{\cA}|_{X}\right).
\end{equation*}
As far as the Feynman rules are concerned, the propagator will arise
from the first term.  This kinetic term is now the same as for the $\U(1)$
theory, and so as discussed in \S\ref{prop-delta} 
the propagator will be given by 
$$\Delta(Z_{1},Z_{2})=\bar{\delta}^{2|4}(Z_{1},Z_\rf, Z_{2}),$$ the
$(0,2)$-form on $\PT$ that imposes the collinearity of its three
arguments.  By our earlier remarks, 
this propagator satisfies the CSW gauge condition $\overline {(*\cdot\p_1)}\,
\lrcorner\,\Delta=\overline {(*\cdot\p_2)}\, \lrcorner \,\Delta=0$.

\subsubsection*{\textit{Vertices}}
We will write out the vertices in such a way as to manifest as much
conformal symmetry as possible.  Thus, we write the
measure on the real contour $\M_{\R}$ as 
\begin{equation}\label{volform}
\d^{4|8}x=\frac{\d^{4|4}Z_{A}\wedge\d^{4|4}Z_{B}}{\vol(\GL(2,\C))}.
\end{equation}
where we have supposed that $X$ is the line joining twistors
$Z_A$ and $Z_B$ and the quotient by $\GL(2,\C)$ is that of the choices of such 
$Z_A$ and $Z_B$.

The vertices can be made explicit by perturbatively expanding the
logarithm of the determinant, yielding \cite{Boels:2006ir,Boels:2007qn}: 
\be{eqn: vert1}
\log\det\left(\dbar_\cA |_{X}\right)=\tr\left(\log\dbar|_{X}\right)+\sum_{n=2}^{\infty}\frac{1}{n}\int_{X^{n}}\tr\left(\dbar |_{X}^{-1}\cA_{1}\dbar |_{X}^{-1}\cA_{2}\cdots\dbar |_{X}^{-1}\cA_{n}\right),
\ee
where $\dbar|_{X}$ is the restriction of the $\dbar$-operator from
$\PT$ to $X\cong\CP^{1}$, and $\cA_{i}$ is a field inserted at a point
$Z(\sigma_i)\in X$ where we have introduced the (inhomogeneous) coordinate $\sigma$ on $X$ by 
\be{sigma-def}
Z(\sigma)=Z_A+\sigma Z_B\,  .
\ee
In terms of this coordinate, the $\dbar|_X^{-1}$ are the Green's
functions for the $\dbar$-operator restricted to $X$ given by
integration against the Cauchy kernel 
\begin{equation*}
\left(\dbar|_X^{-1}\cA\right) (\sigma_{i-1})=\frac{1}{2\pi i}\int\frac{\cA(Z(\sigma_i))\wedge \d\sigma_i }{\sigma_{i}-\sigma_{i-1}},
\end{equation*}
Thus, the $n$th term in our expansion yields the vertex 
\be{eqn: vert2}
\frac{1}{n}\left(\frac{1}{2\pi i}\right)^{n}\int_{\M_{\R}} \d^{4|8}x\int_{X^{n}}\tr\left(\prod _{i=1}^n \frac{\cA_i(Z(\sigma_i))\wedge\d\sigma_i }{\sigma_i-\sigma_{i-1}} \right) \, .  
\ee 
Here the index $i$ is understood cyclically with $\sigma_i=\sigma_{n+i}$ and $\M_{\R}$ denotes a real slice of complexified space-time.

In order to obtain the dualized amplitude as discussed in
\S\ref{elemental-amp}, 
we use  for the our external wave functions 
\be{eqn: trep}
\cA_{i}=\bar{\delta}^{3|4}(Z_{i},Z(\sigma_{i})),\qquad i=1,\ldots, n,
\qquad Z(\sigma_{i})=Z_{A}+\sigma_{i}Z_{B}
\ee
thought of as a $(0,1)$-form in $Z(\sigma_i)$ and a $(0,2)$-form in $Z_i$.  

Integrating over the insertion points on the line $Z_{A}\wedge Z_{B}$
in \eqref{eqn: vert2} reduces each external particle's contribution to
a $(0,2)$-form, so the $n$-valent vertex takes values in the $n$-fold
product of $\Omega^{0,2}_{c}$ (as required by \eqref{eqn: kernel}) and
is supported on a line in $\PT'$.  This is precisely the $n$-particle
MHV vertex in twistor space.

Thus, with \eqref{volform} we can write the $n$-particle MHV vertex on
twistor space as 
\be{eqn: TMHV}
V(Z_{1},\ldots, Z_{n})=\int_{\M_\R} \frac{\d^{4|4}Z_{A}\wedge\d^{4|4}Z_{B}}{\vol(\GL(2,\C))}\int_{(X_{AB})^{n}}\prod_{i=1}^{n}\frac{\bar{\delta}^{3|4}(Z_{i},Z(\sigma_{i}))\d \sigma_{i}}{(\sigma_{i}-\sigma_{i-1})}.
\ee
This manifests the conformal symmetry and cyclic invariance of the
amplitude and corresponds to the twistor-string
path-integral formulation of the tree-level MHV amplitude
\cite{Nair:1988bq,Witten:2003nn,Roiban:2004yf}.

These formulae for the MHV vertices can be recursively built from
lower order vertices and superconformally invariant delta functions. 
An $n+1$-particle vertex can be written in terms of a $n$-particle
vertex and a delta function by transforming from the variable
$\sigma_i$ to $s=(\sigma_i-\sigma_{i-1})/(\sigma_i-\sigma_{i+1})$ and
observing that the $s$ integral decouples from the others to yield
\be{eqn: inverse-soft}
V(Z_{1},\ldots, Z_{n+1})=V(Z_{1},\ldots, \widehat{Z_i},\ldots, Z_{n+1})\bar{\delta}^{2|4}(Z_{i-1},Z_{i},Z_{i+1}).
\ee
where $\widehat{Z_i}$ is omitted from the right-hand-side.  This
operation has become known as the inverse soft limit
\cite{Mason:2009sa,ArkaniHamed:2009si}. Proceeding inductively along
these lines, one arrive at many expressions of the
form \cite{Adamo:2011cb} 
\be{eqn: TMHV2}
V(Z_{1},\ldots, Z_{n})=V(Z_{1},Z_{2})\prod_{i=2}^{n}\bar{\delta}^{2|4}(Z_{1},Z_{i-1},Z_{i}),
\ee 
where $V(Z_{1},Z_{2})$ is the two point vertex.   The two-point vertex
can be written as 
\be{eqn: 2pt}
\V(Z_{1},Z_{2})=\int_{\M\times (\CP^1)^2}{\D^{3}Z_{A}\D^{3}Z_{B}}\, \bar{\delta}^{3}_{0,-4} (Z_{1},Z_A)\, \bar{\delta}^{3}_{0,-4}(Z_{2},Z_B).
\ee
The expression for the MHV vertex given by \eqref{eqn: TMHV2} shows
clear superconformal invariance and has the minimal number of residual
integrals, but of course no longer manifests the cyclic invariance of
the original twistor-string expression (this is due to our choice of
representation of the inverse soft limit).  The equivalence between
cyclic permutations can be recovered by repeatedly using the identity 
\be{eqn: cyclic}
\V (Z_1,Z_{2},Z_3)\,\bar{\delta}^{2|4}(Z_1,Z_{3},Z_4)=\V (Z_2,Z_{3},Z_4)\,\bar{\delta}^{2|4}(Z_2,Z_{4},Z_1)
\ee
for the 4-point vertex.

\section{From the Twistor Action to the MHV Formalism}
\label{twistorspace}

An important output of Witten's twistor-string theory
\cite{Witten:2003nn} is the MHV formalism, a set of momentum-space
Feynman rules for calculating scattering amplitudes in $\cN=4$ SYM
which are significantly simpler than the ordinary Feynman rules that
arise from a space-time action
\cite{Cachazo:2004kj,Cachazo:2004zb,Cachazo:2004by}.  In contrast to
the full twistor-string theory, the MHV formalism can, on its own, be
extended to loop calculations.  Indeed, it has been shown to compute
the correct amplitudes at 1-loop explicitly \cite{Brandhuber:2004yw},
and after being expressed on momentum twistor space (see
\cite{Bullimore:2010pj} and \S \ref{momtwist} of this review) was shown to
be correct to all loop-orders for planar $\cN=4$ SYM
\cite{Bullimore:2010dz} using an all-line recursion relation building
on the work of \cite{Risager:2005vk,Elvang:2008na,Elvang:2008vz}.

In \cite{Boels:2007qn} it was shown that the momentum space MHV
formalism arises naturally as the Feynman rules of the twistor action
\eqref{eqn: TA} when a very simple axial gauge (the CSW gauge) is
chosen, but it was not until recently that the MHV formalism was
derived from the twistor action in a manner that is entirely
self-contained on twistor space \cite{Adamo:2011cb}.  This section
reviews the construction of the twistor space MHV formalism,
emphasizing how amplitudes are represented on twistor space
cohomologically and how tree-level N$^k$MHV amplitudes are calculated
on twistor space.  We also make some remarks about the treatment of
loop amplitudes by the theory.  The Feynman rules dveloped here will
also be those used when we come to discuss correlation functions
later.

On momentum space, an amplitude is said
to be N$^k$MHV if it has homogeneity degree $4(k+2)$ in the fermionic
portion of the super-momenta $P=(p_{AA'},\eta_a)$ (it must be a multiple of $4$ because the amplitude
must be invariant under the $\SU(4)$ R-symmetry acting on the
$a$-index).  Such an amplitude contains the gluon amplitude with $k+2$
negative helicity particles, with the rest positive.  It is easily
seen by inserting momentum eigenstate wave functions into a twistor
amplitude that such an N$^k$MHV amplitude is obtained from one on
twistor space that is a polynomial in the fermionic $\chi$
coordinates of homogeneity degree $4(n-k-2)$ (as far as the fermionic
variables are concerned, it is a Fourier transform).  By counting the
contributions from each propagator and vertex, it can be seen that an $l$-loop
diagram made from MHV vertices and propagators as above will
contribute to an N$^k$MHV amplitude if it has $k+l+1$ vertices.

\subsection{The MHV formalism in twistor space}

Using the axial gauge Feynman rules for the twistor action developed in \S \ref{FR}, we will see that the integrals corresponding to generic diagrams can
be performed explicitly and algebraically against the delta functions.
Here generic is meant in the sense of fixed $k$ and large $n$.  
The reason for this is essentially geometric: each vertex corresponds
to a line in twistor space, and if two of these vertices are connected by a
propagator, then the delta function in the propagator
forces the insertion points to be collinear with the reference
twistor.   However, it is easily seen that when the lines are in
general position, there is a unique transversal line through $Z_\rf$
to the pair of lines corresponding to the vertices.  Thus
geometrically, each diagram will correspond to $k+l+1$ lines for the
vertices connected by one line through $Z_\rf$ for each propagator.

We will divide the diagrams into three categories. The \emph{generic}
diagram will be one with no adjacent propagator insertions on any of
its vertices.  The \emph{boundary}\footnote{The terminology arises from the
  fact that in the summation over diagrams to form the amplitude, the
  boundary diagrams arise at the boundary of the range of summation
  over the indices specifying the location of the external particles
  on the different vertices.} diagrams are those in which one or more
vertices have two or more adjacent propagator insertions, but each
vertex has at least two external particles attached to it.  The
\emph{boundary-boundary} diagrams will be those in which there are MHV
vertices with less than two external legs; in this last case we will not be
able to perform all the integrals against delta functions.  We deal
with the generic case in some detail and refer the reader to
\cite{Adamo:2011cb} for details of the other cases and the particulars of all calculations, summarizing
briefly the results.  

\subsubsection*{\textit{Example: NMHV tree}}

For tree level NMHV amplitudes, the only diagrams that contribute are
those appearing in Figure \ref{NMHVfig} with two vertices connected by one
propagator.  Our twistor space Feynman rules tell us that each term in
this amplitude is computed as
\begin{figure}
\centering
\includegraphics[width=5.5 in, height=1.5 in]{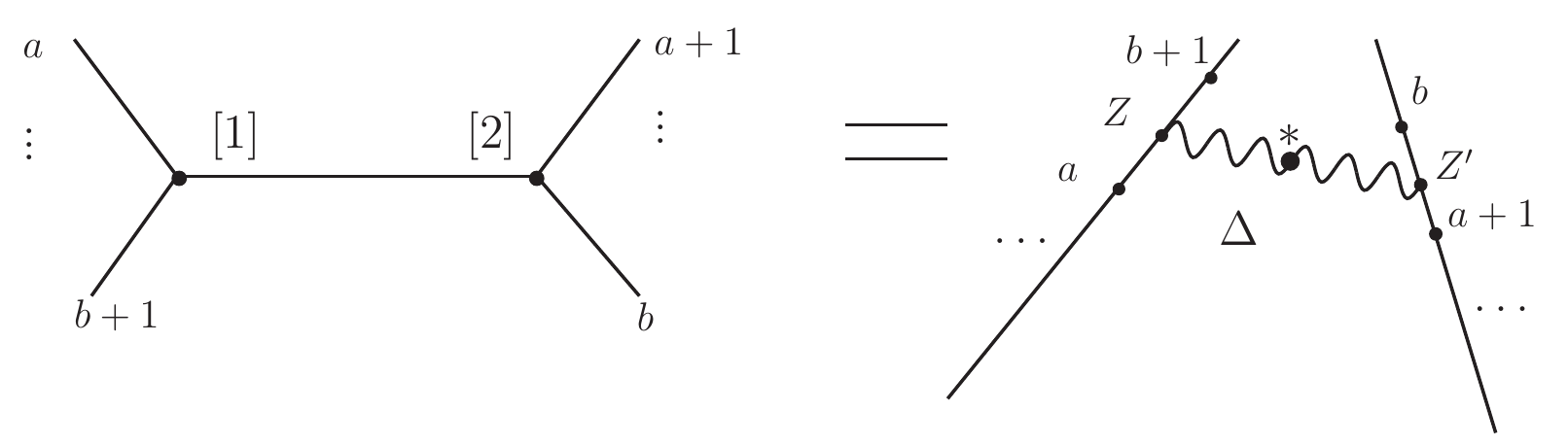}\caption{\emph{Twistor support of a typical NMHV tree diagram}}\label{NMHVfig}
\end{figure}
\begin{equation*}
\int_{\PT\times\PT} \D^{3|4}Z\D^{3|4}Z' \V ({b+1},\ldots, a, Z)\, \bar{\delta}^{2|4}(Z,\rf ,Z') \V ({a+1},\ldots, b, Z') \, .
\end{equation*}
We can separate out the dependence of the vertices on the  internal
twistors using \eqref{eqn: inverse-soft} to obtain
\begin{multline*}
\V ({b+1},\ldots, a)\V ({a+1},\ldots, b)
\times  \hfill\\
\hfill 
\int_{\PT\times\PT}\D^{3|4}Z\,\D^{3|4}Z'\,
\bar{\delta}^{2|4}({a},{b+1},Z)\, \bar{\delta}^{2|4}(Z,\rf ,Z'
)\, \bar{\delta}^{2|4}({b},{a+1},Z')\, .
\end{multline*}
These integrals can be performed algebraically against the delta
functions as in \eqref{composition}  to yield
\begin{equation*}
\V ({b+1},\ldots, a)\V ({a+1},\ldots, b)[{b+1},{a},\rf ,{b},{a+1}].
\end{equation*}
Geoemtrically this diagram corresponds to the lines for each vertex,
and the $R$-invariant then can be thought of as being associated to
the transversal to these two lines through $Z_\rf$.

The full NMHV amplitude is then the sum over $a<b$ of such contributions.

\subsection*{\textit{Generic diagrams}}

The NMHV calculation above extends directly to each propagator of a
generic diagram. 
For such diagrams, there are no adjacent propagator insertions at any
vertex, and so the neighbourhood of a propagator can be depicted as
in Figure \ref{PropR}.  We can use
\eqref{eqn: inverse-soft} to
can strip off a $\bar{\delta}^{2|4}$ at the propagator insertion point
on each vertex, 
leaving  MHV vertices that no longer depends on the propagator
insertion points $Z_{1}$ and $Z_{2}$.  Thus a propagator leads to a 
factor of
\be{}
\int \D^{3|4}Z_1 \, \D^{3|4}Z_2 \, \bar \delta^{2|4}(Z_{1},\rf,Z_{2}) \,
\bar{\delta}^{2|4}(Z_1,\alpha, \beta)\, \bar{\delta}^{2|4}(Z_2,\gamma,
\kappa) =[\alpha,\beta,\rf,\gamma,\kappa]
\ee
multiplied by the diagram with that propagator removed.
Here
$\alpha$ and $\beta$ are the 
two nearest external particles on one side of the progagator, while
$\gamma$ and $\kappa$ are the closest on the other side (see Figure
\ref{PropR}) and we have performed the integral algebraically against
the delta functions using \eqref{composition} as before.  

\begin{figure}
\centering
	\includegraphics[width=75mm]{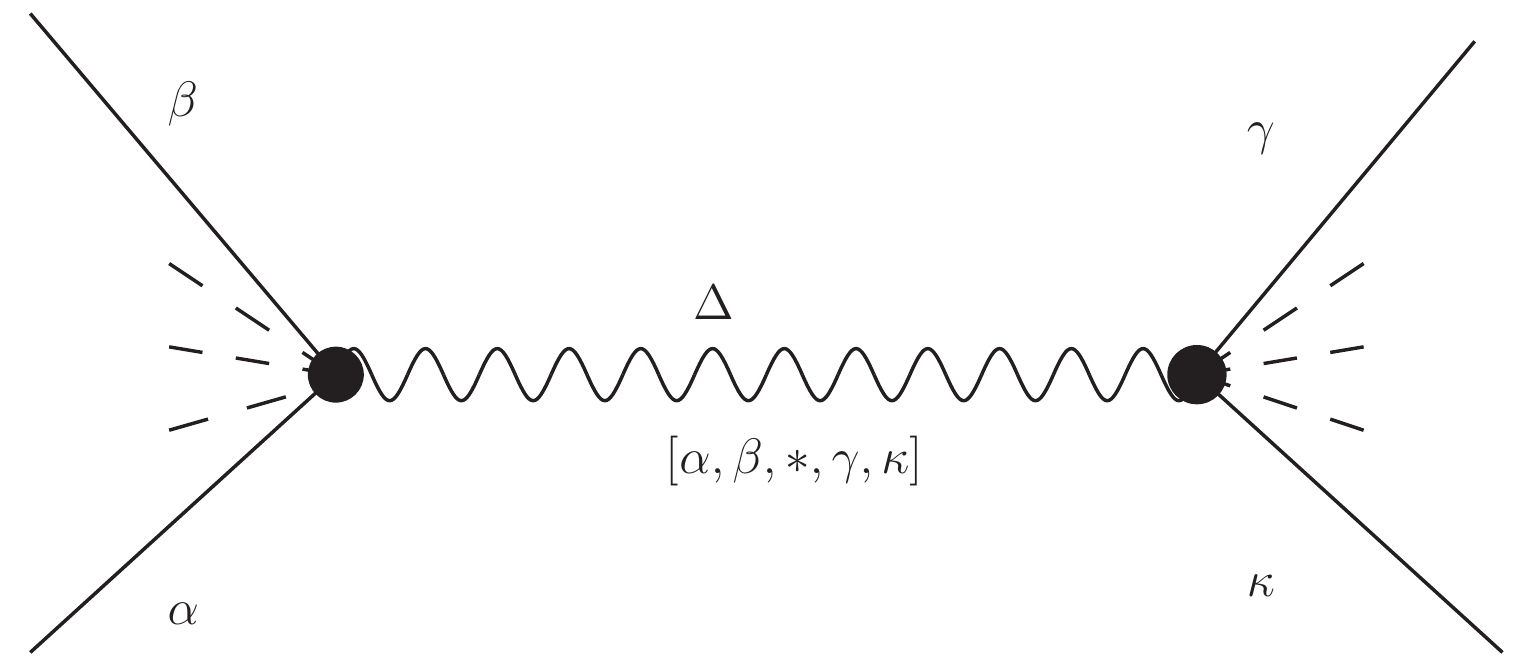}

	\caption{\emph{Propagator contributions}}
\label{PropR} 
\end{figure}

Proceeding inductively, we see that a tree-level generic N$^k$MHV
diagram gives a product of $k$ R-invariants, one for each propagator
depending on $Z_\rf$ and each adjacent external twistor.  These are
multiplies by the $k+1$ MHV vertices that now depend only on the
external particles on the corresponding original 
vertex in the initial MHV diagram.

\subsection*{\textit{Boundary Terms}}

If two or more propagators are adjacent at a vertex, we can still use
\eqref{eqn: inverse-soft} to pull out a $\bar\delta^{2|4}$ factor
containing the only dependence on the insertion
points of the propagator so long as there are at least two external
particles at each vertex.  However, the dependence of the
$\bar\delta^{2|4}$ will yield a slightly more complicated
integral.  This can nevertheless be performed algebraically against
the delta functions to yield an R-invariant, but now the
R-inavriant that corresponds to a propagator will  depend on shifted
external twistors.  For a boundary diagram such as figure \ref{kboundary}
we end up with the
rules \cite{Adamo:2011cb}:
\begin{itemize}    
\item Each vertex in the diagram gives rise to a factor of an MHV vertex in the
answer that depends only on the external legs at that vertex.
\item Each propagator corresponds to an R-invariant $[\widehat{a_1},a_2,\rf,\widehat{b_1},b_2]$ where $a_1$ and $a_2$ are the nearest
  external twistors with $a_1<a_2$ in the cyclic ordering on the
  vertex at one end of the propagator, and similarly for $b_1<b_2$ on
  the vertex at the other end.  Let $p$ be the insertion point on the
  vertex containing $a_1$ and $a_2$.  We have that $\widehat{a_1}$ is
  shifted according to the rule \be{shift-rule}
  Z_{\widehat{a_{1}}}=\left\{
\begin{array}{ll}
Z_{a_{1}} & \mbox{if $p$ is next to $a_1$} \\
X_{a_{1},a_{2}}\cap \langle b_{1},b_{2},\rf\rangle & \mbox{if $p$ is next to the propagator on the $a_1$ side} \\
 & \mbox{that connects to the line $(b_{1},b_{2})$}\, ,
\end{array}\right. 
\ee
where $X_{a_{1},a_{2}}$ is the line defined by $Z_{a_1}$, $Z_{a_2}$ and $\langle b_{1},b_{2},\rf\rangle$ is the plane defined by $Z_{b_1}$, $Z_{b_2}$ and $Z_{\rf}$.  The rule for $\widehat{b_1}$ follows by $a\leftrightarrow b$.

\end{itemize}
\begin{figure}
\centering
	\includegraphics[width=100mm]{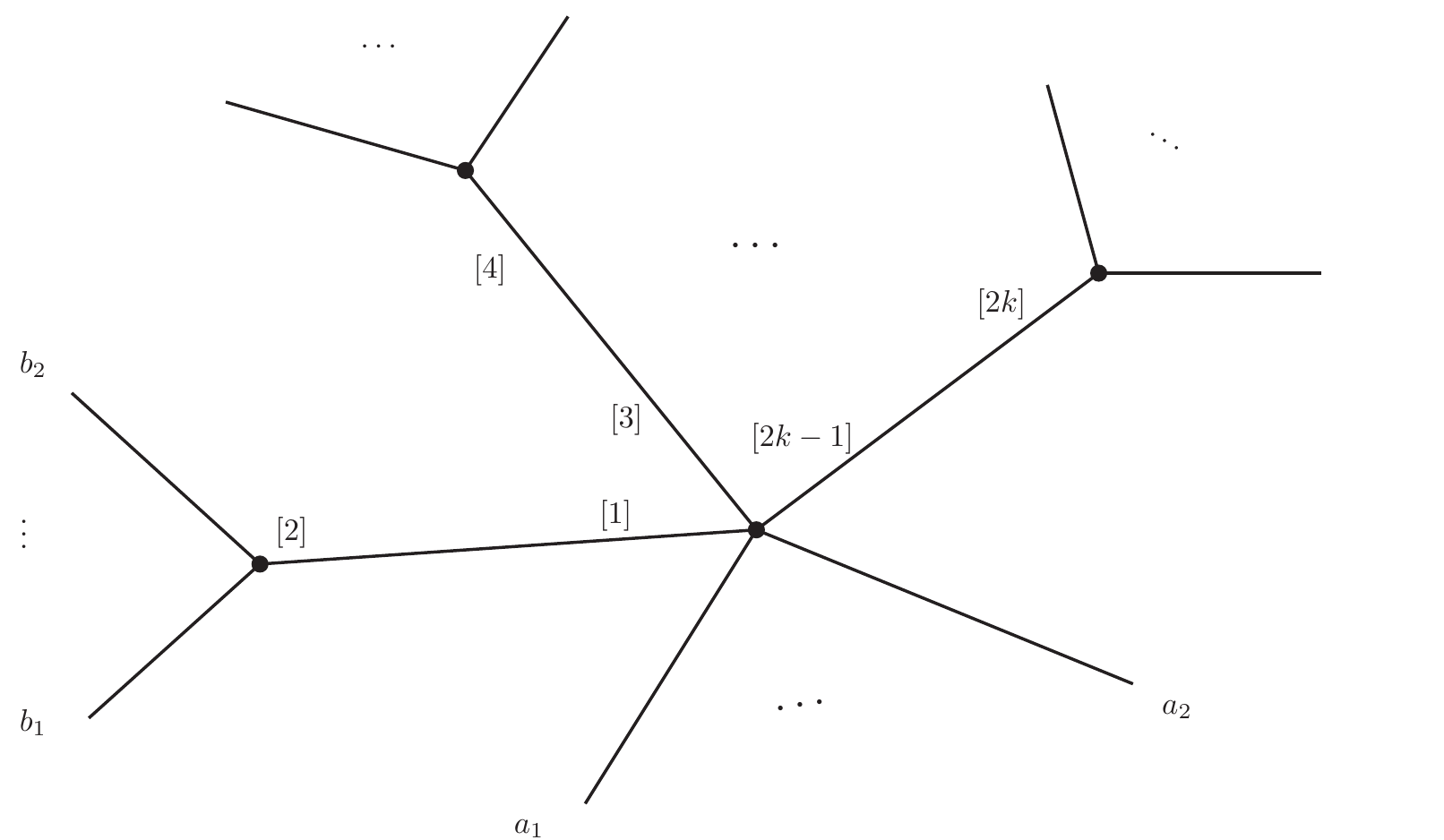}
\caption{\emph{N$^k$MHV boundary term with $k$ adjacent propagators}}
\label{kboundary}
\end{figure}

\subsection*{\textit{Boundary-Boundary Terms}} 

The final class of diagrams are the boundary-boundary contributions
when there are fewer than two external legs on some vertices.  In this
case, the above integration procedure breaks down since the location of
the line corresponding to each MHV vertex is no longer fixed by the
external particles.  There will in general be some residual
integrations associated to those in the MHV vertex to be performed.
Although many of the integrations can be performed (or indeed all by the introduction of signature-dependent machinery \cite{Adamo:2011cb}), we do not obtain
as attractive a formalism as we do for the other diagrams.  

Examination of these diagrams on momentum space suggests that there is
nothing particularly special about these diagrams there. 

\subsection*{\textit{Loop diagrams}}

The rules given above extend readily to loop diagrams leaving no
residual integrals in the generic and boundary cases.  Although all
planar loop diagrams are boundary diagrams, in the non-planar theory
there are many straightforwardly finite loop diagrams, even at 1-loop
MHV.  Indeed in the planar theory, there are straightforwardly
finite (boundary) diagrams at NMHV.  These presumably lead to polylogs
on transform back to momentum space.  Although most of the planar MHV diagrams
ought to be finite, the procedure above leads to $0/0$ and a good 
procedure has not yet been arrived at to evaluate these \cite{Adamo:2011cb}.

\subsection{Derivation of the momentum space MHV formalism}
\label{momentumspace}

As a reality check, we would like to see that this process indeed
gives rise to the MHV formalism when transformed to momentum space.
Since momentum space representations break conformal invariance, we will use a
version of the MHV vertex in which the volume form on $\M$ is
$\rd ^{4|8}x$ and  the $\GL(2)$ symmetry has been fixed by
coordinatizing the line $X$ by its projection to homogeneous
coordinates $\lambda_{A}$ 
as in \eqref{eqn: inc}.  Up to an irrelvant constant factor, this
yields the formula 
\be{eqn: vert-mom}
\int_{\M_{\R}}
\d^{4|8}x\int_{X^{n}}\tr\left(\prod _{i=1}^n
  \frac{\cA_i(-ix^{AA'}\lambda_{A\;i},\lambda_{A\;i},\theta^{Aa}\lambda_{A\;i})\wedge
   \D\lambda_i }{\la\lambda_i\, \lambda_{i-1}\ra } 
\right) \, ,   
\ee 
where as usual $\la\,\ra$ denotes the unprimed spinor inner product.  Most
straightforwardly, we can check that the MHV vertices give the standard MHV 
amplitudes by taking the $\cA_{i}$ to be on-shell
momentum eigenstates with super momentum
$P_{i}=(p_{i\;A},\tilde{p}_{i\;A'},\eta_{i\;a})$  
\be{eqn: on-shell}
\cA_{i}(P_{i};Z)=\int_{\C}\frac{\d s}{s}
e^{s(\mu^{A'}\tilde{p}_{i\,A'}+\chi^{a}\eta_{i\,a})}
\bar{\delta}^{2}(s\lambda_{A}-p_{i\,A})\, ,
\ee
multiplied by some generator of the Lie algebra of the gauge group.
It is now easily seen that the delta functions simply
enforce\footnote{
In this subsection, in order to be able to perform this
integration that gives this identification between $\lambda_A$ and
$p_A$, we must distinguish between the value of the
$\lambda$-coordinate at the twistor insertion point and the
corresponding spinor part $p_A$ of the momenta; in the future we will
exploit the delta function to use the same $\lambda_A$ notation for
both where no ambiguity can arise.} 
$s\lambda_i=p_i$ and the integral $\rd^{4|8}x$ gives the
super-momentum conserving delta function to end up with the 
Parke-Taylor formula \cite{Parke:1986gb,Nair:1988bq} for the MHV tree
amplitude extended to $\cN=4$ SYM: \be{eqn: PT}
A_{\mathrm{MHV}}^{0}(P_{1},\ldots,P_{n})= 
\frac{\delta^{4|8}\left(\sum_{i=1}^{n}P_i
      \right)}{\prod_{i=1}^{n}   \la p_{i}\, p_{i+1}\ra},  .
\ee 
Here 
we have stripped off an overall color trace factor together with an
irrelevant constant factor, and the super\-momentum conserving delta-function is 
$$
\delta^{4|8}\left(\sum P_{i}\right)=\delta^{4|0}\left(\sum
 p_i\tilde{p}_{i}\right)\delta^{0|8}\left(\sum \eta_{i}p_{i}\right), 
$$
where
$$
\delta^{0|8}\left(\sum_i \eta_i p_{i}\right)
=\prod_{a,\,A}\left(\sum_i \eta_{i\;a}\lambda_{i\, A}\right)\, .
$$

We need to also check that our rule for the insertion of the
propagator $\Delta(Z,Z')=\bar\delta^{2|4}(Z,\rf,Z')$ leads to the
rules for the propagator found in the MHV formalism
\cite{Cachazo:2004kj}.  In order to do this we pull $\Delta$ back to
the spin bundle (i.e., express it as a function of $(x,\theta,\lambda,
x',\theta',\lambda')$ using \eqref{eqn: inc}) and Fourier transform
over the $(x,x')$ variables.  This requires the choice of a real slice
of complex Minkowski space to integrate over for the Fourier
transform.  In order to obtain the correct Feynman $i\epsilon$
prescription, we do so over the Euclidean slice, and then analytically
continue the momentum space formula to the Minkowski slice that we
have focused on for the vertices.  Observing that $\Delta$ only
depends on $x-x'$ allows us to express the Fourier transform
$\widetilde \Delta (p,p',\ldots)=
\delta^{4}(p_{AA'}-p_{AA'}')\widetilde\Delta(p,\ldots)$ where, after some
calculation that can be found in \cite{Adamo:2011cb}, 
\be{} \widetilde \Delta(p_{AA'},\chi,\lambda,\chi'\lambda')= \frac
1{p^2} \, \int\frac{\rd s\rd t}{st} \bar\delta^2(\lambda_A+s p_A)
\bar\delta^2(\lambda_A-t\lambda'_A)\delta^{0|4}(\chi-t\chi') \, .\ee
Here, for reference twistor $Z_\rf=(0,\hat\iota^{A'})$, the spinor
part of the off-shell momenta is given by \be{CSWref}
p_A=\iota^{A'}p_{AA'}\, .  \ee If we now use a Fourier representation
of the fermionic delta function, substitute in
$\chi=\theta^{iA}\lambda_A$ etc., and use the support of the delta
functions, we obtain the formula 
\be{} \widetilde
\Delta(p_{AA'},\chi,\lambda, \chi'\lambda')= \frac 1{p^2} \,
\int\rd^4\eta\; \bar\delta^1_0(\lambda_A,p_A)
\, \bar\delta^1_0(\lambda'_A,p_A)\, \e^{i\eta\cdot(\theta-\theta')|p\ra} 
\ee
where $\bar\delta^1_0(\cdot,\cdot)$ are the spinor projective delta functions
defined in \eqref{spin-delta}.  Fourier transforming back, we finally
obtain the Fourier representation for the pullback of $\Delta$ to the
spin bundle 
\be{}
\Delta(x,\chi,\lambda,x', \chi'\lambda')= 
\int \frac {\rd^4p}{p^2} \;\rd^4\eta\; \bar\delta^1_0(\lambda_A,p_A)
\, \bar\delta^1_0(\lambda'_A,p_A)\,
\e^{ip\cdot(x-x')+i\eta\cdot(\theta-\theta')|p\ra}  \, .
\ee
This now yields the Fourier
representation of $\Delta$ whose ends can be substituted into the
twistor MHV vertices \eqref{eqn: vert-mom}.  The
delta functions lead to the CSW prescription in which the spinor $p_A$
associated to an off-shell momentum $p_{AA'}$ is given by 
$p_{A}= \iota^{A'} p_{AA'}$ where now $\iota^{A'}$ is the `reference
spinor' introduced in that formalism \cite{Cachazo:2004kj}, although
we see from the above that it is also 
the Euclidean conjugate of the primary part of the
reference twistor.  The exponential factors guarantee that the ends
introduce the appropriate supermomenta $P=\pm( p_{AA'},p_A\eta^i)$.

Amplitudes are then constructed as a sum of Feynman-like diagrams
built out of such propagators and MHV vertices.  At N$^k$MHV there
will be $k+1$ vertices in each diagram.  One can prove that dependence
on the spinor $\hat \iota^{A'}$ drops out of the final sum of
contributions and this corresponds precisely to the $n$-particle
$\mathrm{N}^{k}$MHV tree-amplitude \cite{Risager:2005vk,
  Elvang:2008na, Elvang:2008vz}.  The N$^k$MHV amplitudes in this
context are then those of homogeneity $4(k+2)$ in the $\eta$s.

Hence, the Feynman rules for the twistor action in the CSW gauge are
precisely those of the MHV formalism: for a N$^k$MHV amplitude,
connect $k+1$ MHV vertices (given by \eqref{eqn: TMHV} or \eqref{eqn:
  TMHV2}) together with $k$ propagators (given by \eqref{propagator}),
and integrate over propagator insertion points on each MHV line in
twistor space.


\section{The Momentum Twistor MHV Formalism}
\label{momtwist}

In their AdS/CFT approach to scattering amplitudes, Alday and
Maldacena introduced a non-compact `T-duality' in order to simplify their
calculations \cite{Alday:2007hr}.  This had the effect of replacing space-time by region
momenta space, an affine version of momentum space with no fixed
origin.  Their work led to the conjecture~\cite{Drummond:2007aua} that the $n$-particle
planar amplitude, at least at MHV, should be given by the correlation
function of a certain Wilson loop in region momentum space, calculated
in $\cN=4$ super Yang-Mills theory, but now defined on region momentum
space as opposed to the original space-time.  In
particular their conjecture implied that there should be a completely
unexpected \emph{dual} (super) conformal symmetry: that acting on
region momentum space.  First introduced by Hodges
\cite{Hodges:2009hk}, momentum twistors are the twistors associated to
this region momentum space and serve to make explicit the invariance
of the formalism, but now under {\em dual} (super) conformal symmetry
which acts linearly on momentum twistor space.
Momentum twistor space is still the Calabi-Yau supermanifold
$\CP^{3|4}$, and has all the properties described in \S \ref{NC}; the
only difference is that the space-time is now \emph{region momentum
  space} rather than the physical chiral Minkowski space-time.

We have illustrated how the momentum space MHV formalism can be
derived from the Feynman rules of the twistor action in CSW axial
gauge.  In this section we show how the momentum-space MHV formalism can be
re-expressed as Feynman rules in terms of momentum twistor data that
will be locally related to the momentum space data.  Whereas the twistor
formalism brought out the superconformal invariance of the amplitudes
and performed all integrals; we shall see that the momentum twistor
formalism emphasizes \emph{dual} superconformal invariance, and
computes the \emph{integrand} of scattering amplitudes.  In the next
section we will see that this momentum twistor version of the MHV
formalism arises as the Feynman diagrams for the calculation of the
correlation function of a Wilson loop in twistor space.

\subsection{Momentum Twistor Space}
Momentum twistors encode the kinematic data for scattering amplitudes
of massless particles in a way that builds in momentum conservation.
The region momentum space and corresponding momentum twistor space are
constructed in the following fashion.

As we have seen, the scattering of incoming massless particles is
described by momenta $p^{AA'}_i$ that are null $p_i^2=0$ and obey the
momentum conservation condition $\sum_ip^{AA'}_i=0$. The null
condition is solved by expressing the four-momenta in terms of
spinors\footnote{We denote on shell momenta in this fashion rather
  than $p_{AA'}=p_{A}\tilde{p}_{A'}$ because in momentum twistor
  space, the homogeneous coordinate $\lambda_{A}$ is directly
  identified with the un-primed portion of the momentum.} 
\begin{equation}
p_{i\;AA'} = \lambda_{i\, A} \tilde{\lambda}_{i\, A'}\, .
\end{equation}
For planar amplitudes, the four-momenta involved in a scattering process may be expressed as the differences of region momenta
\begin{equation}
p_i^{AA'} = x_{i+1}^{AA'}-x_i^{AA'}\, .
\end{equation}
The conditions $p_{i}^2=0$ and $\sum_{i}p_{i}=0$ imply that the region momenta $x_i$ form the vertices of a null polygon. Since the the whole polygon may be translated without changing the kinematic data, these vertices $x^{AA'}_i$ live in affine Minkowski space, to which we can associate a twistor space as described in \S \ref{NC}. This associated twistor space has become known as \emph{momentum} twistor space.  It is simply the standard twistor space of region momentum space.

Via the incidence relation \eqref{eqn: inc}, the $x_i$ correspond to complex lines $X_{i}\cong\mathbb{CP}^1$ in (bosonic) momentum twistor space. Since the points $x_i$ and $x_{i+1}$ are null separated in the region momentum space, the associated lines $X_{i}$ and $X_{i+1}$ intersect in twistor space at some point $Z_i$. Therefore we have a sequence of intersecting lines called the momentum twistor polygon - see figure~\ref{polygons}.

\begin{figure}[t]
\centering
\includegraphics[height=4cm]{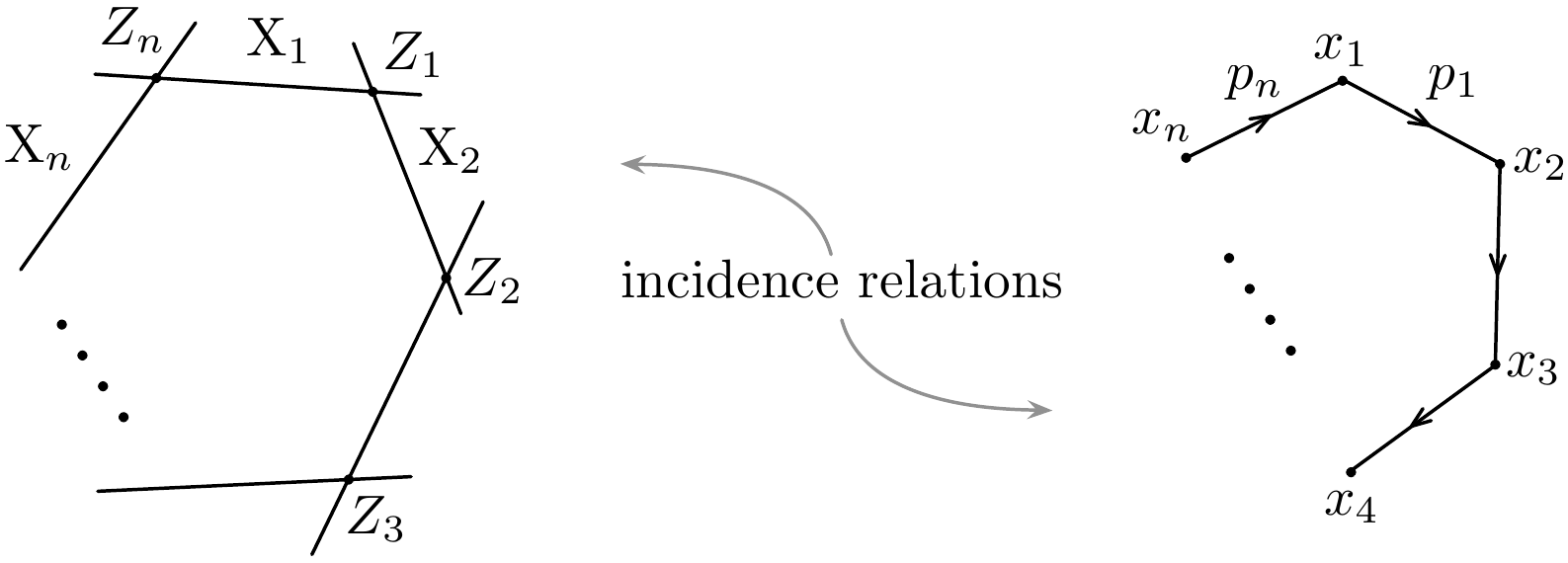}
\caption{\emph{The momentum twistor polygon and its associated null polygon in spacetime.}}
\label{polygons}
\end{figure}

We may now turn the construction around and freely specify any $n$ momentum twistors with components $Z^{\alpha}_i = (\lambda_{i\;A},\mu_i^{A'})$; this determines a sequence of intersecting lines
\begin{equation}
X_{i}\leftrightarrow X_i^{\alpha\beta}  = \frac{Z_{i-1}^{\alpha} Z_i^{\beta} - Z_{i}^{\alpha} Z_{i-1}^{\beta}}{\langle\lambda_{i-1}\lambda_{i}\rangle}\, ,
\end{equation}
which in turn determines a null polygon with cusps
\begin{equation*}
x_i^{AA'} = \frac{\lambda_{i-1}^A\mu_i^{A'}-\lambda_{i}^A \mu_{i-1}^{A'}}{\langle\lambda_{i-1}\lambda_{i}\rangle}\, 
\end{equation*}
in region momentum space.  Note that \emph{any} choice of $n$ momentum twistors defines a polygon with $n$ null edges in region momentum space, so the specification of the kinematic data in terms of momentum twistors is completely unconstrained and therefore solves the momentum conservation condition trivially. 

For planar loop amplitudes, the region momenta and momentum twistors play an important role in defining the \emph{loop integrand}~\cite{ArkaniHamed:2010kv}: the ambiguity in assigning loop momenta between separate Feynman diagrams is absorbed into the overall freedom in choice of origin for the dual spacetime.  Region momenta may then be assigned consistently to all internal regions. For example, in figure \ref{box} the internal region is assigned coordinate $x$ and the four-momenta through internal propagators are expressed as differences, e.g. $l_{1} = x - x_{1}$. The loop integrand is then a rational function $f(x,x_i)$ of the internal and external region momenta.

\begin{figure}[tp]
\centering
\includegraphics[height=5cm]{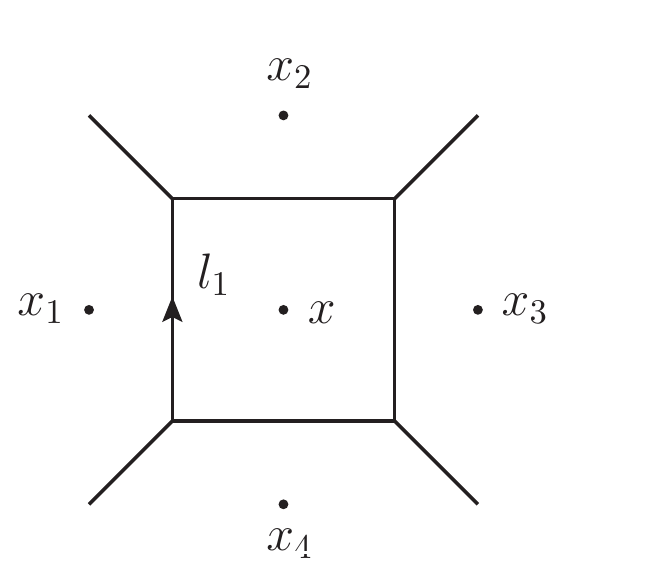}
\caption{\emph{Region momenta may be assigned to internal regions.}}
\label{box}
\end{figure}

In momentum twistor space, internal region momenta $x$ are associated with complex lines $X$, so loop integration is equivalent to integration over these lines in twistor space.  We have already seen a convenient way to express this integral in terms of lines through two auxilliary twistors $Z_{A}$ and $Z_{B}$ via \eqref{volform}: 
\begin{equation}
\rd^{4} x = \frac{\rd^{4}Z_A \wedge \rd^{4}Z_{B}}{\vol(\GL(2,\C)) \la AB\ra^4}
\end{equation}
and $l$-loop integrands $I^{(l)}_n$ become rational functions of momentum twistors
\be{integrand}
I_n^{(l)} = I(Z_1,\ldots, Z_n; (AB)_1,\ldots, (AB)_l)
\ee
The notation $(AB)$ means that the intergand depends only on the skew product $Z_A\wedge Z_B$ or in other words on the line $X$. It is important for consistency that for $l>1$ the integrand is symmetrised over the loop momentum twistors $(AB)_1,\ldots,(AB)_l$.

The extension of this picture to superamplitudes in planar $\cN=4$ SYM follows naturally, again along the lines mentioned earlier in \S \ref{NC}. The momentum twistors now have an additional four fermionic components $Z_i^I = (Z_i^\alpha,\chi_i^a)$ and determine a chiral null polygon with cusps $(x_i,\theta_i)$ through the additional relations
\begin{equation}
\theta^{Aa} = \frac{\lambda_{i-1}^A \chi_i^a - \lambda_i^A \chi_{i-1}^a }{\langle\lambda_{i-1}\lambda_{i}\rangle}\, . 
\end{equation}
Once again, any set of $n$ momentum (super-)twistors corresponds to a null polygon in region momentum space, with $\theta_{i}-\theta_{i-1}=\lambda_{i}\eta_{i}$, where $\eta_{i}$ is now interpreted as the fermionic portion of the super-momentum $P_{i}=(\lambda_{i}\tilde{\lambda}_{i},\eta_{i})$.  Conformal transformations on this region momentum space are called \emph{dual} superconformal transformations.

Internal regions are points in chiral superspace $(x,\theta)$ and again correspond (with a slight abuse of notation) to lines $X\cong\mathbb{CP}^1$ which are now described by two momentum supertwistors $Z^I_A$ and $Z_B^I$. The loop integration measure is extended supersymmetrically as before
\begin{equation*}
\rd^{4|8}x = \frac{\rd^{4|4}Z_A \wedge \rd^{4|4}Z_B}{\vol(\GL(2,\C))},
\end{equation*}
and again we may consider loop integrands as in equation~\eqref{integrand} where all momentum twistors are now super-momentum twistors. The $l$-loop integrand may be expanded in powers of the fermionic components
\begin{equation} 
I_n^{(l)} = I_{n,0}^{(l)} \;+\; I_{n,1}^{(l)} \;+\; \cdots \;+\; I_{n,n-4}^{(l)}\, ,
\end{equation}
where $I^{(l)}_{n,k}$ has Grassmann degree $4k$ and corresponds to N$^k$MHV amplitudes.

It has been shown by use of BCFW recursion (for which see \S\ref{BCFW}) \cite{ArkaniHamed:2010kv,Boels:2010nw} that the loop integrands $I_{n,k}^{(l)}$ in planar $\cN=4$ SYM are annihilated by generators of superconformal and dual superconformal symmetries, which together generate a Yangian symmetry algebra. In terms of momentum twistors, the generators of the dual superconformal and standard superconformal are respectively~\cite{Drummond:2009fd}
\begin{eqnarray}
{J^I}_J & = & \sum\limits_{i=1}^n\; Z_i^I \frac{\partial}{\partial Z_i^J} \\
{J^{(1)I}}_J & = & \sum\limits_{i<j}(-1)^K\left[ Z_i^I\frac{\partial}{\partial Z_i^K} Z_j^K \frac{\partial}{\partial Z_j^J}  - (i \leftrightarrow j)\right]
\end{eqnarray}
The Yangian invariant quantities appearing in amplitudes of planar $\cN=4$ SYM and the numerous non-trivial relations among them are captured by a Grassmannian contour integral~\cite{ArkaniHamed:2009si,ArkaniHamed:2009sx,ArkaniHamed:2009dn,ArkaniHamed:2009dg,ArkaniHamed:2010gh}, as we discuss briefly in \S\ref{furthertopics}.


\subsection{MHV Diagrams in Momentum Twistor Space}

We now explain how the momentum space MHV diagram formalism of \cite{Cachazo:2004kj}, derived in \S \ref{momentumspace}, can be reformulated in momentum twistor space to compute the loop integrands $I_{n,k}^{(l)}$ \cite{Bullimore:2010pj}. As we shall see, the result is that vertices become unity and that each propagator contributes a single dual superconformal\footnote{R-invariants are now \emph{dual} superconformal invariants because their arguments are momentum twistors.} R-invariant $[\,,,,,]$ to the loop integrand, whose arguments are determined by a simple rule. This allows expressions for loop integrands to be quickly generated in a form that manifests cyclic invariance. In addition, the independence of choice of reference twistor $Z_*$ forms a highly non-trivial check on the result.

Because momentum twistors encode supermomentum conservation, we cannot of course encode the delta function part of the amplitude.   In fact it turns out that, in order to obtain dual conformal invariant expressions, we factor out by the MHV amplitude.  Thus, below, we rewrite the diagrams from the MHV formalism in momentum twistor variables, but with this MHV prefactor removed.

\subsubsection*{\textit{Tree Diagrams}}
As discussed in \S \ref{twistorspace} and \ref{momentumspace}, the simplest diagrams have a single propagator connecting two vertices and contribute to the NMHV tree amplitude. For the particular diagram illustrated in figure~\ref{NMHVfig}, we have the following momentum space expression
\begin{equation}
\int \rd^4\eta \; A^{(0)}_{\mathrm{MHV}}(i,\ldots,j-1,\{\lambda,\eta\})\, \frac{1}{(x_i-x_j)^2}\, A^{(0)}_{\mathrm{MHV}}(\{-\lambda,\eta\},j,\ldots,i-1)
\end{equation}
where $(x_i-x_j)=(p_1+\cdots+p_{j-1})$ is the four-momentum though the propagator and the Grassmann integration performs the sum over states propagating in the channel. The momentum spinor for the off-shell propagator momenta is $\lambda_A = p_{AA'}\iota^{A'}$ according to the CSW prescription~\cite{Cachazo:2004kj} where $\iota^{A'}$ is the auxilliary CSW reference spinor of \eqref{CSWref}.

\begin{figure}[htp]
\centering
\includegraphics[height=2.5cm]{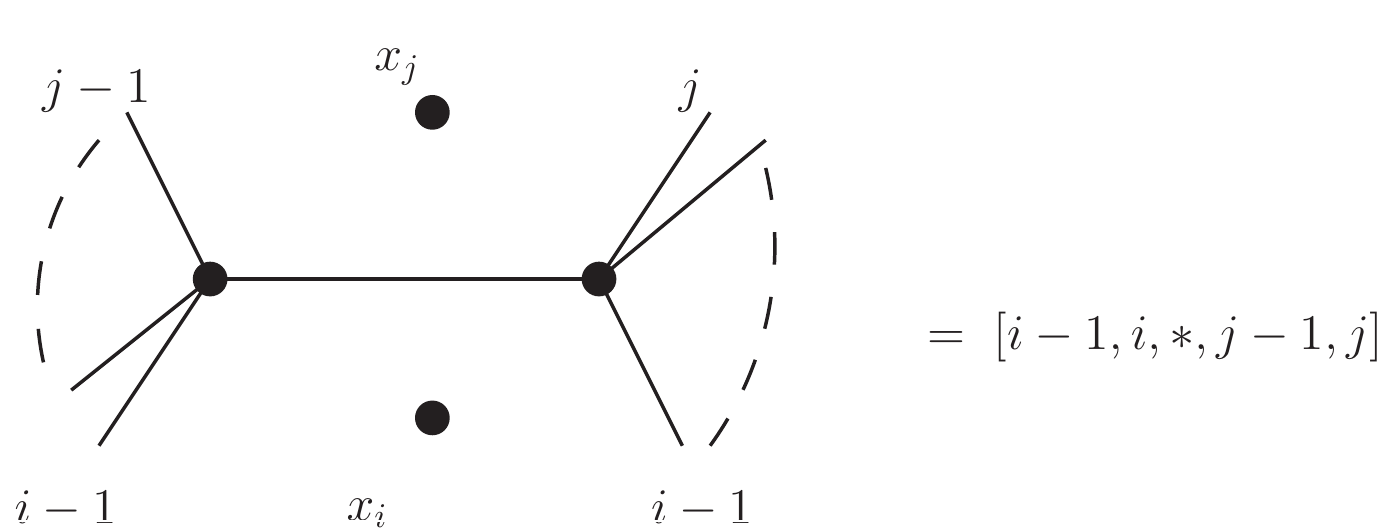}
\caption{\emph{The MHV diagrams contributing to the NMHV tree amplitude.}}
\label{NMHVtree}
\end{figure}

We now extract an overall factor of the MHV amplitude $A^{\mathrm{MHV}}(1,\ldots,n)$ to expose the dual superconformal structure of the diagram, and introduce the momentum twistor $Z^I_*=(0,\iota^{A'},0)$ containing the auxilliary spinor as its only non-zero component. Once this has been done the contribution from the diagram becomes~\cite{Bullimore:2010pj}
\begin{equation}
\left[i-1,i,\rf, j-1,j\right]
\end{equation}
where $[\,,,,,]$ is the dual superconformal R-invariant introduced in \S \ref{prop-delta}. Summing over inequivalent diagrams, the NMHV tree amplitude may then be expressed as follows
\be{NMHV}
	I^{(0)}_{n,1} = \frac{1}{2}\, \sum_{ i<j} \; [i-1,i,\rf, j-1,j] \, .
\ee
This expression manifests cyclic invariance in the external particles but breaks Lorentz invariance beccause of the dependence on the choice of reference momentum twistor $Z_*$.  Note that as in \S \ref{twistorspace}, propagators correspond to R-invariants, but now they are dual conformal and can be thought of as linking the two pairs of external legs adjacent to the propagator via $Z_*$ as in twistor space, but we will see in the next section that the propagator can also be regarded as linking the nearest regions via the reference twistor. 

We can nevertheless demonstrate the dual superconformal symmetry of the tree amplitude. However, given any R-invariant $[a,b,c,d,e]$ we can perform a deformation $Z_a\rightarrow Z_a+tZ_f$ and apply Cauchy's theorem to derive the following linear identity
\begin{equation}
	[a,b,c,d,e] + [b,c,d,e,f] + [c,d,e,f,a] + [d,e,f,a,b] + [e,f,a,b,c] + [f,a,b,c,d] =0\, .
\end{equation}
Performing the deformation $Z_*\rightarrow Z_*+tZ_{*'}$ on each term in equation~\eqref{NMHV}, all dependence on $Z_*$ cancels pairwise in the sum, so $Z_*$ may be replaced by any other reference twistor $Z_{*'}$. In particular, when $Z_*$ is an external momentum twistor, then equation~\eqref{NMHV} reduces immeadiately to the BCFW expression~\cite{ArkaniHamed:2010kv}.

For N$^2$MHV tree amplitudes we have a new phenomenon.  The diagrams contain three vertices connected by two propagators, as illustrated in figure~\ref{NNMHVtree}. For the generic diagram where $j<k$ we associate a dual superconformal R-invariant with each propagator, to obtain
\begin{equation}
[i-1,i,\rf, j-1,j]\;[k-1,k,\rf, l-1,l]\, .
\end{equation}
However, when there are two propagators connected to adjacent sites on a vertex, we have a boundary diagram in the same sense as \S \ref{twistorspace}, one or more of the nearest external legs is the other side of an adjacent propagator and the arguments of the R-invariants are shifted. For the case when $j=k$ the correct expression is
\be{mshift1}
[i-1,i,\rf, j-1,j ]\;[k-1,k,\rf, i-1,\widehat{i}]\, 
\ee
where the shifted momentum twistor is defined as
\be{mshift2}
\widehat{Z_i} = (\rf, j-1,j,i-1)\, Z_i - (\rf , j-1,j,i)\, Z_{i-1}\, .
\ee
Geometrically, the shifted twistor is the intersection of the line $X_{i}$ with the plane through the line $X_{j}$ and the auxilliary twistor $Z_*$.  Note that this shifted geometry is different from the shifts required for boundary terms in the twistor space picture of \S\ref{twistorspace}. 
\begin{figure}[htp]
\centering
\includegraphics[height=2.3cm]{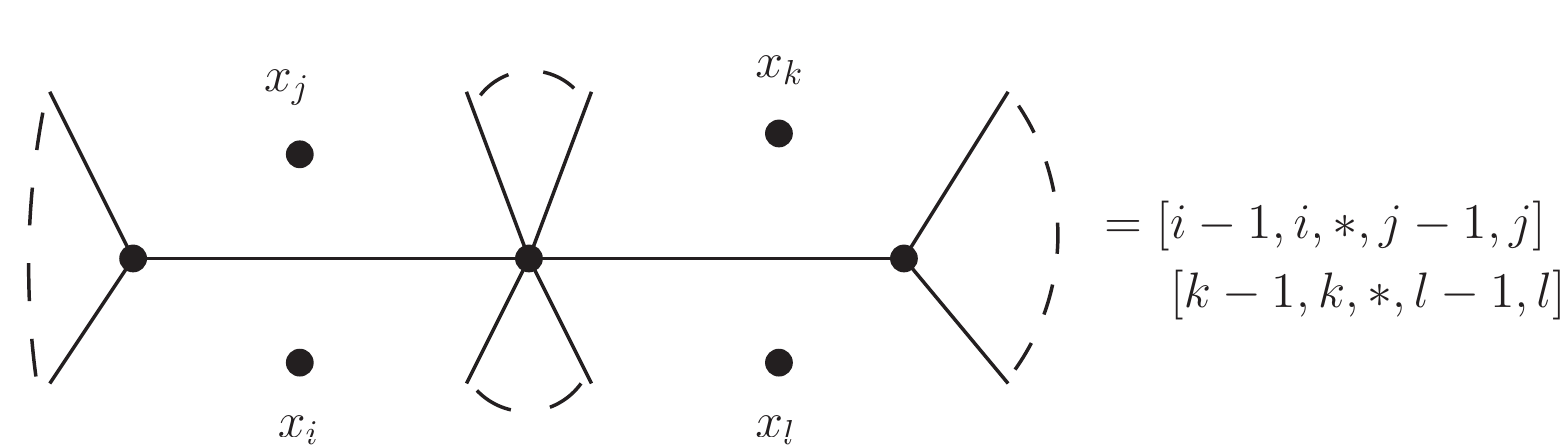}
\caption{\emph{The MHV diagrams contributing to the N$^2$MHV tree amplitude.}}
\label{NNMHVtree}
\end{figure}

A general tree-level N$^k$MHV amplitude corresponds to a product of $k$ dual superconformal R-invariants built in this way, with boundary terms defined in terms of shifted momentum twistors \cite{Bullimore:2010pj}.  Because the geometry of these shifts does not rely on our ability to define a line from the external particles at any one vertex, boundary-boundary terms are not a distinguished class of diagram in the momentum twistor formalism.

\subsection*{\textit{Loop integrands}}
Let 
$I^{(1)}_{n,0}$ denote  the one-loop contribution to the planar MHV amplitude. The relevant MHV diagrams have two propagators connecting two vertices and are illustrated in figure~\ref{MHV1loop}. Once again we assign a R-invariant to each propagator; however, now the arguments involve the momentum twistors $(AB)$ associated with the loop integration. The correct result for the diagram in figure is
\begin{equation}\label{Loop1}
[i-1,i,\rf, A,B']\;[j-1,j,\rf, A,B'']
\end{equation}
where 
\begin{equation}
B' = (AB)\cap\langle\rf ,j-1,j\rangle, \qquad  B''=(AB)\cap\langle\rf ,i-1,i\rangle\, .
\end{equation}
We remark that although the $B'$ and $B''$ are determined by  the geometry, $A$ is an arbitrary twistor on the line $(AB)$ and the expression is actually independent of that choice.  The fermionic integrations may be performed easily, leading to the following expression for the bosonic integrand \cite{Bullimore:2010pj}:
\begin{multline*}
\frac{1}{2}\sum\limits_{i<j} \\ \frac{(\rf ,i-1,i,[A\ra\la B],j-1,j,\rf)^2}{(A,B,i-1,i)(A,B,j-1,j)(\rf ,i-1,i,j-1)(\rf , i-1,i,j)(\rf ,i-1,j-1,j)(\rf ,i,j-1,j)}
\end{multline*}
Each term in this expansion contains four spurious propagators of the form $(\rf,,,)$ and is highly chiral. However, the parity even part of this expression has been shown to agree with the standard box expansion at the level of the integrand, and the BCFW expression is again found immediately by choosing the auxilliary twistor $Z_*$ to be an external momentum twistor.
\begin{figure}[htp]
\centering
\includegraphics[height=3.5cm]{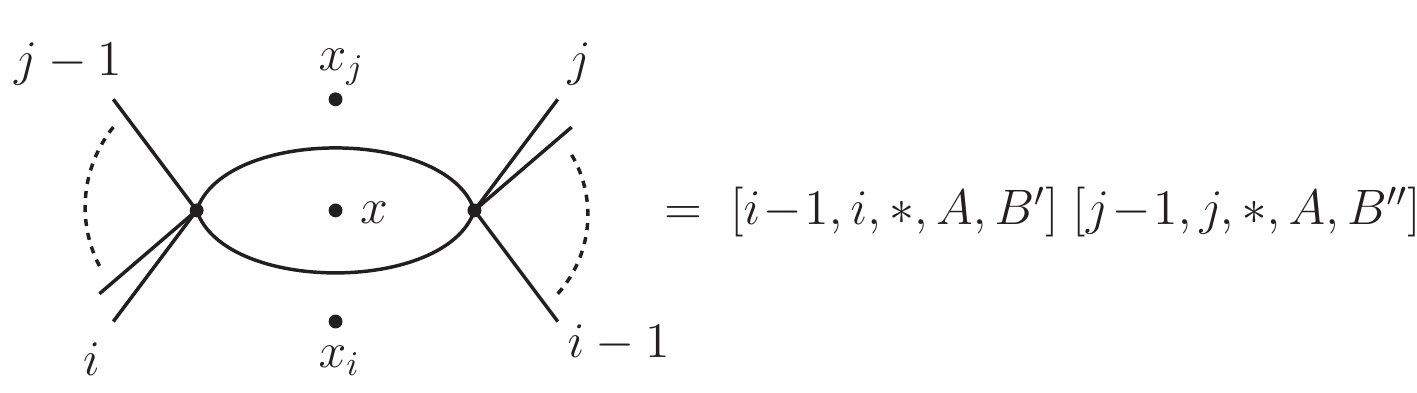}
\caption{\emph{The MHV diagrams contributing to the MHV one-loop integrand.}}
\label{MHV1loop}
\end{figure} 

\subsection*{\textit{Generic Diagrams}}

For a general diagram at arbitrary Grassmann degree and loop level, the vertices are unity and each propagator contributes a dual superconformal R-invariant $[,,\rf ,,]$. There is a simple local rule for assigning the momentum twistor arguments of this R-invariant for any propagator. Let the region momenta $u,v,\ldots$, (which may be external or internal regions) correspond to lines $U$, $V,\ldots$ through pairs of momentum twistors $(U_1U_2), (V_1V_2),\ldots$. The general assignment of momentum twistors for the invariant may then be summarised by figure \ref{Assignment}.
\begin{figure}[htp]
\centering
\includegraphics[height=6cm]{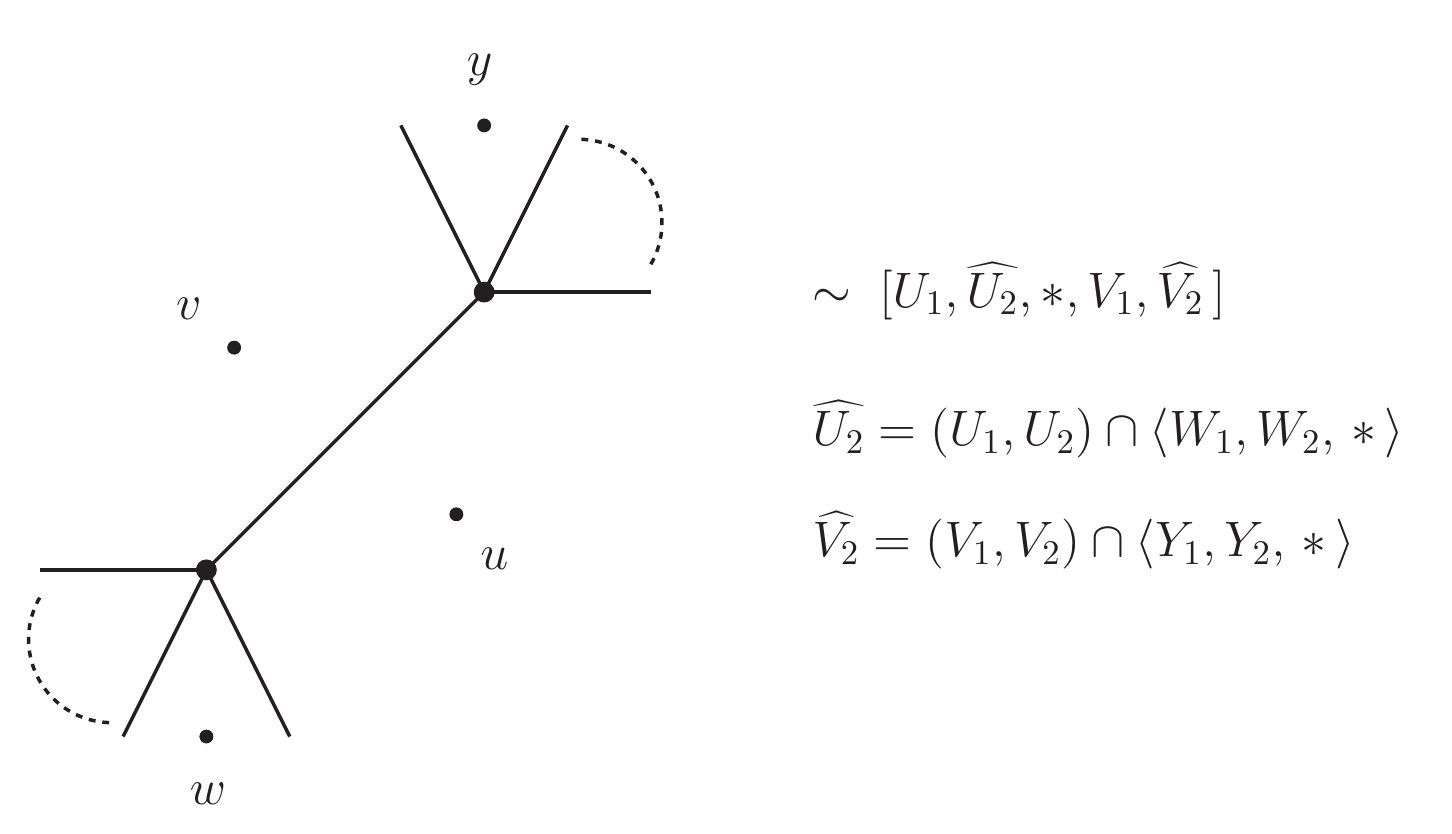}
\caption{\emph{Assignment of momentum twistors for a propagator in a generic diagram}}
\label{Assignment}
\end{figure}
(This prescription requires a choice of orientation of the diagram, but the expression arrived at turns out to be independent of the choice.)


\section{Holomorphic Wilson Loops in Twistor Space}
\label{HWL}

Although we have a fundamental derivation of the MHV diagrams as Feynman diagrams for the twistor action in ordinary twistor space, now that we have reformulated them in momentum twistor space, we would like to know how they arise from some construction in momentum twistor space directly.  In this section we explain how the amplitude reformulated in momentum twistor space arises as the correlation function of the holomorphic Wilson loop in momentum twistor space defined by the polygon we have been using, computed in the theory defined by the twistor action.  The Feynman integrals for the computation of this correlation function are precisely those for the MHV formalism that we have just reformulated in momentum twistor space.  However, the Feynman diagrams for the amplitude are not those for the correlation function of the Wilson loop, but are related by planar duality.  That this is meaningful relies on the fact that the contributions from the vertices is 1, and the only nontrivial contributions are from the propagators.

This correspondence was motivated by and extends the duality between space-time Wilson loops and the MHV sector of  scattering amplitudes.  This was first discovered at strong coupling~\cite{Alday:2007hr} through AdS/CFT and subsequently observed at weak coupling~\cite{Brandhuber:2007yx,Drummond:2007aua,Drummond:2007cf,Drummond:2007bm,Drummond:2008aq}. The holomorphic Wilson loop in momentum twistor space gives the full amplitude, and not just the MHV sector.  In the twistor formulation, the correspondence is clear, and easy to prove, whereas on space-time the integral representations for the amplitude are quite different.   It leads directly to many approaches for computing the loops integrands, including the MHV diagram formalism~\cite{Mason:2010yk} and the BCFW recursion relations~\cite{Bullimore:2011ni} \S\ref{BCFW}.

We will first consider the abelian theory where the expectation value can be computed exactly and then indicate the general structure of the Feynman diagrams in the non-abelian case.



\subsection{Motivation and the abelian theory}

To motivate the holomorphic Wilson loop, we return to the simplest MHV diagram with a single propagator. In momentum twistor space, this diagram contributes a single dual superconformal R-invariant $[i-1,i,\rf, j-1,j]$ from \eqref{NMHV}. By \S \ref{prop-delta}, we see that this R-invariant may be expressed as an integral of a collinearity delta-function:
\be{Rdef}
[\ie, i,\rf,\je,j] = \int \frac{\d s}{s}\frac{\d t}{t} \; \odel^{2|4}(Z(s),\rf,Z(t))
\ee
for momentum twistors  
\begin{eqnarray*}
Z(s) &=& Z_{i-1}+sZ_i \\
Z(t) &=& Z_{j-1}+tZ_j 
\end{eqnarray*}
parameterising the complex lines $X_{i}$ and $X_{j}$ in the momentum twistor polygon.  The delta-function appearing on the right hand side in equation~\eqref{Rdef} is the propagator for the twistor action in the axial gauge from \eqref{propagator}:
\begin{equation}
\odel^{2|4}(\rf,Z(s),Z(t))=\Delta(Z(s),Z(t))\, .
\end{equation}
Therefore, this contribution is that of a momentum twistor Feynman diagram where a single propagator connects the lines $X_{i}$ and $X_{j}$; this would arise from the expectation value 
\be{}
\left\la \tr \left(\int_{X_i}\omega_i\wedge \cA\right) \left(\int_{X_j}\omega_j\wedge \cA\right)\right\ra
\ee
for forms $\omega_i$, $\omega_j$ are meromorphic 1-forms defined on the lines $X_i$, $X_j$ by
\be{}
\omega_i=\int_\PT \bar\delta^{2|4}(i-1,Z,i)\wedge \D^{3|4}Z \, .
\ee
The diagram for this correlator is the planar dual graph of the momentum space MHV diagram - see figure~\ref{Dualgraph1}.  We will see that this duality between momentum space MHV diagrams and momentum twistor Feynman diagrams remains true for all tree amplitudes and planar loop integrands.

\begin{figure}[t]
\centering 
\includegraphics[height=3cm]{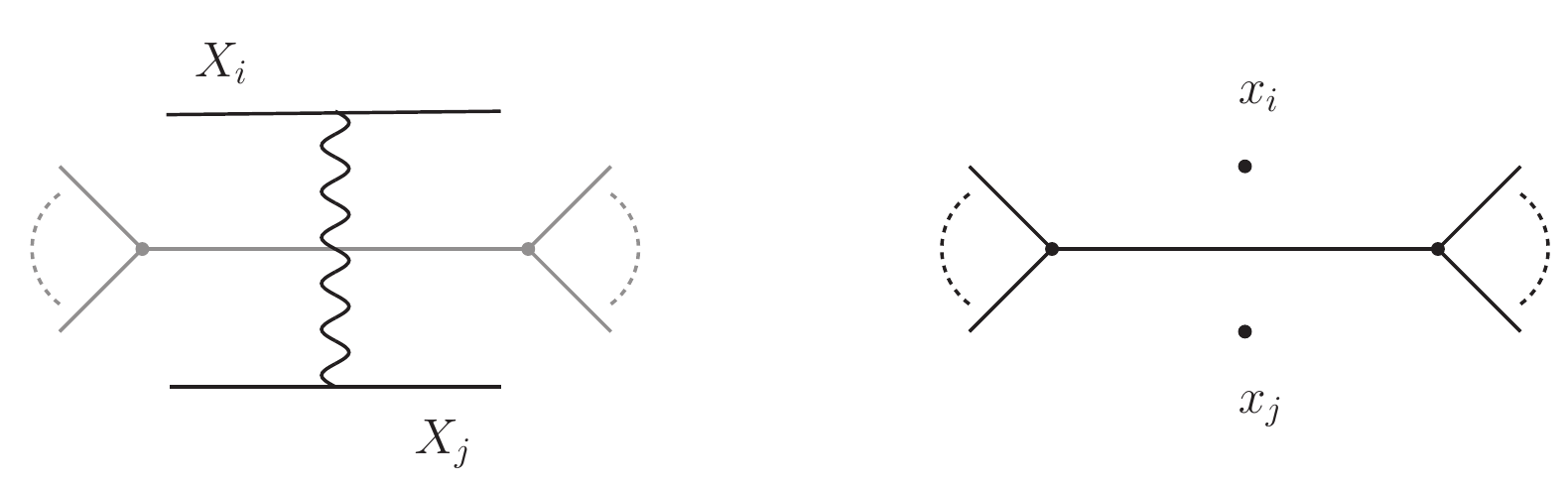}
\caption{\emph{The momentum twistors Feynman diagram and planar dual MHV diagram corresponding to $[\ie,i,\rf,\je,j]$.}}
\label{Dualgraph1}
\end{figure}

The full tree-level NMHV amplitude is then given by
\be{}
A^0_{\mathrm{NMHV}}= \left\la \left(\int_{C} \omega_C \wedge \cA \right)^2\right\ra\, .
\ee
where $C=\cup_i X_i$ is our twistor space polygon and the meromorphic $1$-form $\omega_{C}$ is be defined by
\begin{equation*}
	\omega_{C}=\int_\PT \bar{\delta}_{C}\wedge\D^{3|4}Z\ ,\qquad
	\bar{\delta}_{C}(Z)=\sum_{i=1}^{n}\bar{\delta}^{2|4}(Z_{i-1},Z,Z_{i})\ ,
\end{equation*}
where $\bar{\delta}_{C}$ is dual to $C$, and clearly $\omega_C|_{X_i}=\omega_i$. 

More geometrically, in the abelian theory, we can consider this NMHV amplitude to be the first nontrivial term in the expectation value of the holomorphic Wilson loop:
\begin{equation}
\big\la W[C] \big\ra = \left\langle  \exp \int_{C}\omega_C\wedge \cA \, \right\rangle \, .
\end{equation}
This can be evaluated exactly in the abelian theory.  
Let us first concentrate on the self-dual abelian theory, where there are no MHV vertices and the cubic holomorphic Chern-Simons term vanishes.  Our path integral expression becomes:
\begin{multline}\label{EWLSD}
\left\langle W[C] \right\rangle =\mathcal{Z}^{-1}\int \mathcal{D}\cA e^{-S_{\mathrm{SD}}[\cA]}\exp\left(\int_{\PT}\cA\wedge\bar{\delta}_{C}\wedge\D^{3|4}Z\right) \\
=\mathcal{Z}^{-1}\int \mathcal{D}\cA \exp\left(-\int_{\PT}\D^{3|4}Z\wedge\cA\wedge(\dbar\cA-\bar{\delta}_{C})\right).
\end{multline}

This can be evaluated explicitly by introducing a background gauge field $\cA_{0}$ which satisfies $\dbar\cA_{0}=\bar{\delta}_{C}$, and expanding about $\cA=\mathcal{B}-\cA_{0}/2$.  The dependence on $\cA_{0}$ can then be pulled outside the path integral to get
\be{EWLSD2}
\left\langle W[C]\right\rangle = \exp\left(-\frac{1}{4}\int \D^{3|4}Z\wedge\cA_{0}\wedge\dbar\cA_{0}\right).
\ee
Solving for the background field $\cA_0$ using the twistor propagator, we find an expression for the \emph{logarithm} of the momentum twistor Wilson loop:
\be{}
\log \big\la W(C) \big\ra = \frac{1}{4} \int \Delta(Z,Z')\; \odel_{C}(Z)\; \odel_{C}(Z')\; \D^{3|4}Z\; \D^{3|4}Z'\, .
\ee
In axial gauge this becomes
\be{}
\log \big\la W(C) \big\ra = \sum\limits_{i<j}[i\!-\!1,i,\rf, j\!-\!1,j],
\ee
which is exactly the NMHV tree amplitude of \eqref{NMHV}.

To see what happens when MHV vertices are included, we can expand around the self-dual sector, to include the simplest 2 point MHV vertex in the abelian theory. Hence, the path integral is again Gaussian, and may be evaluated exactly with the result
\be{logabelian}
\log \big\la W(C) \big\ra =  \sum\limits_{i<j} \left( [i\!-\!1,i,\rf, j\!-\!1,j] + \int \rd^{4|8}x\;  [i\!-\!1,i,\rf, A,B']\;[j\!-\!1,j,\rf, A,B''] \right)
\ee
where $X$ is the complex line through the momentum twistors $Z_A$ and $Z_B$. In the following section we will see that the second term in equation~\eqref{logabelian} is precisely the MHV one-loop amplitude. Hence, in the full abelian theory the \emph{logarithm} of the momentum twistor Wilson loop is the sum of the NMHV tree and MHV one-loop amplitudes.

This already demonstrates a striking feature of the amplitude Wilson loop correspondence, that the tree level amplitude for the full $\cN=4$ SYM is computed in the self-dual sector of the Wilson loop, and the expansion around that self-dual sector will be the loop expansion for the amplitude.  However, it is clear that the full tree amplitude is not simply the exponential of the NMHV amplitude (nor is the multiloop amplitude the expontential of the 1-loop one) and we will see that to extend this correspondence further, we will need to introduce nonabelian holomorphic Wilson loops.  However, this is not something that can be done for a general curve $C$, as in the complex there is not in general a notion of parallel transport for sections of a holomorphic vector bundle on a complex curve, and we will see that we need to use its special structure as a nodal curve with rational components.


\subsection{Holomorphic Wilson Loops}
For a curve in space-time, a Wilson loop is simply (the expectation of) the trace of the holomony around a closed curve, and we are in particular interested in the case of our polygon in region momentum space.  A  $\dbar$-operator on a bundle over a complex curve doesn't generally lead to a notion of parallel transport around the curve.  However, the Ward transform allows one to reformulate the space-time holonomy into twistor space as a functional of the $(0,1)$-form gauge field $\cA$.  These concepts can all be defined quite naturally in the bosonic setting, where the momentum twistor polygon becomes a nodal curve in $\CP^{3}$ and the gauge field is reduced to the leading component of $\cA$, namely $a\in\Omega^{0,1}(\PT_{b}',\mathrm{End}(E))$ where $\cA$ (or $a$) defines a $\dbar$-operator on the bundle $E\rightarrow \PT$ (or its bosonic `body').  The supersymmetric Wilson loop is then found by passing to the immediate supersymmetric extension of these constructions.  Here, we present only the final supersymmetric results as we are primarily concerned with $\cN=4$ SYM; see \cite{Mason:2010yk,Bullimore:2011ni} for full details.

The momentum twistor polygon $C$ is a (nodal) complex curve whose components are Riemann spheres.  It is a standard fact from complex geometry that homorphic bundles on the Riemann sphere that are topologically trivial are also generically holomorphically trivial. In particular it will be trivial for any choice of $\dbar$-operator $\dbar_\cA$ that is not too far from the trivial one, as will be the case in the perturbative context in which we are working.
\begin{figure}[htp]
\centering 
\includegraphics[height=3cm]{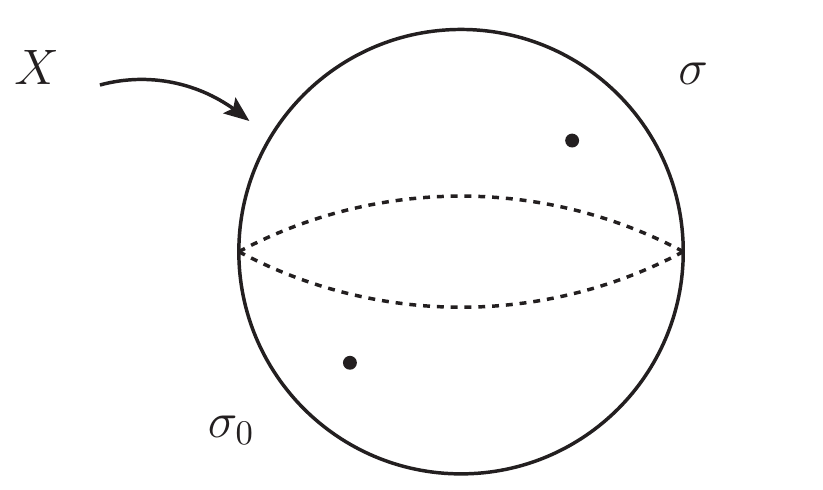}
\caption{\emph{The monodromy operator $U(\sigma,\sigma_0)$ propagates from an initial point $\sigma_0$ to a final marked point $\sigma$ along the line $X$.}}
\label{Line}
\end{figure}
Consider first a single component $X\cong\mathbb{CP}^1$ of the momentum twistor polygon and choose a local inhomogeneous coordinate $\sigma$. The first step is to introduce a global holomorphic frame.  To emphasize the anology with parallel propagation, we will work with an operator $U$ that propagates from an initial point $\sigma_0$ to a final marked point $\sigma$ on the Riemann sphere; see figure~\ref{Line}.  $U$ will be the identity in a global holomorphic frame on the Riemann sphere and can be made explicit by first solving for a frame of $E|_X$ that holomorphically trivializes $E|_{X}$.  This is a gauge transformation $h$ such that
\begin{equation*}
h^{-1}\dbar_{\cA}|_{X}\;h=\dbar |_{X},
\end{equation*}
or equivalently
\be{Udef}
(\dbar |_{X}+\cA(\sigma))h(\sigma)=0,
\ee 
where the notation $\cA(\sigma)$ indicates that the gauge field is pulled back from twistor space to the point $\sigma\in X$.  Now, define
\be{Udef2}
U(\sigma, \sigma_{0}):=h(\sigma)h^{-1}(\sigma_{0}).
\ee
Clearly $U(\sigma, \sigma_{0}):E|_{\sigma_{0}}\rightarrow E|_{\sigma}$ and is subject to the initial condition $U(\sigma_{0},\sigma_{0})=\mathbb{I}$.  Furthermore, \eqref{Udef} makes it clear that $U$ satisfies
\be{conc}
 U(\sigma_{3},\sigma_{2})U(\sigma_{2},\sigma_{1}) = U(\sigma_{3},\sigma_{1}) 
\ee
and under gauge transformations $\dbar_{\cA}\rightarrow g\dbar_{\cA}g^{-1}$ transforms according to
\be{gt}
U(\sigma_{2},\sigma_{1}) \longrightarrow g(\sigma_{2})U(\sigma_{2},\sigma_{1})g(\sigma_{1})^{-1}\, .
\ee
This will play the role in our holomorphic context of the parallel propagator along a real curve in real Chern-Simons or Yang-Mills theory.  We can concatenate these together from node to node around the complex polygon in twistor space to obtain a holonomy around the polygon.  This is precisely the holonomy that one finds for the gauge field obtained on space-time via the Ward correspondence, see \cite{Mason:2010yk}.  The expectation value of its trace will be the holomorphic Wilson loop that we are after.

We now find an explicit perturbative expression for $U$ by reformulating equations~\eqref{Udef}, \eqref{Udef2} as the boundary value problem
\be{Udef3}
(\dbar +\cA)U=0\, , \qquad U(\sigma_0)=\mathbb{I} \, .
\ee
Formally we can solve for this by rewriting the equation as $U=\mathbb{I} + \dbar^{-1} (\cA U)$
and iterating.  The operator $\dbar^{-1}$ is integration against the
 Green's function for the $\dbar$-operator on the complex line $X$ with zero boundary condition at $\sigma_0$ which is the meromorphic one-form
\begin{equation}
\omega(\sigma) = \frac{(\sigma_1 -\sigma_0)\d \sigma}{(\sigma_1 -\sigma)(\sigma-\sigma_0)}\, , \qquad \dbar \omega(\sigma)= \left(\delta^2(\sigma-\sigma_0)- \delta^2(\sigma-\sigma_1)\right)\d^2\sigma\, .
\end{equation}
This leads to the perturbative solution for the momodromy operator:
\begin{eqnarray}
\label{pathorder}
U_X(\sigma,\sigma_0) &=& \mathbb{I} + \int_{X} \frac{(\sigma-\sigma_0)\d\sigma_1 \wedge\cA(\sigma_{1})}{(\sigma-\sigma_1)(\sigma_1-\sigma_0)} + \int_{X^2} \frac{(\sigma-\sigma_0)\d\sigma_{2}\d\sigma_{1}\wedge\cA(\sigma_{2})\cA(\sigma_{1})}{(\sigma -\sigma_2)(\sigma_2-\sigma_1)(\sigma_1-\sigma_0)} + \cdots \nonumber\\
&=& \mathbb{I}+  \sum\limits_{n=1}^{\infty}\int_{X^n} \prod\limits_{i=1}^n \omega(\sigma_i) \wedge \cA(\sigma_i)\, .
\end{eqnarray}
The meromorphic form $\omega(\sigma_i)$ has simple poles at the initial point $\sigma_0$ and the next point $\sigma_{i+1}$ in the ordering (which is $\sigma$ when $i=n$). Note that all of the intermediate points $\{\sigma_1,\ldots,\sigma_n\}$ in each term are integrated over the whole line $X$. Equation~\eqref{pathorder} defines the holomorphic analogue of path-ordering in the sense that the matrix multiplication follows the ordering of the sequence of points $\sigma_i$ $i=0,1,2,\ldots$ (although these points are of course not themselves ordered on the Riemann sphere) and we therefore write this perturbative solution as
\begin{equation}
U_X(\sigma,\sigma_0) = \mathrm{P} \exp \int_{X} \omega \wedge \cA 
\end{equation}
by analogy with the parallel propagator along a real curve.

To form the Wilson loop, we now consider the whole momentum twistor polygon. This consists of a sequence of intersecting lines
\begin{equation*}
C = X_1\cup\cdots\cup X_n
\end{equation*}
whose images intersect pairwise at the momentum twistor points $Z_i=X_{i}\cap X_{i+1}$. The monodromy operator on the component $X_i$ which propagates from momentum twistor $Z_{i-1}$ to momentum twistor $Z_i$ is denoted by $U_{X_i}(\sigma_i,\sigma_{i-1})$, where $\sigma_{i}$ labels the momentum twistor $Z_{i}$ on the curve. The momentum twistor Wilson loop is now defined by propagating from point to point around the momentum twistor polygon $C$ and taking the trace \cite{Mason:2010yk}:
\begin{eqnarray}
\label{holWL}
W[C] &=& \tr \left[ U_{X_n}(\sigma_{n},\sigma_{n-1})\ldots U_{X_2}(\sigma_{2},\sigma_{1})U_{X_1}(\sigma_{1},\sigma_{n})\right] \nonumber\\
&=& \tr\mathrm{P}\, \exp \int_{C} \omega\, \wedge \cA\, .
\end{eqnarray}
The properties~\eqref{conc} and~\eqref{gt} of the monodromy ensure the gauge invariance of this operator and independence of the choice of initial point. This operator is our definition of the holomorphic Wilson loop in momentum twistor space.  Classically, it  is also a transcription of the corresponding Wilson loop around the space-time polygon. 


\subsection{The Expectation Value}

We would now like to compute the quantum expectation value of the holomorphic Wilson loop with respect to the twistor action:
\be{expWL}
\left\langle W[C]\right\rangle = \mathcal{Z}^{-1}\int \mathcal{D}\cA\;\; W[C] \, \exp(-S[\cA])\, ,
\ee
where $S[\mathcal{A}]$ is the complete twistor action for $\cN=4$ SYM defined in \eqref{eqn: TA}, and $\mathcal{Z}$ is its partition function. 

Recall that the twistor action has two components: $S[\cA]=S_{\mathrm{SD}}[\cA]-\frac{\varepsilon}{2}I[\cA]$, where $\varepsilon$ is a parameter expanding the theory around the SD sector.  Hence, by expanding \eqref{expWL} in $\varepsilon$, the loop corrections will be equivalent to computing expectation values of $W[C]$ in a holomorphic Chern-Simons theory but with insertions from the interactions in $I[\cA]$.  From \eqref{eqn: TASD}, we know that these interactions contain a space-time integral $\int \d^{4|8}x$, which can be pulled outside to define the \emph{Wilson loop integrand} in the same manner as described for scattering amplitudes earlier in \S \ref{momtwist}.  

The momentum twistor Wilson loop integrands may also be expanded in the fermionic momentum twistor components, and we will use the notation
\be{}
W_{n,k}^{(l)}(Z_1,\ldots,Z_n,(AB)_1,\ldots,(AB)_l)
\ee
to denote the $l$-loop integrand with Grassmann degree $O(\chi^{4k})$. It has now been proven that the integrands $W_{n,k}^{(l)}$ are equal to the corresponding amplitude integrands $I_{n,k}^{(l)}$ by demonstrating that they satisfy and all-loop generalisation of BCFW recursion relations~\cite{Bullimore:2011ni} \S\ref{BCFW}.

\subsubsection{Perturbative Expansion and MHV Diagrams}

We now consider the non-abelian case and expand the expectation value~\eqref{expWL} perturbatively around the self-dual sector of the theory in the large $N$ (planar) limit.  We will find that axial gauge Feynman diagrams for the expectation value~\eqref{expWL} are in one-to-one correspondence with MHV diagrams for scattering amplitudes. 

As before, in our axial gauge, the cubic term in the holomorphic Chern-Simons action vanishes, and the log det term expands to give the infinite sum of additional interaction terms given in
\eqref{eqn: vert2} as
\begin{equation*}
\int_{\M_{\R}} \d^{4|8}x\sum_{m=2}^{\infty}\int_{X^{m}}\tr\left(\prod _{i=1}^m \frac{\cA(\sigma_i)\wedge\d\sigma_i }{\sigma_i-\sigma_{i-1}} \right).
\end{equation*}
Each term corresponds to an interaction connecting the line $X$ (which in this context will become a supersymmetric loop integration variable) to $m$ insertions of the twistor gauge field $\cA$ either on other vertices, or on $C$. 

We first start  by working in the self-dual sector omitting the MHV vertices.  The simplest Feynman diagram is a single propagator connecting two components $X_i$ and $X_j$ of the momentum twistor polygon; we have already seen that these diagrams are in one-to-one correspondence by planar duality with the MHV diagrams for the NMHV tree amplitude $I_{n,1}^{(0)}$.

The next simplest case is of two propagators connecting two pairs of lines, $X_i$ to $X_j$ and $X_k$ to $X_l$.  It is a standard fact that in the planar limit, these two lines must not cross (the crossing diagrams are suppressed by $1/N$).  Thus we can order cyclically $i<j\leq k<l\leq i$.  The generic contribution will then be the product of R-invariants $[i-1,i,\rf,j-1,j] [k-1,k,\rf,l-1,l]$.  However, when $j=k$ or $l=i$, we must perform the calculation  more carefully, and we find that the answer is again a product of R-invariants, but now the twistors are shifted according to the prescription given in \eqref{mshift2}.  These diagrams are again easily seen to be the planar duals to the N$^2$MHV  tree-level MHV diagrams. 

This pattern can easily be seen to extend to the full tree-level sector of the theory: the Feynman diagrams for the Wilson loop in the planar limit calculated in the self-dual sector of the theory are precisely the planar duals to the MHV tree diagrams for the scattering amplitude. 

The pattern easily extends to the loop integrands when we include the MHV vertices.  We start with a single, quadratic (MHV) vertex connected to two components $X_i$ and $X_j$ of the momentum twistor polygon. Such Feynman diagrams contribute to the Wilson loop integrand $W_{n,0}^{(1)}$. The contribution of the diagram illustrated in figure~\ref{Dualgraph2} is
\be{} 
\int\frac{\d s}{s}\frac{\d t}{t}\frac{\d \sigma_1 \d \sigma_2}{(\sigma_1-\sigma_2)^2} \Delta(Z_i(s), Z(\sigma_1,\sigma_2))\; \Delta(Z_j(t),Z(\sigma_1,\sigma_2))
\ee
where we have defined the momentum twistors $Z_i(s) = Z_{i-1}+sZ_i$ and $Z(\sigma_1,\sigma_2) = \sigma_1Z_A+\sigma_2Z_B$ which parametrise respectively the component $X_i$ of the momentum twistor polygon and the interaction line $X$. The integrals are performed against the delta-functions to give
\be{}
[i\!-\!1, i, \rf, A,B']\;[j\!-\!1, j,\rf A,B'']
\ee
where 
\begin{equation*}
B' = (AB)\cap\langle\rf ,j-1,j\rangle, \qquad  B''=(AB)\cap\langle\rf ,i-1,i\rangle\, .
\end{equation*}
This is identical to the contribution to the MHV one-loop integrand $I_{n,0}^{(1)}$ from an MHV diagram which is planar dual to the momentum twistor Feynman diagram, as illustrated in figure~\ref{Dualgraph2} and calculated in \eqref{Loop1}.
\begin{figure}[htp]
\centering 
\includegraphics[height=4cm]{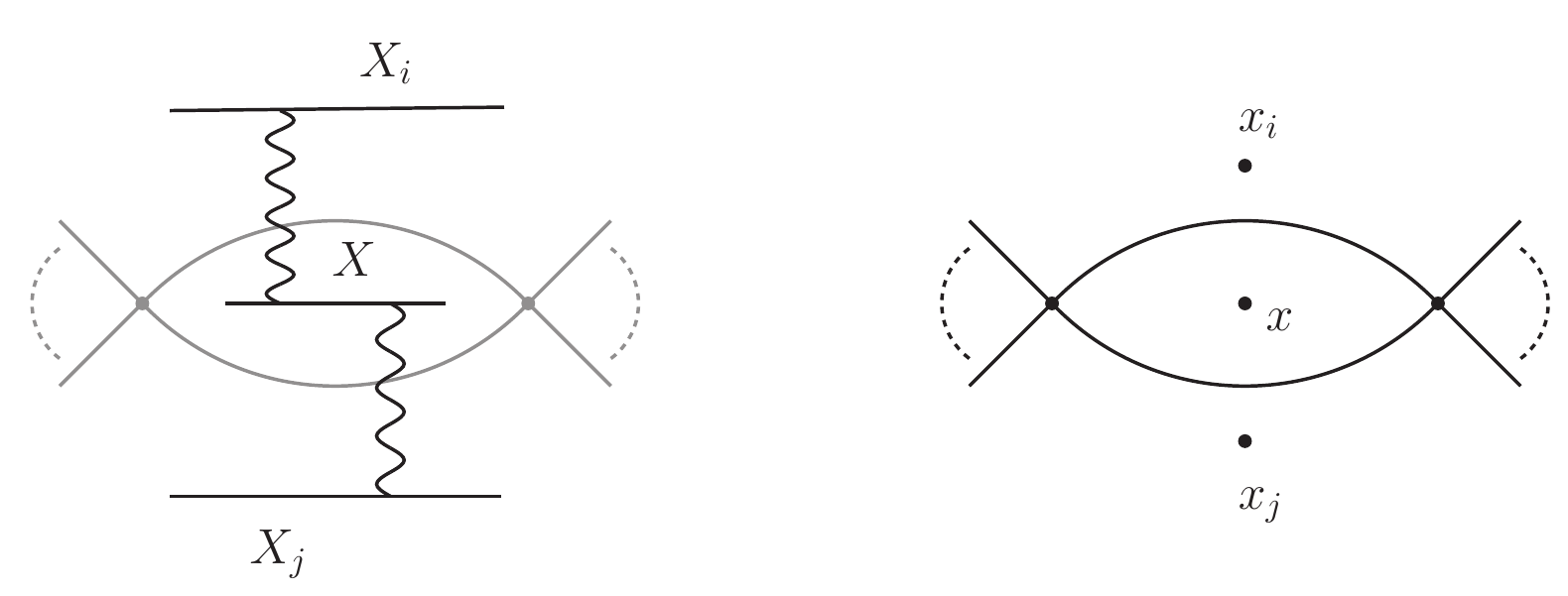}
\caption{\emph{The momentum twistor Feynman diagrams and planar dual MHV diagrams for MHV one-loop integrand.}}
\label{Dualgraph2}
\end{figure}

In~\cite{Mason:2010yk} further computations of momentum twistor Feynman diagrams were performed. In all cases checked, the Feynman diagrams contributing to the integrand $W_{n,k}^{(l)}$ are identical to the planar dual MHV diagrams for the amplitude integrand $I_{n,l}^{(l)}$, leading to the conjecture that the momentum twistor Wilson loop computes all planar integrands of scattering amplitudes in $\cN=4$ SYM. 

The fact that this is consistent arises from the fact that the diagram rules only associate a nontrivial contribution to the propagators, and these are essentially the same for the dual diagram, except for being rotated by a right angle, but this too is  a symmetry of the propagator contribution.


\subsubsection{A proof via loop equations and BCFW recursion}

We can understand why the expectation value of the momentum twistor Wilson loop should compute scattering amplitudes directly from its fundamental properties without resorting to a perturbative analysis. This follows by examining how the expectation value changes when the momentum twistor polygon $C$ is deformed.  We can then show that the correct singularities of scattering amplitudes as functions of the kinematic variables emerge from the properties of holomorphic Wilson loops. This  allows us to show that the Wilson loop satisfies BCFW recursion relations (which are briefly described in \S\ref{BCFW}) and therefore leads to a proof of the duality for the integrands between the amplitude  and the Wilson loop.

We consider a one-parameter holomorphic deformations of the momentum twistor polygon $C_t$, labelled by a complex parameter $t$.   In practice we will simply shift just one of the twistors using \eqref{mtwisBCFWshift} although the following applies to general deformations.  We would like to know whether the expectation value $\big\la W(C_t) \big\ra$ changes holomorphically. The action of the $\bar{\partial}$-operator with respect to the complex parameter $t$ on the classical holomorphic Wilson loop is given by~\cite{Bullimore:2011ni}
\be{}
\overline{\partial}_t W[C_t] = \int_{C_t} \omega(\sigma)\, \wedge \mathrm{tr}( \mathrm{Hol}_{\sigma}(C_t) \mathcal{F}^{0,2}(\sigma))
\ee
where $\mathrm{Hol}_{\sigma}(C_t)$ is the holonomy around the momentum twistor polygon $C_t$ starting and finishing at the point $\sigma$, and $\mathcal{F}^{0,2}(\sigma)$ is the $(0,2)$-form component of the curvature of $\cA$ pulled back to the infinitesimal surface swept out by  $C_t$ as $t$ is perturbed. This is the holomorphic analogue of the result that the change in a real Wilson loop under a small deformation of the curve is given by the flux through the area swept out. 

In holomorphic Chern-Simons theory, the equations of motion are $\mathcal{F}^{0,2}=0$ and hence we conclude that $\overline{\partial}_tW=0$ on-shell in the self-dual gauge theory. However, there are corrections to this picture both from quantum anomalies and from the additional MHV vertices in the twistor action. These corrections imply that the expectation value $\big\la W[C] \big\ra$ has the right singularity structure to compute tree amplitudes and loop amplitudes respectively. Here we summarise the results; details may be found in~\cite{Bullimore:2011ni}.

\begin{itemize}
\item There is an anomaly for the equation $\bar{\partial}_t\big\la W[C_t]\big\ra=0$ when the momentum twistor polygon $C_t$ intersects itself. This occurs when two components $X_i$ and $X_j$ intersect and corresponds to the momentum space factorisation channel $(p_i+\cdots+p_{j-1})^2\rightarrow 0$. This intersection pinches the momentum twistor polygon into two nodal curves $C_{L}$ and $C_{R}$ which meet at the intersection.  In the planar limit, we can deduce that the anomaly is proportional to the product $\big\la W[C_L] \big\ra \big\la W[C_R] \big\ra$.  This implies that the integrands $W_{n,k}^{(l)}$ have the same factorisation behaviour along multi-particle channels as the integrands $I_{n,l}^{(l)}$ of scattering amplitudes.


\item The interaction terms in the twistor action provide another correction to the equation $\bar{\partial}_t\big\la W[C_t]\big\ra=0$. This occurs when a component $X_i$ of the momentum twistor polygon intersects the interaction line $X$, and corresponds to an internal propagator in the loop integrand going on shell $(x-x_i)^2\rightarrow 0$. These corrections imply that the integrands $W_{n,k}^{(l)}$ have the same factorisation behaviour along internal channels as the loop integrands $I_{n,k}^{(l)}$ of scattering amplitudes.

\end{itemize}

By directly substituting in the standard BCFW~\cite{Britto:2004ap,Britto:2005fq} deformation in momentum space \eqref{BCFWshift} into the coordinate transformation into momentum twistors, we arrive at the particular deformation
\be{mtwisBCFWshift}
Z_n(t) = Z_n - tZ_{n-1}\, .
\ee
For this deformation, the loop equations can be integrated over the deformation parameter $t$. Once expanded in Grassmann degree and loop order, we find the all-loop BCFW recursion relations for the loop integrand of the momentum twistor Wilson loop:
\be{all-loopBCFW}
\begin{aligned}
	W_{n,k}^{(l)}(1,\ldots,n) &= W_{n-1,k}^{(l)}(1,\ldots,{n\!-\!1}) \\
	&\hspace{-1cm} + \sum\limits_{j} \sum\limits_{n_i,k_1,l_1}\ [n\!-\!1,n,1,j\!-\!1,j]\, W_{n_1,k_1}^{(l_1)}(1,\ldots,j\!-\!1,I_j)\,
	W_{n_2,k_2}^{(l_2)}(I_j,j,\ldots,n\!-\!1,\hat{n}_j) \\
	&\hspace{-1cm}+
	\int\limits_{\M_{\R}\times S^1\times S^1}\hspace{-0.4cm} \D^{3|4}Z_A\wedge\D^{3|4}Z_B\ [n\!-\!1,n,1,A,B\,]\,
	W_{n+2,k+1}^{(l-1)}(1,\ldots,n\!-\!1,\hat n_{AB},\hat{A},B)\, ,
\end{aligned}
\ee
where the hatted momentum twistors are
\be{}
\begin{aligned}
	& \hat{A} = (A,B)\cap ( n\!-\!1,n,1) \qquad\qquad \hat{n}_{AB} = (n\!-\!1,n)\cap( A,B,1)\\
	& \hat{n}_j = (n\!-\!1,n)\cap( j\!-\!1,j,1) \qquad\, I_j = (j\!-\!1,j)\cap( n\!-\!1,n,1)\, ,
\end{aligned}
\ee
and the summation ranges are for $n_1+n_2=n+2$,$k_1+k_2=k-1$ and $l_1+l_2=l$. This recursion is precisely the same as that obtained  for the amplitude integrands $I_{n,k}^{(l)}$ via BCFW recursion in momentum twistor space in~\cite{ArkaniHamed:2010kv}.  Since this recursion determines the loop integrand, this gives a proof of the duality $I_{n,k}^{(l)} = W_{n,k}^{(l)}$ between the amplitude and momentum twistor Wilson loop integrands.

\subsection{Correlation Functions}

The duality between MHV scattering amplitudes and Wilson loops has recently been extended further to include light-cone limits of certain $n$-point correlation functions in work by Alday, Eden, Korchemsky, Maldacena and Sokatchev~\cite{Alday:2010zy,Eden:2010zz,Eden:2010ce}. This has recently been extended to the supersymmetric setting~\cite{Adamo:2011dq,Eden:2011} encoding the whole amplitude as opposed to just the MHV sector. This correspondence can be understood as arising from the presence of the operators $U_X(\sigma,\sigma')$  in the twistor   definition of a field.

We consider super-correlation functions of local gauge invariant operators $\mathcal{O}(x,\theta)$ on chiral superspace. We can for example choose the Konishi multiplet, whose superconformal primary operator is 
\be{}
\mathcal{O}(x) = \frac{1}{2}\epsilon^{abcd}\mathrm{Tr}\left( \Phi_{ab}(x)\Phi_{cd}(x) \right)\, ,
\ee
or some 1/2 BPS relative.  As we  saw in the Abelian theory \eqref{Susy-Pint2}, the operators $\phi_{ab}(x)$ have an extension $\cF_{ab}(x,\theta)$ to chiral superspace that is given by a straightforward integral formula  in terms of $\cA$ over $X$.  In the non-abelian case, we must trivialize the bundle $E$ over $X$ and the abelian integral formula generalizes to become
\be{}
\cF_{ab}=\int_{X} U_X(\sigma',\sigma)\frac{\p^2\cA}{\p\chi^a\p\chi^b} U_X (\sigma,\sigma')\d\sigma\, .
\ee
Thus, in twistor space, these superfields become non-local operators based on the complex line $X$. For the Konishi multiplet we have~\cite{Adamo:2011dq}
\be{}
O(x,\theta) = \frac{1}{2}\epsilon^{abcd} \int\limits_{X\times X} \rd \sigma \rd \sigma' \; \mathrm{Tr}\left( \frac{\partial^2\cA}{\partial\chi^a \partial\chi^b}(\sigma)\, U_X(\sigma,\sigma')\,  \frac{\partial^2\cA}{\partial\chi^c \partial\chi^d}(\sigma')\, U_X(\sigma',\sigma)\right) \, .
\ee
This is illustrated schematically in figure~\ref{fig:nonAbPhi2}. Correlation functions of such operators can then be computed in twistor space using the twistor action for $\cN=4$ SYM. 

\begin{figure}[tp]
\begin{centering}	
	\includegraphics[width=55mm]{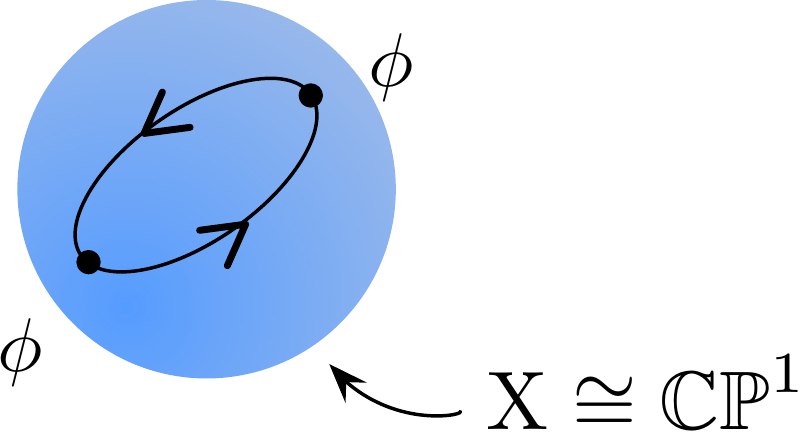}
	\caption{\emph{In a non-Abelian theory, the twistor space form of the local space-time operator ${\rm Tr}\,\Phi^2$ involves 
	holomorphic Wilson lines on the Riemann sphere $X$.}}
	\label{fig:nonAbPhi2}
\end{centering}
\end{figure}

The duality is obtained by examining correlation functions of $n$ such operators
\be{}
G_n(x_1,\theta_1; \ldots; x_n,\theta_n) = \big\la \mathcal{O}(x_1,\theta_1) \ldots \mathcal{O}(x_n,\theta_n) \big\ra 
\ee
in the limit as the chiral superspace points become null separated to form the vertices of our null polygon so that the lines $X_{i+1}$ and $X_i$ intersect in twistor space. Working with the twistor action and at the level of the loop \emph{integrand}, the most singular contribution comes from the contraction of two scalar component fields $\phi$ in the expansion of $\partial^2\cA/\partial\chi^1\partial\chi^2$ on adjacent lines $X_{i}$ and $X_{i+1}$. Once the singular part $1/x_{12}^2\ldots x_{n1}^2$ has been extracted, the $U$ operators on each line are now multiplied together around the polygon, so that the remainder is a momentum twistor Wilson loop in the \emph{adjoint} representation, as illustrated schematically in figure~\ref{fig:limit} (so we have two copies of our former Wilson loop). The duality may be expressed as
 \be{}
 \label{duality}
\lim\limits_{x_{i,i+1}^2\rightarrow0} \, \frac{G_n}{G_{n,0}^{(0)}} = \big\la W^{\mathrm{adj}}_n \big\ra = \big\la W_n \big\ra^2
\ee 
where the final equality is valid only in the planar limit. 

\begin{figure}[tp]
\begin{centering}   
    \includegraphics[width=150mm]{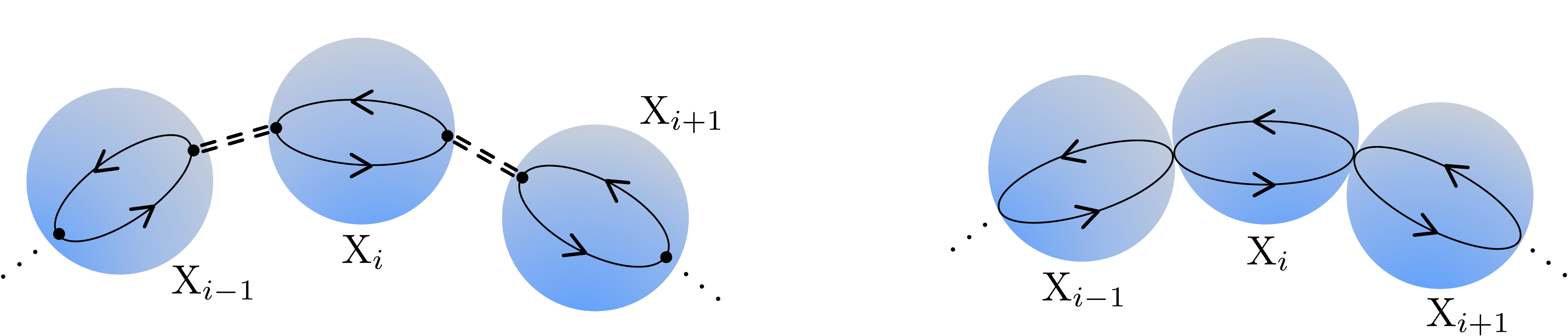}
    \caption{The only non-vanishing contribution to the integrand ratio in the null limit comes from
    direct contractions between $\phi$s on adjacent Riemann spheres. The remaining operator is the supersymmetric momentum twistor Wilson loop, acting in the adjoint representation.}
    \label{fig:limit}
\end{centering}
\end{figure}
 
We have already discussed in some detail the duality $I_{n,k}^{(l)} = W_{n,k}^{(l)}$ between scattering amplitude and Wilson loop integrands. The new duality with correlation functions implies further relationships with the integrands of the correlation function. Expanding equation~\eqref{duality} in the momentum twistor Grassmann degree and number of loops, we find relationships involving the correlation function integrands $G_{n,k}^{(l)}$. For example, in the self-dual theory we find the tree-level result
\be{}
\lim\limits_{x_{i,i+1}^2\rightarrow0}\  \frac{G_{n,1}^{(0)}}{G_{n,0}^{(0)}}\  = \ 2\, W_{n,1}^{(0)} \hspace{1cm} \lim\limits_{x_{i,i+1}^2\rightarrow0}\  \frac{G_{n,2}^{(0)}}{G_{n,0}^{(0)}}\  =\  2\, W_{n,2}^{(0)} + \left( W_{n,1}^{(0)} \right)^2\, ,
\ee
and similarly for integrands of Grassmann degree zero we find
\be{}
\lim\limits_{x_{i,i+1}^2\rightarrow0}\ \frac{G_{n,0}^{(1)}}{G_{n,0}^{(0)}} \ =\  2\, W_{n,0}^{(1)} \hspace{1cm} \lim\limits_{x_{i,i+1}^2\rightarrow0}\  \frac{G_{n,2}^{(0)}}{G_{n,0}^{(0)}} \ =\  2\, W_{n,0}^{(2)} + \left( W_{n,0}^{(1)} \right)^2\, ,
\ee
corresponding to agreement with the all-loop integrand of the non-supersymmetric adjoint Wilson loop in space-time.


\section{Further topics}
\label{furthertopics}

There are many important twistor developments that we have not been able to cover in detail in 
this review.  To make up for this, we give a brief introduction to some of the most important with 
pointers to the literature.

\subsection{Twistor-string theory}
Witten's twistor-string~\cite{Witten:2003nn} was the breakthrough that stimulated these recent developments. Although still not yet satisfactorily realized, it remains one of the most ambitious visions for how scattering amplitudes should be understood.   The idea is that the full $l$-loop N$^k$MHV amplitude should be expressed as a path-integral over the space of holomorphic maps from algebraic curves of degree $k+1+l$ and genus $l$ into twistor space. 
Twistor-string theory, as first introduced by Witten, was a twisted B-model coupled to D1-instantons
supported on holomorphic curves in twistor space. The B-model is more usually found in the study 
of Calabi-Yau compactifications and, through a worldsheet `twisting' procedure, is sensitive only to 
the holomorphic structure.  Soon after, Berkovits introduced a simpler model essentially of 
holomorphic strings in twistor space \cite{Berkovits:2004hg} and later a definition was given as a 
half-twisted heterotic model in \cite{Mason:2007zv}.  All these models have essentially the same 
physical content and their spectra includes both $\cN=4$ SYM and also $\cN=4$ conformal 
supergravity.  The conformal supergravity conformal supergravity corrupts gauge theory  
amplitudes at loop level and there is currently no obvious mechanism for decoupling.  None of 
these models has yet been shown to be fully consistent and there remain a number of interesting 
unanswered questions about their anomalies~\cite{Berkovits:2004hg}.  There has been little work 
on building twistor-string theories with different physical content: the one attempt to build a twistor-string adapted to Einstein supergravity \cite{AbouZeid:2006wu} was shown to contain only self-
dual interactions \cite{Nair:2007md,Brodel:2009ep}.  

As far as gauge theory amplitudes are concerned, then, the main direct impact of twistor-string 
theory thus far is to provide a remarkable formula for tree-level amplitudes. In many ways, this 
remains the best and most compact formula available, manifesting more symmetries than any
other.  It gives the $n$-point tree-level N$^k$MHV amplitude as an integral over the moduli space 
of degree $k+1$ maps from the $\CP^1$ worldsheet to $\PT$.  Letting $\sigma$ be an affine 
coordinate on the world-sheet,  a general degree $d$ map can be written
\begin{equation*}
	Z(\sigma)=\sum_{r=0}^{d}Y_{r}\sigma^{r}\ ,
\end{equation*}
where the $Z_{j}\in\C^{4|4}$ are thought of as coordinates on a $4(d+1)|4(d+1)$-dimensional 
space.  As in the MHV case discussed earlier,  fields are inserted at points $\sigma_i$ on the 
worldsheet.   If we insert  elemental states $\cA_{i}=\bar{\delta}^{3|4}(Z_{i},Z(\sigma_{i}))$ we 
obtain the path integral formula for the tree-level amplitudes (generalizing a form used by 
Roiban, Spradlin and Volovich \cite{Roiban:2004yf} in split signature):
\be{treesmatrix}
A^{\mathrm{tree}}(1,\ldots,n)=\int \frac{\rd^{4(d+1)|4(d+1)}Y_r}{\vol(\GL(2,\C))}\int_{(\CP^1)^{\otimes n}}\prod_{i=1}^{n}\frac{\bar{\delta}^{3|4}(Z_{i},Z(\sigma_{i}))\wedge\d\sigma_{i}}{\sigma_{i}-\sigma_{i-1}}.
\ee
where the vol(GL$(2;\C)$) that we must quotient by consists of  Mobius transformations on the 
worldsheet together with an overall scaling of the $Z_i$s.  In this Dolbeault framework,
the integral should be taken over the full $n$ copies of $\CP^1$, but over a middle dimensional 
cycle in the space of holomorphic maps; the construction of this cycle is not yet properly 
understood except for $(++--)$ signature. This path integral formula is easily seen to reduce to the 
expression \eqref{eqn: TMHV} for the MHV vertex by setting $d=1$. In this case, 
$Z(\sigma_{i})=Z_{A}+Z_{B}\sigma_{i}$, and the integral over the $(8|8)$-dimensional moduli 
space of degree one maps (divided by the volume of $\GL(2)$ tranformations) becomes an 
integral over $\D^{4|4}Z_{A}\wedge \D^{4|4}Z_{B}$. At higher degree~\eqref{treesmatrix} was 
shown to satisfy the tree-level BCFW recursion relations in \cite{Skinner:2010cz}, building on 
earlier work~\cite{Dolan:2009wf,Spradlin:2009qr,Bullimore:2009cb,ArkaniHamed:2009dg,Dolan:2010xv,Bourjaily:2010kw}.

At higher genus, some 1-loop amplitudes for the original twistor string (thus containing conformal 
supergravity modes running around the loop) were studied in~\cite{Dolan:2007vv}. It was shown 
in~\cite{Bullimore:2009cb} that leading singularities of multi-loop amplitudes have exactly the 
twistor support expected from a putative higher-genus twistor string, with the cut momenta of the 
leading singularity corresponding to degenerations of the worldsheet. It was then conjectured\cite{Bullimore:2009cb} that it might be possible to use an extension of the Grassmannian formula 
(discussed below) to determine a meromorphic form on the space of genus $g$, degree $d$, 
$n$-pointed maps to $\PT$ (analogous to the $g=0$ measure in~\eqref{treesmatrix}) by requiring 
this measure to have poles corresponding to leading singularties.  In a separate line,  Yangian 
charges on the worldsheet of the twistor-string have been studied in \cite{Corn:2010uj}.

\subsection{Recursion relations}\label{BCFW}
Recursion techniques have been a powerful tool for generating amplitudes and proving 
conjectures concerning momentum space amplitudes. They are reviewed in more detail 
in~\cite{Brandhuber:2011ke}.  

The starting point (is to deform the external momenta in such a way that total 
momentum is still conserved and each of the external particles remains on-shell.  For BCFW 
recursion~\cite{Britto:2004ap,Britto:2005fq,ArkaniHamed:2008gz,Brandhuber:2008pf}, this is typically done by replacing
\be{BCFWshift}
\begin{aligned}
	(\lambda_1,\tilde\lambda_1,\eta_1) 
	&\rightarrow (\lambda_1 +t\lambda_n,\tilde \lambda_1,\eta_1)\\
	(\lambda_n,\tilde\lambda_n,\eta_n) 
	&\rightarrow (\lambda_n,\tilde\lambda_n-t\tilde\lambda_1,\eta_n-t\eta_1)\ ,
\end{aligned}
\ee
in the momentum space amplitude, where $t$ is a complex parameter. These shifts are generated by the vector
\be{}
	\lambda_n\frac{\del}{\del\lambda_1} -\tilde\lambda_1\frac{\del}{\del\tilde\lambda_n}-
	\eta_1\frac{\del}{\del\eta_n}\ ,
\ee
which becomes
\be{}
	\lambda_n\frac{\del}{\del\lambda_1} +\mu_n\frac{\del}{\del\mu_1}
	+\chi_n\frac{\del}{\del\chi_1}= Z_n\frac{\del}{\del Z_1}
\ee
in twistor space, showing that the twistor amplitude is similarly deformed by replacing 
$Z_1\to Z_1+tZ_n$. In particular, this makes it manifest that the BCFW shift procedure is 
superconformally invariant.

Upon transforming to twistor space, the momentum space BCFW recursion formula
\be{BCFW-mom}
	A(1,\ldots,n)
	= \sum_i \frac{A(1(t_i);\ldots,i,-P_i(t_i))\,A(P_i(t_i);i+1;\ldots;n(t_i))} {P_i^2}\ , 
\ee
(where $P_i=p_1+\ldots +p_i$ and $P_i(t_i)=p_1(t_i)+\cdots+p_i$, with $t_i=P_i^2 / [1|P_i|n\ra$) becomes~\cite{Mason:2009sa}
\be{BCFW-twis}
	A(Z_1,\ldots,Z_n)= 
	\sum_i \int \D^{3|4}Z \, \frac{\rd t}t\, A(Z_1+tZ_n,\ldots,Z_i,Z)\, A(Z,Z_{i+1},\ldots,Z_n) \, .
\ee  
in twistor space. The recursion needs to be seeded by the three point amplitudes.  Clearly the 3-
point MHV  amplitude is simply $A_{\mathrm{MHV}}(1,2,3)=V(1,2,3)=V(1,2)\bar\delta^{2|4}(1,2,3)$ 
as in section~\ref{FR}, while the 3-point $\overline{\rm MHV}$ amplitude is \be{}
\begin{aligned}
	A_{\overline{\mathrm{MHV}}}(Z_1,Z_2,Z_3)
	&=\int_\PT\D^{3|4}Z\wedge \bar\delta^{3|4}(Z,Z_1)\wedge\bar\delta^{3|4}(Z,Z_2)
	\wedge\bar\delta^{3|4}(Z,Z_3)\\
 	&=\bar\delta^{3|4}(Z_1,Z_2)\,\bar\delta^{3|4}(Z_1,Z_3)\ ,
\end{aligned}
\ee 
coming from the Chern-Simons vertex.

The BCFW representation of twistor space tree amplitudes was first studied 
in~\cite{Mason:2009sa}, working in split signature\footnote{In the formalism of this paper, the 
expressions given in~\cite{Mason:2009sa} simplify as all the conformal symmetry breaking $\sgn$ 
functions are omitted.  In split signature, they were required to give the formulae the correct 
symmetries, but here those symmetries are provided by the algebra of differential forms.  However, 
in the present formalism, factors of the two-point MHV vertex  (coming with the 3-point MHV vertex 
as above) are required instead;  these were absent in the split signature formalism.}, and twistor 
BCFW representations were further investigated in~\cite{Korchemsky:2009jv,Bullimore:2009cb}.
The transform of BCFW recursion into twistor variables was actually first studied by Hodges in 
\cite{Hodges:2005aj,Hodges:2005bf} in a formalism based on the use of both twistors and dual 
twistors.  It was used to make contact with twistor diagram theory and to define twistor diagrams for 
arbitrary tree amplitudes.  This approach was taken further by \cite{ArkaniHamed:2009si}, but now 
more systematically using Witten's half-Fourier transform. This led to a new representation -- the 
{\em link representation} -- that was a precursor of the Grassmannian formalism given below.  
Whereas the MHV formalism remains controversial for gravity amplitudes, BCFW recursion does 
work for gravity and can also be used to construct gravity amplitudes in twistor space~\cite{Mason:2009sa,ArkaniHamed:2009si}.
 
\smallskip

As well as being superconformally invariant, the BCFW procedure was also shown to be \emph{dual} superconformally invariant in~\cite{Brandhuber:2008pf} where it was used to prove 
dual superconformal invariance of all tree amplitudes.  This invariance can most easily be seen by 
reformulating the shift~\eqref{BCFWshift} in momentum twistor space where it becomes
\be{}
	Z_n \rightarrow Z_n - tZ_{n-1}
\ee
as in \eqref{mtwisBCFWshift}. (Note that the $\la1 n]$ shift in momentum space here translates 
$Z_n$ along the momentum twistor line $(n\!-\!1,n)$).  Using momentum twistors, BCFW recursion 
was extended to the all-loop integrand for planar supersymmetric gauge theries by 
Arkani-Hamed et.\ al.\ \cite{ArkaniHamed:2010kv}, yielding the recursion 
formula~\eqref{all-loopBCFW} for the scattering amplitude (see also \cite{Boels:2010nw} for a 
discussion in momentum space).  

\smallskip

Another recursion procedure that is naturally expressed in momentum twistor space is that due to Risager \cite{Risager:2005vk}.  This gave the first proof of MHV rules at tree level \cite{Elvang:2008na}.  In momentum space it involved a choice of reference spinor $\iota^{A'}$ and in its most general form was the shift $\tilde\lambda_{iA'}\rightarrow \tilde \lambda_{iA'}+a_it\iota_{A'}$ where the $a_i$ are chosen so as to conserve momentum.  
In momentum twistor space in can be expressed easily as the shift $Z_i\rightarrow Z_i+c_itZ_\rf$ with $c_i$ arbitrary.  In this form it was used to prove the MHV rules for the construction of the planar loop integrand in \cite{Bullimore:2010dz,He:2010ju}.

\subsection{The Grassmannian formula and leading singularities}

The Grassmannian formula generates rational Yangian invariants of the kinematic data.  These 
include not only all possible BCFW terms in the decomposition of tree amplitudes but also all 
possible leading singularities of multiloop amplitudes.  Leading singularities are invariants of 
multiloop scattering amplitudes obtained by performing the $4l$ loop integrations not over the 
physical contour ($l$ copies of momentum space), but by residues. This has the effect of putting  
propagators on-shell. The result is expressed as a product of on-shell tree amplitudes, glued 
together at the legs corresponding to the on-shell propagators.  

The Grassmannian formula comes in two forms: the original one on twistor space 
\cite{ArkaniHamed:2009dn} that makes superconformal invariance manifest, and a subsequent 
one on momentum twistor space \cite{Mason:2009qx} that manifests dual superconformal 
invariance.   These were shown to be essentially equivalent in \cite{ArkaniHamed:2009vw}.
The first papers to show that all leading singularities live in the Grassmannian were 
\cite{Bullimore:2009cb, Kaplan:2009mh}, while the Yangian invariance of the formula 
was investigated in~\cite{ArkaniHamed:2009vw,Drummond:2010uq}. See 
also~\cite{Bargheer:2011mm} in this journal issue for a review of the Grassmannian formula 
emphasising its symmetries.   Our treatment here will empahsize the connection with twistor-string theory following \cite{Bullimore:2009cb}.

The Grassmannian $G(k,n)$ is the space of $k$-planes $\C^k\subset \C^n$. It can be 
parametrized by a $k\times n$ matrix $C_{ri}$, (with $r=1,\ldots,k$ and $i=1,\dots,n$) where we 
must quotient by the action of GL($k;\C$) on the $r$-index.  For an $n$-particle N$^k$MHV 
amplitude, the twistor space formula involves the Grassmannian $G(k+2,n)$ whereas the 
momentum twistor formula involves $G(k,n)$. Both formulae take a similar form\footnote{We have 
expressed $\cL_{n,k}$ in a form that makes the relation to twistor-string theory (discussed below) 
more direct. $\cL_{n,k}$ can be made to more closely resemble $\cR_{n,k}$ (or its dual) by 
integrating out the $Y$s.}   
\be{Grassmannian}
\begin{aligned}
	\cL_{n,k} (Z_1,\ldots,Z_n)&=
	\oint_\Gamma \frac {\rd^{(k+2)n} C}{\mathrm{vol}\, {\rm GL}(k+2)}
	\, \frac{ \left(\prod_{r=1}^{k+2}\rd ^{4|4} Y_r\right)
	\left(\prod_{i=1}^n\,\bar\delta^{4|4}(Z_i-C_{qi}Y_q)\right)}
	{ (1\, 2\ldots  k+2)(2\, 3\dots k+3)\ldots (n\, 1 \, 2\ldots k+1)}\\
	\cR_{n,k} (Z_1,\ldots,Z_n)&=
	\oint_{\widetilde \Gamma} \frac {\rd^{k\, n} C}{\mathrm{vol}\, {\rm GL}(k)}
	 \, \frac{ \prod_{r=1}^k\bar\delta^{4|4}(C_{ri}Z_i)}
	 { (1\, 2\ldots  k)(2\, 3\dots k+1)\ldots (n\, 1 \, 2\ldots k-1)}\ ,
\end{aligned}
\ee
where the summation convention is understood on $q$ and 
\be{}
	(i\  i\!+\!1\cdots i\!+\!k):=\det (C_{ri},C_{r\, i+1},\ldots, C_{r\, i+l})
\ee
are the Pl\"ucker coordinates on the Grassmannian $G(k,n)$ that respect the cyclic ordering.  In 
these formulae,  $\Gamma$ and $\widetilde \Gamma$ are contours that allow 
integration by residues down to a cycle that should have dimension $2n-4$ in the case of 
$\Gamma$ and dimension $4k$ in the case of $\widetilde\Gamma$.  We will abuse notation and 
denote the cycles that support these residues also by $\Gamma$ and $\widetilde \Gamma$, respectively. 

\smallskip

Although the two formulae in~\eqref{Grassmannian} are superficially identical there are key 
differences.  For the momentum twistor $\cR_{n,k}$, once the contour integral over 
$\widetilde\Gamma$ has been performed,  there are precisely enough bosonic delta functions to 
saturate the integrals over the remaining parameters. Thus, $\cR_{n,k}$ yields an algebraic 
function of the momentum twistors $Z_1,\ldots, Z_n$ with Grassman degree $4k$.  This therefore 
gives an N$^k$MHV invariant that turns out to be some leading singularity; momentum twistor 
space is essentially a coordinate transformation of ordinary momentum space and all tree 
amplitudes and leading singularities are rational or algebraic functions on momentum space.  In 
the momentum twistor formulation, the MHV amplitude is 1, so the simplest nontrivial case is at 
$k=1$, and then $G(1,n)$ is just projective space $\CP^{n-1}$.  At $n=5$, the quotient by GL(1) 
can be implemented by setting one of the $C_i$ coordinates equal to 1, and the formula reduces 
to our original formula~ \eqref{superconf} for the basic R-invariant -- the 5-point NMHV 
amplitude.  For higher $n$ our cycle must be a linear combination of cycles that set all but four 
components of $C$ to zero (including the GL(1) gauge fixing), so we obtain a linear combination 
of R-invariants.  

The $\cL_{n,k}$ formula gives the same leading singularities (or Yangian invariants) as 
$\cR_{n,k}$, but expressed in original twistor space.  In this case delta functions remain after all  
integrations have been performed, leading to the fact that leading singularities have restricted 
support in twistor space.  The $\cL_{n,k}$ formula makes it clear that this support occurs where 
$Z_i=C_{ri}Y_r$ for $C\in\Gamma$ and $Y_r$ aribtrary.  For example, when $k=0$, $\Gamma$ is 
the whole of $G(2,n)$, and setting 
\be{}
Y_1=Z_A\, , \quad Y_2= Z_B\, , \quad C_{ri}=\sigma^{r-1}_i s_i\ ,
\ee
where $\sigma_i$ is a parameter on the line and $s_i$ the parameter that we use to integrate a 
$\bar\delta^{4|4}$ down to give the projective delta function $\bar\delta^{3|4} (Z_i,Z(\sigma_i))$,  
the Grassmannian formula can be seen to give our twistor formula for the MHV amplitude \eqref{eqn: TMHV}.

All the Yangian invariants generated by the Grassmannian correspond to `primitive' leading 
singularities -- those made up from gluing tree subamplitudes that are either MHV or 
$\overline{\mathrm{MHV}}$. A priori, this may seem to be a restriction on the leading singularities 
that can be obtained. However, BCFW recursion decomposes N$^k$MHV tree amplitudes into 
leading singularities of higher loop amplitudes whose components all have lower degree. (In fact, 
this was the original approach to the recursion relation by Britto, Cachazo and 
Feng~\cite{Britto:2004ap}.)  Thus, all leading singularities can be expressed as residues of the 
Grassmannian.

In twistor space, primitive leading singularities are supported on sets of intersecting lines~\cite{Bullimore:2009cb}. This was first observed for terms in the BCFW expansion of a tree 
amplitude by \cite{Korchemsky:2009jv} and extended to all leading singularities and 
Grassmannian residues  in \cite{Bullimore:2009cb}. The twistor geometry underlying leading 
singularities thus leads to a partial classification of Grassmannian residues.   For example, at MHV 
we only ever obtain lines in twistor space for leading singularities, whatever the loop order.  At 
NMHV, the BCFW terms are triangles (i.e., genus one), but the most general NMHV leading 
singularity is an arrangement of five lines with genus 3 and so on. These ideas have now been 
developed into a systematic classification of residues in the Grassmannian~\cite{Bullimore:2010pa,Bourjaily:2010kw,Ashok:2010ie}.

\smallskip

Individual residues of the Grassmannian correspond to individual leading singularities, and one 
would like to understand how to combine residues so as to compute the tree amplitudes 
themselves. One way to achieve this uses a correspondence between the Grassmannian and the 
twistor-string moduli space.  The line bundle $\cO(1)$ on twistor space, when pulled back to the 
string worldsheet $\Sigma$ gives a line bundle $L\cong\cO(k+1)$ whose holomorphic sections 
can be described by polynomials of degree $k+1$ in the worldsheet coordinates.  Given 
trivialisations of $L$ at each of the marked points $\sigma_i\in\Sigma$ (i.e., preferred coordinates 
on each of the fibres $L|_{\sigma_i}$), we can obtain $n$ complex numbers from any holomorphic 
section of $L$ by looking at its values at each marked point. Thus the space $H^0(\Sigma,L)\cong\C^{k+2}$ of holomorphic sections of $L$ is naturally a subspace of $\oplus_i L|_{\sigma_i}\cong\C^n$. In other words, a choice of $n$ marked points on the worldsheet, together with a degree 
$k+1$ line bundle $L$ and trivialisations at each marked point, naturally corresponds to a point in 
$G(k+2,n)$. If we vary the location of our marked points on $\Sigma$ (subject to SL$(2;\C)$ 
transformations) and vary the choice of trivialisation at each marked point, we obtain a cycle in 
$G(k+2,n)$ of dimension
\be{}
	(n-3) + n - 1= 2n-4\ ,
\ee
where the $-1$ comes from an overall rescaling. The twistor-string moduli space thus naturally 
provides a cycle $\Gamma\subset G(k+2,n)$ of dimension $2n-4$, exactly as needed in $\cL_{n,k}$. Explicitly, the embedding is given by the Veronese map
\be{Veronese}
	C_{ri}=\sigma^{r-1}_is_i\ ,
\ee
as observed by a number of authors \cite{Dolan:2009wf,Dolan:2010xv,Spradlin:2009qr, 
Bullimore:2009cb,ArkaniHamed:2009dg}. (As at MHV above, the $s_i$ scaling 
parameter can be interpreted as defining the trivialisation of $L|_{\sigma_i}$.) Although this 
defines a map from the twistor-string moduli space to cycles in the Grassmannian, the resulting 
cycles can be obtained as residues of $\cL_{n,k}$ only when $k=0, -1$. Various tricks based 
on global residue theorems in the Grassmannian have been used to convert this embedding into 
the precise cycle that generates the tree amplitude~\cite{ArkaniHamed:2009dg, 
Dolan:2010xv,Bourjaily:2010kw}.   It has also been seen how to make contact with the MHV 
formalism \cite{ArkaniHamed:2009sx}. 

The construction above also extends to nodal curves, provided each component is 
rational~\cite{Bullimore:2009cb}, thus providing a `twistor-string theory for leading singularities' 
based on higher genus, but nodal curves. It is hoped that this construction for leading singularities 
provides a good starting point for constructing a twistor-string theory proper at higher genus.

\subsection{Spurious singularities, polytopes and local forms}

Hodges' paper introducing momentum twistors \cite{Hodges:2009hk} also contained the 
suggestive insight that the momentum twistor formula for the R-invariants -- the principle 
ingredients of the BCFW expansion of gauge theory NMHV tree amplitudes -- could be interpreted 
as the volume of a simplex in a dual momentum twistor space defined by the arguments of the 
R-invariant. The sum of R-invariants that form an amplitude then correspond to fitting these 
basic simplices together into a polytope, whose volume is the amplitude. He further emphasised 
that the BCFW decomposition of tree amplitudes, whilst both superconformally and dual 
superconformally invariant, contains spurious singularities that could not have arisen from a 
physical propagator.  The different choices of BCFW decomposition were interpreted as different 
decompositions of the polytope for the full amplitude, while the spurious singularities were 
associated to vertices of the constituent simplices that do not end up being vertices of the final 
polytope.   

These ideas have been explored further in \cite{ArkaniHamed:2010gg,ArkaniHamed:2010gh}.  In 
particular they find a new basis for NMHV tree and certain multi-loop MHV and NMHV amplitudes 
that, in contrast to the BCFW expressions, have only local poles -- ones which correspond to a 
sum of momenta (compatible with the cyclic ordering) going on-shell.   The MHV formulae are 
particularly striking, being suggestive of some kind of integrable structure underlying at least 
the MHV sector. 

A similar idea was taken up in a one loop context in \cite{Mason:2010pg}. There, it was shown that 
Hodges' procedure \cite{Hodges:2010kq} for studying 1-loop box functions using momentum 
twistors has the geometric interpretation of computing 3-volumes of tetrahedra in AdS$_5$. The 
vertices of these tetrahedra lie `at infinity' on conformal Minkowski space, where their location 
defines the region momenta associated to the box integral.  When the whole 
MHV ampitude is computed, these tetrahedra join together to give a closed 3-polytope in 
AdS$_5$.  Once again,  the various spurious singularities or branch cuts of individual box 
integrals cancel when the whole polytope is considered.


\section*{Acknowledgements}
TA is supported by the NSF Graduate Research Fellowship (USA) and Balliol College; MB is supported by an STFC Postgraduate Studentship; LM is 
supported by the EPSRC grant number EP/F016654; DS is supported by the Perimeter Institute for Theoretical Physics. Research at the Perimeter Institute is supported by the Government of Canada through Industry Canada and by the Province of Ontario through the Ministry of Research $\&$ Innovation.   Both LM and DS gratefully acknowledge the KITP for hospitality while this review was being finalised.


\bibliography{review}

\providecommand{\href}[2]{#2}\begingroup\raggedright\begin{thebibliography}{10%
0}

\bibitem{ArkaniHamed:2009dn}
N.~Arkani-Hamed, F.~Cachazo, C.~Cheung, and J.~Kaplan, {\it {A Duality For The
  S Matrix}},  \href{http://xxx.lanl.gov/abs/0907.5418}{{\tt arXiv:0907.5418}}.

\bibitem{Mason:2009qx}
L.~Mason and D.~Skinner, {\it {Dual Superconformal Invariance, Momentum
  Twistors and Grassmannians}},  {\em JHEP} {\bf 11} (2009) 045,
  [\href{http://xxx.lanl.gov/abs/0909.0250}{{\tt arXiv:0909.0250}}].

\bibitem{Alday:2010ku}
L.~F. Alday, D.~Gaiotto, J.~Maldacena, A.~Sever, and P.~Vieira, {\it {An
  Operator Product Expansion for Polygonal Null Wilson Loops}},
  \href{http://xxx.lanl.gov/abs/1006.2788}{{\tt arXiv:1006.2788}}.

\bibitem{Mason:2010yk}
L.~Mason and D.~Skinner, {\it {The Complete Planar S-matrix of $\cN=4$ Super
  Yang-Mills from a Wilson Loop in Twistor Space}},  {\em JHEP} {\bf 12} (2010)
  018, [\href{http://xxx.lanl.gov/abs/1009.2225}{{\tt arXiv:1009.2225}}].

\bibitem{Caron-Huot:2010ek}
S.~Caron-Huot, {\it {Notes on the Scattering Amplitude / Wilson Loop Duality}},
   \href{http://xxx.lanl.gov/abs/1010.1167}{{\tt arXiv:1010.1167}}.

\bibitem{Bullimore:2011ni}
M.~Bullimore and D.~Skinner, {\it {Holomorphic Linking, Loop Equations and
  Scattering Amplitudes in Twistor Space}},
  \href{http://xxx.lanl.gov/abs/1101.1329}{{\tt arXiv:1101.1329}}.

\bibitem{Drummond:2010zv}
J.~Drummond, L.~Ferro, and E.~Ragoucy, {\it {Yangian Symmetry of Light-Like
  Wilson Loops}},  \href{http://xxx.lanl.gov/abs/1011.4264}{{\tt
  arXiv:1011.4264}}.

\bibitem{Goncharov:2010jf}
A.~Goncharov, M.~Spradlin, C.~Vergu, and A.~Volovich, {\it {Classical
  Polylogarithms for Amplitudes and Wilson Loops}},
  \href{http://xxx.lanl.gov/abs/1006.5703}{{\tt arXiv:1006.5703}}.

\bibitem{Alday:2010vh}
L.~F. Alday, J.~Maldacena, A.~Sever, and P.~Vieira, {\it {Y-system for
  Scattering Amplitudes}},  \href{http://xxx.lanl.gov/abs/1002.2459}{{\tt
  arXiv:1002.2459}}.

\bibitem{ArkaniHamed:2010kv}
N.~Arkani-Hamed, J.~Bourjaily, F.~Cachazo, S.~Caron-Huot, and J.~Trnka, {\it
  {The All-Loop Integrand for Scattering Amplitudes in Planar $\cN=4$ SYM}},
  \href{http://xxx.lanl.gov/abs/1008.2958}{{\tt arXiv:1008.2958}}.

\bibitem{Adamo:2011dq}
T.~Adamo, M.~Bullimore, L.~Mason, and D.~Skinner, {\it {A Proof of the
  Supersymmetric Correlation Function / Wilson Loop Correspondence}},
  \href{http://xxx.lanl.gov/abs/1103.4119}{{\tt arXiv:1103.4119}}.

\bibitem{Eden:2011}
B.~Eden, P.~Heslop, G.~Korchemsky, and E.~Sokatchev, {\it {The Super-Correlator
  / Super-Amplitude Duality: Parts I \& II}},
  \href{http://xxx.lanl.gov/abs/{1103.3714 \& 1103.4353}}{{\tt {1103.3714 \&
  1103.4353}}}.

\bibitem{Belitsky:2011zm}
A.~Belitsky, G.~Korchemsky, and E.~Sokatchev, {\it {Are scattering amplitudes
  dual to super Wilson loops?}},  \href{http://xxx.lanl.gov/abs/1103.3008}{{\tt
  arXiv:1103.3008}}. * Temporary entry *.

\bibitem{Penrose:1989pb}
R.~Penrose, {\it {On the Origins of Twistor Theory}}, .

\bibitem{Mason:2005zm}
L.~Mason, {\it {Twistor Actions for Non Self-Dual Fields}},  {\em JHEP} {\bf
  10} (2005) 009, [\href{http://xxx.lanl.gov/abs/hep-th/0507269}{{\tt
  hep-th/0507269}}].

\bibitem{Boels:2006ir}
R.~Boels, L.~Mason, and D.~Skinner, {\it {Supersymmetric Gauge Theories in
  Twistor Space}},  {\em JHEP} {\bf 02} (2007) 014,
  [\href{http://xxx.lanl.gov/abs/hep-th/0604040}{{\tt hep-th/0604040}}].

\bibitem{Boels:2007qn}
R.~Boels, L.~Mason, and D.~Skinner, {\it {From Twistor Actions to MHV
  Diagrams}},  {\em Phys. Lett.} {\bf B648} (2007) 90--96,
  [\href{http://xxx.lanl.gov/abs/hep-th/0702035}{{\tt hep-th/0702035}}].

\bibitem{Mason:2007sd}
L.~Mason and M.~Wolf, {\it {Twistor Actions for Self-Dual Supergravities}},
  {\em Commun. Math. Phys.} {\bf 288} (2009) 97--123,
  [\href{http://xxx.lanl.gov/abs/0706.1941}{{\tt arXiv:0706.1941}}].

\bibitem{Mason:2008jy}
L.~Mason and D.~Skinner, {\it {Gravity, Twistors and the MHV Formalism}},  {\em
  Commun. Math. Phys.} {\bf 294} (2010) 827--862,
  [\href{http://xxx.lanl.gov/abs/0808.3907}{{\tt arXiv:0808.3907}}].

\bibitem{Jiang:2008xw}
W.~Jiang, {\em {Aspects of Yang-Mills Theory in Twistor Space}}.
\newblock PhD thesis, Oxford University, 2008.
\newblock \href{http://xxx.lanl.gov/abs/0809.0328}{{\tt arXiv:0809.0328}}.

\bibitem{Bullimore:2010pj}
M.~Bullimore, L.~Mason, and D.~Skinner, {\it {MHV Diagrams in Momentum Twistor
  Space}},  {\em JHEP} {\bf 12} (2010) 032,
  [\href{http://xxx.lanl.gov/abs/1009.1854}{{\tt arXiv:1009.1854}}].

\bibitem{Adamo:2011cb}
T.~Adamo and L.~Mason, {\it {MHV Diagrams in Twistor Space and the Twistor
  Action}},  \href{http://xxx.lanl.gov/abs/1103.1352}{{\tt arXiv:1103.1352}}.

\bibitem{Alday:2007hr}
L.~F. Alday and J.~Maldacena, {\it {Gluon Scattering Amplitudes at Strong
  Coupling}},  {\em JHEP} {\bf 06} (2007) 064,
  [\href{http://xxx.lanl.gov/abs/arxiv 0705.0303 [hep-th]}{{\tt arxiv 0705.0303
  [hep-th]}}].

\bibitem{Drummond:2007aua}
J.~Drummond, G.~Korchemsky, and E.~Sokatchev, {\it {Conformal Properties of
  Four-Gluon Planar Amplitudes and Wilson Loops}},  {\em Nucl. Phys.} {\bf
  B795} (2007) 385--408, [\href{http://xxx.lanl.gov/abs/0707.0243}{{\tt
  arXiv:0707.0243}}].

\bibitem{Korchemsky:1985xj}
G.~Korchemsky and A.~Radyushkin, {\it {Loop Space Formalism and Renormalisation
  Group for the Infrared Asymptotics of QCD}},  {\em Phys. Lett.} {\bf B171}
  (1986) 459--467.

\bibitem{Collins:1989bt}
J.~Collins, {\it {Sudakov Form Factors}},  {\em Adv. Ser. Direct. High Energy
  Phys.} {\bf 5} (1989) 573--614,
  [\href{http://xxx.lanl.gov/abs/hep-ph/0312336}{{\tt hep-ph/0312336}}].

\bibitem{Magnea:1990zb}
L.~Magnea and G.~Sterman, {\it {Analytic Continuation of the Sudakov
  Form-Factor in QCD}},  {\em Phys. Rev.} {\bf D42} (1990) 4222--4227.

\bibitem{Brandhuber:2007yx}
A.~Brandhuber, P.~Heslop, and G.~Travaglini, {\it {MHV Amplitudes in $\cN=4$
  Super Yang-Mills and Wilson Loops}},  {\em Nucl. Phys.} {\bf B794} (2008)
  231--243, [\href{http://xxx.lanl.gov/abs/0707.1153}{{\tt arXiv:0707.1153}}].

\bibitem{Drummond:2007cf}
J.~Drummond, J.~Henn, G.~Korchemsky, and E.~Sokatchev, {\it {On Planar Gluon
  Amplitudes / Wilson Loops Duality}},  {\em Nucl. Phys.} {\bf B795} (2008)
  52--68, [\href{http://xxx.lanl.gov/abs/0709.2368}{{\tt arXiv:0709.2368}}].

\bibitem{Drummond:2007bm}
J.~Drummond, J.~Henn, G.~Korchemsky, and E.~Sokatchev, {\it {The Hexagon Wilson
  Loops and the BDS Ansatz for the Six-Gluon Amplitude}},  {\em Phys. Lett.}
  {\bf B622} (2008) 456--460, [\href{http://xxx.lanl.gov/abs/0712.4138}{{\tt
  arXiv:0712.4138}}].

\bibitem{Drummond:2008aq}
J.~Drummond, J.~Henn, G.~Korchemsky, and E.~Sokatchev, {\it {Hexagon Wilson
  Loop = Six-Gluon MHV Amplitude}},  {\em Nucl. Phys.} {\bf B815} (2009)
  142--173, [\href{http://xxx.lanl.gov/abs/0803.1466}{{\tt arXiv:0803.1466}}].

\bibitem{Alday:2010zy}
L.~F. Alday, B.~Eden, G.~Korchemsky, J.~Maldacena, and E.~Sokatchev, {\it {From
  Correlation Functions to Wilson Loops}},
  \href{http://xxx.lanl.gov/abs/1007.3243}{{\tt arXiv:1007.3243}}.

\bibitem{Alday:2009zm}
L.~F. Alday, J.~Henn, J.~Plefka, and T.~Schuster, {\it {Scattering Into the
  Fifth Dimension of $\mathcal{N}=4$ Super Yang-Mills}},  {\em JHEP} {\bf 01}
  (2010) 077, [\href{http://xxx.lanl.gov/abs/0908.0684}{{\tt
  arXiv:0908.0684}}].

\bibitem{Henn:2010bk}
J.~Henn, S.~Naculich, H.~Schnitzer, and M.~Spradlin, {\it {Higgs-regularized
  three-loop four-gluon amplitude in $\cN=4$ SYM: exponentiation and Regge
  limits}},  \href{http://xxx.lanl.gov/abs/1001.1358}{{\tt arXiv:1001.1358}}.

\bibitem{Hodges:2010kq}
A.~Hodges, {\it {The Box Integrals in Momentum Twistor Geometry}},
  \href{http://xxx.lanl.gov/abs/1004.3323}{{\tt arXiv:1004.3323}}.

\bibitem{Mason:2010pg}
L.~Mason and D.~Skinner, {\it {Amplitudes at Weak Coupling as Polytopes in
  AdS$_5$}},  {\em {\it Submitted to} JPhys A} (2010)
  [\href{http://xxx.lanl.gov/abs/1004.3498}{{\tt arXiv:1004.3498}}].

\bibitem{Bullimore:2010dz}
M.~Bullimore, {\it {MHV Diagrams from an All-Line Recursion Relation}},
  \href{http://xxx.lanl.gov/abs/1010.5921}{{\tt arXiv:1010.5921}}.

\bibitem{He:2010ju}
S.~He and T.~McLoughlin, {\it {On All-Loop Integrands of Scattering Amplitudes
  in Planar $\mathcal{N}=4$ SYM}},
  \href{http://xxx.lanl.gov/abs/1010.6256}{{\tt arXiv:1010.6256}}.

\bibitem{Brandhuber:2011ke}
A.~Brandhuber, W.~Spence, and G.~Travaglini, {\it {Tree-Level Formalism}},
  \href{http://xxx.lanl.gov/abs/1103.3447}{{\tt arXiv:1103.3447}}.

\bibitem{Eden:2010zz}
B.~Eden, G.~Korchemsky, and E.~Sokatchev, {\it {From Correlation Functions to
  Scattering Amplitudes}},  \href{http://xxx.lanl.gov/abs/1007.3246}{{\tt
  arXiv:1007.3246}}.

\bibitem{Eden:2010ce}
B.~Eden, G.~Korchemsky, and E.~Sokatchev, {\it {More on the Duality Correlators
  / Amplitudes}},  \href{http://xxx.lanl.gov/abs/1009.2488}{{\tt
  arXiv:1009.2488}}.

\bibitem{Witten:2003nn}
E.~Witten, {\it {Perturbative Gauge Theory as a String Theory in Twistor
  Space}},  {\em Commun. Math. Phys.} {\bf 252} (2004) 189--258,
  [\href{http://xxx.lanl.gov/abs/hep-th/0312171}{{\tt hep-th/0312171}}].

\bibitem{Berkovits:2004hg}
N.~Berkovits, {\it {An Alternative String Theory in Twistor Space for $\cN = 4$
  Super-Yang-Mills}},  {\em Phys. Rev. Lett.} {\bf 93} (2004) 011601,
  [\href{http://xxx.lanl.gov/abs/hep-th/0402045}{{\tt hep-th/0402045}}].

\bibitem{Mason:2007zv}
L.~Mason and D.~Skinner, {\it {Heterotic Twistor-String Theory}},  {\em Nucl.
  Phys.} {\bf B795} (2008) 105--137,
  [\href{http://xxx.lanl.gov/abs/0708.2276}{{\tt arXiv:0708.2276}}].

\bibitem{Dolan:2007vv}
L.~Dolan and P.~Goddard, {\it {Tree and Loop Amplitudes in Open Twistor-String
  Theory}},  {\em JHEP} {\bf 06} (2007) 005,
  [\href{http://xxx.lanl.gov/abs/hep-th/0703054}{{\tt hep-th/0703054}}].

\bibitem{Dolan:2009wf}
L.~Dolan and P.~Goddard, {\it {Gluon Tree Amplitudes in Open Twistor String
  Theory}},  \href{http://xxx.lanl.gov/abs/0909.0499}{{\tt arXiv:0909.0499}}.

\bibitem{Spradlin:2009qr}
M.~Spradlin and A.~Volovich, {\it {From Twistor-String Theory To Recursion
  Relations}},  {\em Phys. Rev.} {\bf D80} (2009) 085022,
  [\href{http://xxx.lanl.gov/abs/0909.0229}{{\tt arXiv:0909.0229}}].

\bibitem{Bullimore:2009cb}
M.~Bullimore, L.~Mason, and D.~Skinner, {\it {Twistor-Strings, Grassmannians
  and Leading Singularities}},  {\em JHEP} {\bf 03} (2010) 070,
  [\href{http://xxx.lanl.gov/abs/0912.0539}{{\tt arXiv:0912.0539}}].

\bibitem{ArkaniHamed:2009dg}
N.~Arkani-Hamed, J.~Bourjaily, F.~Cachazo, and J.~Trnka, {\it {Unification of
  Residues and Grassmannian Dualities}},
  \href{http://xxx.lanl.gov/abs/0912.4912}{{\tt arXiv:0912.4912}}.

\bibitem{Dolan:2010xv}
L.~Dolan and P.~Goddard, {\it {General Split Helicity Gluon Tree Amplitudes in
  Open Twistor-String Theory}},  {\em JHEP} {\bf 05} (2010) 044,
  [\href{http://xxx.lanl.gov/abs/1002.4852}{{\tt arXiv:1002.4852}}].

\bibitem{Bourjaily:2010kw}
J.~Bourjaily, J.~Trnka, A.~Volovich, and C.~Wen, {\it {The Grassmannian and the
  Twistor-String: Connecting All Trees in $\cN=4$ SYM}},
  \href{http://xxx.lanl.gov/abs/1006.1899}{{\tt arXiv:1006.1899}}.

\bibitem{Corn:2010uj}
J.~Corn, T.~Creutzig, and L.~Dolan, {\it {Yangian in the Twistor String}},
  \href{http://xxx.lanl.gov/abs/1008.0302}{{\tt arXiv:1008.0302}}.

\bibitem{Skinner:2010cz}
D.~Skinner, {\it {A Direct Proof of BCFW Recursion for Twistor-Strings}},  {\em
  {\it Submitted to} JHEP} (2010)
  [\href{http://xxx.lanl.gov/abs/1007.0195}{{\tt arXiv:1007.0195}}].

\bibitem{Mason:2009sa}
L.~Mason and D.~Skinner, {\it {Scattering Amplitudes and BCFW Recursion in
  Twistor Space}},  {\em JHEP} {\bf 01} (2010) 064,
  [\href{http://xxx.lanl.gov/abs/0903.2083}{{\tt arXiv:0903.2083}}].

\bibitem{ArkaniHamed:2009si}
N.~Arkani-Hamed, F.~Cachazo, C.~Cheung, and J.~Kaplan, {\it {The S-Matrix in
  Twistor Space}},  \href{http://xxx.lanl.gov/abs/0903.2110}{{\tt
  arXiv:0903.2110}}.

\bibitem{Kaplan:2009mh}
J.~Kaplan, {\it {Unraveling $\mathcal{L}_{n,k}$: Grassmannian Kinematics}},
  \href{http://xxx.lanl.gov/abs/0912.0957}{{\tt arXiv:0912.0957}}.

\bibitem{ArkaniHamed:2009vw}
N.~Arkani-Hamed, F.~Cachazo, and C.~Cheung, {\it {The Grassmannian Origin Of
  Dual Superconformal Invariance}},
  \href{http://xxx.lanl.gov/abs/0909.0483}{{\tt arXiv:0909.0483}}.

\bibitem{ArkaniHamed:2009sx}
N.~Arkani-Hamed, J.~Bourjaily, F.~Cachazo, and J.~Trnka, {\it {Local Space-time
  Physics from the Grassmannian}},  {\em JHEP} {\bf 01} (2011) 108,
  [\href{http://xxx.lanl.gov/abs/0912.3249}{{\tt arXiv:0912.3249}}].

\bibitem{Nandan:2009cc}
D.~Nandan, A.~Volovich, and C.~Wen, {\it {A Grassmannian Etude in NMHV
  Minors}},  \href{http://xxx.lanl.gov/abs/0912.3705}{{\tt arXiv:0912.3705}}.

\bibitem{Drummond:2010uq}
J.~Drummond and L.~Ferro, {\it {The Yangian Origin of the Grassmannian
  Integral}},  {\em JHEP} {\bf 12} (2010) 010,
  [\href{http://xxx.lanl.gov/abs/1002.4622}{{\tt arXiv:1002.4622}}].

\bibitem{Ashok:2010ie}
S.~Ashok and E.~Dell'Aquila, {\it {On the Classificatino of Residues of the
  Grassmannian}},  \href{http://xxx.lanl.gov/abs/1012.5094}{{\tt
  arXiv:1012.5094}}.

\bibitem{Bargheer:2011mm}
T.~Bargheer, N.~Beisert, and F.~Loebbert, {\it {Exact Superconformal and
  Yangian Symmetry of Scattering Amplitudes}},
  \href{http://xxx.lanl.gov/abs/1104.0700}{{\tt arXiv:1104.0700}}.

\bibitem{Penrose:1986ca}
R.~Penrose and W.~Rindler, {\em Spinors and Space-Time}, vol.~2.
\newblock Cambridge University Press, 1986.

\bibitem{WardWells}
R.~Ward and R.~Wells, {\em {Twistor Geometry and Field Theory}}.
\newblock CUP, 1990.

\bibitem{HuggettTod}
S.~Huggett and P.~Tod, {\em {An Introduction to Twistor Theory}}.
\newblock Student Texts 4. London Mathematical Society, 1985.

\bibitem{MWbook}
L.~Mason and N.~Woodhouse, {\em {Integrability, Self-Duality and Twistor
  Theory}}.
\newblock OUP, 1997.

\bibitem{Dunajski:2009}
M.~Dunajski, {\em Solitons, instantons and Twistors}.
\newblock Oxford University Press, 2009.

\bibitem{Manin:1988ds}
Y.~Manin, {\em {Gauge Field Theory and Complex Geometry}}.
\newblock No.~289 in Grundlehren Der Mathematischen Wissenschaften. Springer,
  1988.

\bibitem{Cachazo:2005ga}
F.~Cachazo and P.~Svrcek, {\it {Lectures on twistor strings and perturbative
  Yang-Mills theory}},  {\em PoS} {\bf RTN2005} (2005) 004,
  [\href{http://xxx.lanl.gov/abs/hep-th/0504194}{{\tt hep-th/0504194}}].

\bibitem{Wolf:2010av}
M.~Wolf, {\it {A First Course on Twistors, Integrability and Gluon Scattering
  Amplitudes}},  {\em J.Phys.A} {\bf A43} (2010) 393001,
  [\href{http://xxx.lanl.gov/abs/1001.3871}{{\tt arXiv:1001.3871}}].

\bibitem{Atiyah:1978wi}
M.~F. Atiyah, N.~Hitchin, and I.~M. Singer, {\it {Self-duality in
  Four-Dimensional Riemannian Geometry}},  {\em Proc. Roy. Soc. Lond.} {\bf
  A362} (1978) 425--461.

\bibitem{Atiyah:1978ri}
M.~F. Atiyah, N.~Hitchin, V.~Drinfeld, and Y.~Manin, {\it {Construction of
  Instantons}},  {\em Phys. Lett.} {\bf A65} (1978) 185--187.

\bibitem{Atiyah:1981ey}
M.~F. Atiyah, {\it {Green's Functions for Self-Dual Four Manifolds}},  {\em
  Adv. Math. Supp.} {\bf 7A} (1981) 129--158.

\bibitem{Drummond:2008vq}
J.~M. Drummond, J.~Henn, G.~P. Korchemsky, and E.~Sokatchev, {\it {Dual
  Superconformal Symmetry of Scattering Amplitudes in $\cN=4$ Super-Yang-Mills
  Theory}},  {\em Nucl. Phys.} {\bf B828} (2010) 317--374,
  [\href{http://xxx.lanl.gov/abs/0807.1095}{{\tt arXiv:0807.1095}}].

\bibitem{Chalmers:1996rq}
G.~Chalmers and W.~Siegel, {\it {The Self-Dual Sector of {QCD} Amplitudes}},
  {\em Phys. Rev.} {\bf D54} (1996) 7628--7633,
  [\href{http://xxx.lanl.gov/abs/hep-th/9606061}{{\tt hep-th/9606061}}].

\bibitem{Sokatchev:1995nj}
E.~Sokatchev, {\it {An Action for N=4 supersymmetric selfdual Yang-Mills
  theory}},  {\em Phys.Rev.} {\bf D53} (1996) 2062--2070,
  [\href{http://xxx.lanl.gov/abs/hep-th/9509099}{{\tt hep-th/9509099}}].

\bibitem{Abe:2004ep}
Y.~Abe, V.~Nair, and M.-I. Park, {\it {Multigluon amplitudes, N =4 constraints
  and the WZW model}},  {\em Phys.Rev.} {\bf D71} (2005) 025002,
  [\href{http://xxx.lanl.gov/abs/hep-th/0408191}{{\tt hep-th/0408191}}].

\bibitem{Woodhouse:1985id}
N.~Woodhouse, {\it {Real Methods in Twistor Theory}},  {\em Class. Quant.
  Grav.} {\bf 2} (1985) 257--291.

\bibitem{Cachazo:2004kj}
F.~Cachazo, P.~Svrcek, and E.~Witten, {\it {MHV Vertices and Tree Amplitudes in
  Gauge Theory}},  {\em JHEP} {\bf 09} (2004) 006,
  [\href{http://xxx.lanl.gov/abs/hep-th/0403047}{{\tt hep-th/0403047}}].

\bibitem{Nair:1988bq}
V.~P. Nair, {\it {A Current Algebra for Some Gauge Theory Amplitudes}},  {\em
  Phys. Lett.} {\bf B214} (1988) 215.

\bibitem{Roiban:2004yf}
R.~Roiban, M.~Spradlin, and A.~Volovich, {\it {On the Tree-Level S-Matrix of
  Yang-Mills Theory}},  {\em Phys. Rev.} {\bf D70} (2004) 026009,
  [\href{http://xxx.lanl.gov/abs/hep-th/0403190}{{\tt hep-th/0403190}}].

\bibitem{Cachazo:2004zb}
F.~Cachazo, P.~Svrcek, and E.~Witten, {\it {Twistor Space Structure of One-Loop
  Amplitudes in Gauge Theory}},  {\em JHEP} {\bf 10} (2004) 074,
  [\href{http://xxx.lanl.gov/abs/hep-th/0406177}{{\tt hep-th/0406177}}].

\bibitem{Cachazo:2004by}
F.~Cachazo, P.~Svrcek, and E.~Witten, {\it {Gauge Theory Amplitudes in Twistor
  Space and Holomorphic Anomaly}},  {\em JHEP} {\bf 10} (2004) 077,
  [\href{http://xxx.lanl.gov/abs/hep-th/0409245}{{\tt hep-th/0409245}}].

\bibitem{Brandhuber:2004yw}
A.~Brandhuber, W.~Spence, and G.~Travaglini, {\it {One-loop Gauge Theory
  Amplitudes in $\cN=4$ Super Yang-Mills from MHV Vertices}},  {\em Nucl.
  Phys.} {\bf B706} (2005) 150--180,
  [\href{http://xxx.lanl.gov/abs/hep-th/0407214}{{\tt hep-th/0407214}}].

\bibitem{Risager:2005vk}
K.~Risager, {\it {A Direct Proof of the CSW Rules}},  {\em JHEP} {\bf 12}
  (2005) 003, [\href{http://xxx.lanl.gov/abs/hep-th/0508206}{{\tt
  hep-th/0508206}}].

\bibitem{Elvang:2008na}
H.~Elvang, D.~Z. Freedman, and M.~Kiermaier, {\it {Recursion Relations,
  Generating Functions and Unitarity Sums in $\cN=4$ SYM Theory}},  {\em JHEP}
  {\bf 04} (2009) 009, [\href{http://xxx.lanl.gov/abs/0808.1720}{{\tt
  arXiv:0808.1720}}].

\bibitem{Elvang:2008vz}
H.~Elvang, D.~Z. Freedman, and M.~Kiermaier, {\it {Proof of the MHV Vertex
  Expansion for All Tree Amplitudes in $\cN=4$ SYM Theory}},  {\em JHEP} {\bf
  06} (2009) 068, [\href{http://xxx.lanl.gov/abs/0811.3624}{{\tt
  arXiv:0811.3624}}].

\bibitem{Parke:1986gb}
S.~Parke and T.~Taylor, {\it {An Amplitude for $n$ Gluon Scattering}},  {\em
  Phys. Rev. Lett.} {\bf 56} (1986) 2459.

\bibitem{Hodges:2009hk}
A.~Hodges, {\it {Eliminating Spurious Poles from Gauge-Theoretic Amplitudes}},
  \href{http://xxx.lanl.gov/abs/0905.1473}{{\tt arXiv:0905.1473}}.

\bibitem{Boels:2010nw}
R.~Boels, {\it {On BCFW Shifts of Integrands and Integrals}},
  \href{http://xxx.lanl.gov/abs/1008.3101}{{\tt arXiv:1008.3101}}.

\bibitem{Drummond:2009fd}
J.~Drummond, J.~Henn, and J.~Plefka, {\it {Yangian Symmetry of Scattering
  Amplitudes in $\mathcal{N}=4$ Super Yang-Mills}},  {\em JHEP} {\bf 05} (2009)
  046, [\href{http://xxx.lanl.gov/abs/0902.2987}{{\tt arXiv:0902.2987}}].

\bibitem{ArkaniHamed:2010gh}
N.~Arkani-Hamed, J.~Bourjaily, F.~Cachazo, and J.~Trnka, {\it {Local Integrals
  for Planar Scattering Amplitudes}},
  \href{http://xxx.lanl.gov/abs/1012.6032}{{\tt arXiv:1012.6032}}.

\bibitem{Britto:2004ap}
R.~Britto, F.~Cachazo, and B.~Feng, {\it {New Recursion Relations for Tree
  Amplitudes of Gluons}},  {\em Nucl. Phys.} {\bf B715} (2005) 499--522,
  [\href{http://xxx.lanl.gov/abs/hep-th/0412308}{{\tt hep-th/0412308}}].

\bibitem{Britto:2005fq}
R.~Britto, F.~Cachazo, B.~Feng, and E.~Witten, {\it {Direct Proof Of Tree-Level
  Recursion Relation In Yang-Mills Theory}},  {\em Phys. Rev. Lett.} {\bf 94}
  (2005) 181602, [\href{http://xxx.lanl.gov/abs/hep-th/0501052}{{\tt
  hep-th/0501052}}].

\bibitem{AbouZeid:2006wu}
M.~Abou-Zeid, C.~Hull, and L.~Mason, {\it {Einstein Supergravity and New
  Twistor-String Theories}},  {\em Commun. Math. Phys.} (2008)
  [\href{http://xxx.lanl.gov/abs/hep-th/0606272}{{\tt hep-th/0606272}}].

\bibitem{Nair:2007md}
V.~P. Nair, {\it {A Note on Graviton Amplitudes for New Twistor String
  Theories}},  {\em Phys. Rev.} {\bf D78} (2008) 041501,
  [\href{http://xxx.lanl.gov/abs/0710.4961}{{\tt arXiv:0710.4961}}].

\bibitem{Brodel:2009ep}
J.~Broedel and B.~Wurm, {\it {New Twistor String Theories Revisited}},  {\em
  Phys. Lett.} {\bf B675} (2009) 463--468,
  [\href{http://xxx.lanl.gov/abs/0902.0550}{{\tt arXiv:0902.0550}}].

\bibitem{ArkaniHamed:2008gz}
N.~Arkani-Hamed, F.~Cachazo, and J.~Kaplan, {\it {What is the Simplest Quantum
  Field Theory?}},  \href{http://xxx.lanl.gov/abs/0808.1446}{{\tt
  arXiv:0808.1446}}.

\bibitem{Brandhuber:2008pf}
A.~Brandhuber, P.~Heslop, and G.~Travaglini, {\it {A Note on Dual
  Superconformal Symmetry of the $\mathcal{N}=4$ Super-Yang-Mills $S$-Matrix}},
   {\em Phys. Rev.} {\bf D78} (2008) 125005,
  [\href{http://xxx.lanl.gov/abs/0807.4097}{{\tt arXiv:0807.4097}}].

\bibitem{Korchemsky:2009jv}
G.~P. Korchemsky and E.~Sokatchev, {\it {Twistor Transform of All Tree
  Amplitudes in $\cN=4$ SYM Theory}},
  \href{http://xxx.lanl.gov/abs/0907.4107}{{\tt arXiv:0907.4107}}.

\bibitem{Hodges:2005aj}
A.~P. Hodges, {\it {Twistor Diagrams for All Tree Amplitudes in Gauge Theory: A
  Helicity-Independent Formalism}},
  \href{http://xxx.lanl.gov/abs/hep-th/0512336}{{\tt hep-th/0512336}}.

\bibitem{Hodges:2005bf}
A.~P. Hodges, {\it {Twistor Diagram Recursion for All Gauge-Theoretic Tree
  Amplitudes}},  \href{http://xxx.lanl.gov/abs/hep-th/0503060}{{\tt
  hep-th/0503060}}.

\bibitem{Bullimore:2010pa}
M.~Bullimore, {\it {Inverse Soft Factors and Grassmannian Residues}},  {\em
  JHEP} {\bf 01} (2011) 055, [\href{http://xxx.lanl.gov/abs/1008.3110}{{\tt
  arXiv:1008.3110}}].

\bibitem{ArkaniHamed:2010gg}
N.~Arkani-Hamed, J.~Bourjaily, F.~Cachazo, A.~Hodges, and J.~Trnka, {\it {A
  Note on Polytopes for Scattering Amplitudes}},
  \href{http://xxx.lanl.gov/abs/1012.6030}{{\tt arXiv:1012.6030}}.

\end{thebibliography}\endgroup
\bibliographystyle{JHEP}
\end{document}